\documentclass[11pt]{article}
\usepackage{microtype}
\usepackage[algo2e,ruled	,vlined]{algorithm2e}
\usepackage{amsmath,amssymb}
\usepackage{amsfonts,bm,bbm} 
\usepackage{amssymb,amsthm,enumitem}
\usepackage{tikz}
\usetikzlibrary{decorations.pathmorphing,positioning,patterns,snakes}
\usepackage{xcolor,graphicx,float,booktabs}
\usepackage{tabularx}
\usepackage{multirow}
\newcolumntype{L}[1]{>{\raggedright\arraybackslash}m{#1}}
\newcolumntype{C}[1]{>{\centering\arraybackslash}m{#1}}
\newcolumntype{R}[1]{>{\raggedleft\arraybackslash}m{#1}}

\usepackage{epstopdf}

\usepackage{verbatim}
\allowdisplaybreaks[4]
\usepackage{authblk}

\usepackage{hyperref}
\hypersetup{hypertex=true,
colorlinks=true,
linkcolor=blue,
anchorcolor=blue,
citecolor=blue}

\usepackage[round]{natbib}
\usepackage[
margin=1in,
includefoot,
footskip=30pt,
]{geometry}

\newcommand{\cE}{\mathcal{E}}

\newcommand{\cP}{\mathcal{P}}
\newcommand{\E}{\mathbb{E}}
\newcommand{\R}{\mathbb{R}}
\newcommand{\Var}{\mathrm{Var}}
\newcommand{\Cov}{\mathrm{Cov}}
\newcommand{\MSE}{\mathrm{MSE}}
\newcommand{\Bias}{\mathrm{Bias}}
\renewcommand{\P}{\mathbb{P}}

\newcommand{\1}{\mathbbm{1}}
\newcommand{\bbeta}{\boldsymbol{\beta}}

\newtheorem{theorem}{Theorem}
\newtheorem{proposition}[theorem]{Proposition}
\newtheorem{lemma}{Lemma}
\newtheorem{corollary}[theorem]{Corollary}
\newtheorem{example}{Example}
\newtheorem{assumption}{Assumption}

\title{A Correlation-induced Finite Difference Estimator
}

\author[1]{Guo Liang\footnote{liangguo000221@ruc.edu.cn}}
\author[2]{Guangwu Liu\footnote{msgw.liu@cityu.edu.hk}}
\affil[1]{Institute of Statistics and Big Data\protect\\Renmin University of China\protect\\Beijing, China}

\author[1]{Kun Zhang\footnote{kunzhang@ruc.edu.cn}}

\affil[2]{Department of Management Sciences\protect\\City University of Hong Kong\protect\\Tat Chee Avenue, Kowloon, Hong Kong, China}

\date{}

\begin{document}
\normalsize
\maketitle

\begin{abstract}

Finite difference (FD) approximation is a classic approach to stochastic gradient estimation when only noisy function realizations are available. In this paper, we first provide a sample-driven method via the bootstrap technique to estimate the optimal perturbation, and then propose an efficient FD estimator based on correlated samples at the estimated optimal perturbation. Furthermore, theoretical analyses of both the perturbation estimator and the FD estimator reveal that, {\it surprisingly}, the correlation enables the proposed FD estimator to achieve a reduction in variance and, in some cases, a decrease in bias compared to the traditional optimal FD estimator. Numerical results confirm the efficiency of our estimators and align well with the theory presented, especially in scenarios with small sample sizes. Finally, we apply the estimator to solve derivative-free optimization (DFO) problems, and numerical studies show that DFO problems with 100 dimensions can be effectively solved.
	
\end{abstract}

\section{Introduction}\label{sec:intro}

Stochastic gradient estimation is an active field in simulation optimization and machine learning \citep{Mohamed2020}. In realistic scenarios, as the systems become more complex, their closed-form expressions are often unknown, and systems' outputs can only be assessed using their stochastic realizations. Therefore, the derivative information about these systems is typically unavailable, characterizing this scenario as a black-box or zeroth-order model. In such settings, the FD approximation is a classic approach to gradient estimation \citep{Asmussen2007Stochastic,L1991Overview,Zazanis1993Convergence}.

FD approximation finds important applications in operations research and machine learning. A well-known application is sensitivity analysis for stochastic systems, such as queuing systems \citep{Fu2015Gradient}. In queueing systems, it is of interest that the sensitivity analysis for the average system  with respect to (w.r.t.) the input parameters, such as ones in the arrival and service time distributions \citep{Li2020Optimally,Lam2023Enhanced}.
 
Besides applications in sensitivity analysis, other important applications include solving the DFO problem where derivative information is unavailable. In practice, increasingly complex and diverse problems lack gradient information and only offer stochastic realizations of a function. Optimizing these problems has driven the development of DFO \citep{LarsonWild2019DFO}. For example, Google \citep{Golovin2017GoogleVizier} has established a Google-internal service called {\it Google Vizier} based on DFO methods for optimizing many machine learning methods and complex systems. In resent years, the use of FD estimators as surrogates for gradients in descent algorithms yields valid DFO algorithms, which has gained considerable attention because of the simplicity of building them and their of ease of parallelization \citep{berahas2019derivative,shi2023numerical}. 
In addition, a related research line is the simulation optimization via FD estimators \cite[see, e.g.,][]{Fu2002OvS}, which is originated from \cite{Kiefer1952Stochastic}. 
In this paper, we aim to establish an efficient FD estimator, servicing as a powerful tool for sensitivity analysis, applications in DFO, and so on.

FD methods encompass forward, backward, and central FD (CFD) methods. The primary challenge across all FD methods lies in determining the suitable perturbation. For example, increasing the perturbation simultaneously increases the bias and decreases the variance, making its selection crucial for minimizing the MSE ($= \mbox{bias}^2 + \mbox{variance}$). \cite{Fox1989Replication} point out that without employing common random numbers (CRNs), the optimal perturbation size of the CFD technique is $c n^{-1/6}$, corresponding to the optimal MSE of order $n^{-2/3}$, where $c$ is a constant, and $n$ is the number of sample pairs in the simulation. 
Although the order of the optimal perturbation corresponding to the various FD methods is known, the frontal constant, i.e., $c$, remains elusive. 
 
The constant $c$ in particular is notoriously difficult to estimate, due to their reliance on additional model information, such as higher-order derivatives and simulation noises \citep{Li2020Optimally}. These quantities are typically unknown, and even estimating them is a more challenging task than the gradient estimation itself. Empirical evidence suggests that different constants may lead to substantially different MSEs. 

In the vast literature on DFO and simulation optimization, where the FD estimator is used as a surrogate of the gradient, research on how to establish an asymptotically valid estimator of the optimal perturbation, and then the FD estimator itself, is rare. Typically, researchers tend to avoid estimating the constant $c$. Instead, in DFO literature, they find upper or lower bounds for higher-order derivatives and construct surrogates for these constants using these bounds \cite[see, e.g.,][]{shi2023numerical}. Meanwhile, in simulation optimization literature, only reasonable convergence rates of the perturbation and the FD estimator are guaranteed. Therefore, the former results in the FD estimator not converging to the true derivative, while the latter yields a suboptimal FD estimator \cite[see, e.g.,][]{Fu2015Gradient}.

In this paper, our aim is to establish an {\it efficient} FD estimator, in the sense of convergence to the true value with the optimal rate, sample-saving and practical implementation. Two crucial issues arise. The first is the estimation of $c$, which is fundamental for estimating the optimal perturbation. To estimate the constant consistently, the domain knowledge or prior knowledge, e.g., their upper or lower bounds, is insufficient. Therefore, some pilot samples need to be generated, and sample-driven methods should be proposed to estimate the constant based on these samples. However, since the ultimate goal is to obtain an FD estimator rather than a perturbation estimator, the number of pilot samples cannot be large. Thus, it has to be considered how to estimate the constants with relatively few samples. The second issue is establishing an efficient FD estimator. Given the constant estimator, the estimator of the optimal perturbation is obtained. A natural approach is to generate samples at the estimated optimal perturbation and average these samples to form the FD estimator. Intuitively, this estimator has the same asymptotic MSE as the traditional FD estimator. However, this approach is less efficient due to the under-utilization of samples, such as discarding pilot samples not generated at the estimated optimal perturbation. Therefore, effectively reusing these samples is crucial. In this paper, we propose novel methods to address these two issues.

\subsection{Main Contributions}

Specifically, the main contributions of this paper are as follows:

1. \textit{We propose a sample-driven method to estimate unknown constants and analyze their asymptotic properties.} We estimate the constants primarily by leveraging the functional relationships between the bias and variance of the FD estimator and the perturbation. To achieve this, we propose a sample-driven method that combines the bootstrap technique with these functional relationships. Using pilot samples, we establish the bootstrap bias and variance of the FD estimator, proving that they maintain the functional relationships. Therefore, the constants are estimated based on the bootstrap bias and variance, which further lead to an estimator of the optimal perturbation. In addition, we establish the convergence rate of these estimators, serving as a guideline for generating the pilot samples.

2. \textit{We construct correlated samples at the estimated optimal perturbation to establish an efficient FD estimator and prove its theoretical properties.} We innovatively transform pilot samples to resemble those generated at the estimated optimal perturbation, yielding correlated samples. Using these samples, we establish the efficient FD estimator, which we refer to as the {\it correlation-induced FD estimator} (Cor-FD). Although the correlation complicates theoretical analyses, it {\it surprisingly} results in the proposed estimator having a control variate configuration, thereby reducing variance and, in some scenarios, decreasing bias compared to the traditional optimal FD estimator. Notably, the total samples can be used as pilot samples, which leads to the estimator's potential availability even in cases of small sample sizes. 

3. \textit{We establish algorithms for the proposed estimator's practical implementation and application in solving DFO problems, demonstrating its practical benefits through numerical experiments.} Theoretical properties provide a guiding framework for practical implementation. Based on this, we propose a detailed algorithm to derive the proposed estimator. Furthermore, we apply the estimator in various numerical examples. Numerical results are consistent with the theoretical analysis, and illustrate that both the perturbation estimator and the FD estimator preform well compared to other approaches, particularly with small sample sizes. Additionally, by integrating the proposed Cor-FD estimator into a quasi-Newton algorithm, we develop an algorithm to solve DFO problems. This algorithm is compared numerically with some popular algorithms, highlighting the advantages of our algorithm over others.

\subsection{Literature Review}\label{sec:review}

In stochastic gradient estimation, there are direct and indirect approaches \citep{Fu2015Gradient}. The former includes infinitesimal perturbation analysis \citep{Ho1983Infinitesimal,Heidelberger1988Convergence}, likelihood ratio or score function \citep{Glynn1990Likelihood,Rubinstein1986Score,Reiman1989Sensitivity}, weak differentiation \citep{Heidergott2006Measure,Heidergott2010Gradient} and generalized likelihood ratio method \citep{Peng2018new}. These gradient estimation methods need to understand the underlying stochastics of the system. However, as systems become increasingly complex, it is often easier to experiment with them than to fully understand them \citep{Golovin2017GoogleVizier}. In such cases, indirect approaches, including the FD estimator we study in this paper, are more suitable.

The indirect gradient estimation means that the gradient is estimated solely from noisy function evaluations. These approaches include traditional FD \cite[see, e.g., Chapter 7 in][]{Glasserman2013Monte}\citep[Chapter 7 in][]{Asmussen2007Stochastic,L1991Overview,Fu2015Gradient}, Gaussian smoothing FD \citep{Nesterov2015Random}, random FD (RFD) \citep[see, e.g.,][]{salimans2017evolution,shamir2017optimal,fazel2018global}, etc. In these methods, all estimators are established via the average of independent samples, and the crucial aspect is determining the perturbation size (also known as the differencing interval in some literature like \cite{berahas2019derivative}). Typically, as the perturbation decreases, the variance of the estimator increases while the bias decreases. To select a suitable perturbation for the traditional FD estimator, \cite{Fox1989Replication} and \cite{Zazanis1993Convergence} tradeoff its bias and variance by minimizing its MSE, and then provide the optimal orders of the perturbation. However, constants in the perturbation are typically unknown, making the FD estimator optimal only in the sense of convergence rate. In contrast, our work proposes an FD estimator based on correlated samples, which not only achieves the same MSE convergence rate as the optimal traditional FD estimator but also exhibits variance reduction compared to the latter. 

Recently, to circumvent the estimation of the optimal perturbation, \cite{Lam2023Enhanced} propose minimizing the worst-case asymptotic ratio of the MSE of a proposed estimator to that of the traditional FD estimator w.r.t. model characteristics, including the perturbation. In this framework, a general weighted FD estimator is constructed by introducing a hyperparameter. In contrast, our work attempts to establish an FD estimator by directly estimating the optimal perturbation first.

The closest work to ours is \cite{Li2020Optimally}, which suggest a two-stage technique. In the first stage, pilot samples are used to estimate the optimal perturbation; in the second stage, additional samples are generated according to this estimated perturbation to construct an FD estimator. Although this two-stage technique yields an FD estimator with asymptotically optimal MSE, the under-utilization of samples, such as discarding the pilot samples, results in reduced efficiency. In our work, we propose a sample-driven method using the bootstrap technique based on some pilot samples inspired by by \cite{zhang2022bootstrap} to estimate the optimal perturbation. Then we indicate that all samples for establishing our proposed estimator can be used as pilot samples by transforming them to those at the estimated optimal perturbation. This twice sample-reusing procedure not only enhances the efficiency of samples but also potentially allows our proposed estimator to be practically implemented even with a small sample size.

An important study of FD approximation is its application in stochastic optimization \citep{ChauFu2014,Fu2015Gradient}. \cite{Kiefer1952Stochastic} introduce the earliest FD-based stochastic approximation (KW) algorithm, where the step lengths and perturbations must be carefully chosen. To improve the KW algorithm, \cite{broadie2011general} provide a scaled-and-shifted (for the step length and perturbation, respectively) KW algorithm. Recently, \cite{Chen2023OnlineLearning} apply this algorithm to solve an optimization problem in a queueing system. In multivariate settings, \cite{Spall1992Multivariate,Spall1997Multivariate} propose the simultaneous perturbation stochastic approximation (SPSA) algorithm. Recently, \cite{XuZheng2023JOC} propose a KW-type algorithm using the RFD estimator for solving multidimensional simulation optimization problems. \cite{hu2024convergence} demonstrate that to ensure the convergence of these algorithms, the rate of convergence of step lengths to zero must be faster than the increase in the variance of the FD estimator caused by the decrease in perturbations. 

Another research line of stochastic optimization via FD approximation is the so called DFO area \citep{berahas2019derivative,LarsonWild2019DFO,shi2023numerical}, where the step lengths are not required to tend to 0, and the bounded error of the FD estimator is typically assumed, so the algorithms’ results only converge to a neighborhood of the optimums. In contrast, the algorithm we establish for solve DFO problems is based on our proposed CFD estimator. 
To avoid digression, we only provide numerical results, rather than theoretical analyses, to demonstrate the algorithm's good performance, without assuming the convergence of step lengths to zero.

\subsection{Organization}

The rest of this paper is organized as follows. Section \ref{sec:background} gives some background on using the finite difference approximation for gradient estimation. In Section \ref{sec:bootstrap}, we propose a sample-driven method, apply it to perturbation estimation, and provide theoretical guarantees for it. Section \ref{sec:DSR} constructs correlated samples for establishing the efficient FD estimator, and provides theoretical evidence supporting its effectiveness. For practical implementation, a detailed algorithm is proposed in Section \ref{sec:implementation}, and we verify the theoretical results through numerical experiments in Section \ref{sec:experiments}. Section \ref{sec:DFO} applies our proposed FD estimator to solve DFO problems, followed by conclusions in Section \ref{sec:conclusions}. Proofs, discussions and additional results are provided in the appendix.

\section{Background}\label{sec:background}

In this section, we introduce the background of the finite difference method for stochastic gradient estimation. More details can be found in \cite{Asmussen2007Stochastic}, \cite{Fu2006Gradient} and \cite{Glasserman2013Monte}. Consider the models that depend on a single parameter $\theta$, where $\theta$ varies within some range\footnote{When $\Theta \in \R^d, d > 1$, we can repeat the same operation for each direction. Therefore, without loss of generality, we assume $\Theta \in \R$.} $\Theta \subset \R$. Denote that $\alpha(\cdot)$ is a performance measure of interest and assume that it can only be estimated by simulation. Within the simulation for any chosen $\theta \in \Theta$, each trail gives an unbiased but noisy estimator of $\alpha(\theta)$, denoted by $Y(\theta)$, i.e., $\alpha(\theta) = \E[Y(\theta)]$. Suppose that we do not use the CRNs, i.e., for $\theta_{1} \neq \theta_{2} \in \Theta$, $Y(\theta_{1})$ and $Y(\theta_{2})$ are independent. We would like to estimate the first-order derivative $\alpha'(\theta_0)$, where $\theta_0 \in \Theta$ is the point of interest. To estimate $\alpha'(\theta_{0})$, we prefer the CFD method over the forward (backward) FD method because compared to the latter, the CFD method yields an improvement in the convergence rate of the bias. It is worth mentioning that the methods in Sections \ref{sec:bootstrap} and \ref{sec:DSR} are seamlessly adapted to the forward (backward) FD estimator.

The CFD scheme is similar to the definition of the derivative and utilizes the information at the neighboring points on both sides of $\theta_0$. That is, $\alpha'(\theta)$ is approximated by 
\begin{align}\label{eq:numerical_derivative}
	\widetilde{\alpha}_h'(\theta_0) = \frac{\alpha(\theta_0 + h) - \alpha(\theta_0 - h)}{2h},
\end{align}
where $h$ is a perturbation parameter. As $h$ tends to 0, $\widetilde{\alpha}_h'(\theta_0)$ tends to $\alpha'(\theta_0)$. Denote
\begin{align}\label{def:Delta}
	\Delta(h) = \frac{Y(\theta_0 + h) - Y(\theta_0 - h)}{2h}.
\end{align}
Evidently, $\Delta(h)$ is a noisy estimator of $\alpha'(\theta_0)$ and this estimator is the output of the CFD method.

Specifically, the CFD scheme sets the perturbation parameter $h>0$. At $\theta_{0} + h$, the simulation is repeated independently, and $n$ independently and identically distributed (i.i.d.) observations $\{Y_{1}(\theta_{0} + h),...,Y_{n}(\theta_{0} + h)\}$ are obtained. Also, $n$ i.i.d. observations $\{Y_{1}(\theta_{0} - h),...,Y_{n}(\theta_{0} - h)\}$ are simulated at $\theta_{0} - h$. Using the sample average, we construct a CFD estimator of $\alpha'(\theta_{0})$, denoted by
\begin{align}\label{eq:finite_difference_estimate}
	\widehat{\Delta}_{n,h} = \frac{1}{n}\sum_{i=1}^{n}\Delta_i(h) = \frac{1}{n}\sum_{i=1}^{n} \frac{Y_{i}(\theta_{0} + h) - Y_{i}(\theta_{0} - h)}{2h}.
\end{align}
For any $i = 1,...,n$, $Y_i(\theta_0 + h)$ and $Y_i(\theta_0 - h)$ are the unbiased estimators of $\alpha(\theta_0 + h)$ and $\alpha(\theta_0 - h)$, respectively, so $\Delta_i(h)$ and $\widehat{\Delta}_{n,h}$ are both the unbiased estimators of $\widetilde{\alpha}_h'(\theta_0)$. 
Let $Z_i(h)$ denote the zero-mean error associated with $\Delta_i(h)$, and let $\bar{Z}_{n,h} = \sum_{i=1}^{n} Z_i(h)/n$ be the zero-mean error associated with $\widehat{\Delta}_{n,h}$. Then,
\begin{align}\label{eq:estimate_expression}
	\widehat{\Delta}_{n,h} = \widetilde{\alpha}_h'(\theta_0) + \bar{Z}_{n,h}.
\end{align}

The choice of $h$ is crucial when constructing $\widehat{\Delta}_{n,h}$. If $h$ is too large, the bias of $\widehat{\Delta}_{n,h}$ will be large, else if $h$ is too small, the variance of $\widehat{\Delta}_{n,h}$ will explode. A proper $h$ should balance the squared bias and variance of $\widehat{\Delta}_{n,h}$. To calculate the bias and variance elaborately, we make the following assumptions.
\begin{assumption}\label{ass:derivative3}
	$\alpha(\theta)$ is thrice continuously differentiable in a neighborhood of $\theta_0$ and $\alpha^{(3)}(\theta_0) \neq 0$.
\end{assumption}

\begin{assumption}\label{ass:continuity}
	The standard deviation of $Y(\theta)$, denoted by $\sigma(\theta)$, is continuous at $\theta_{0}$ and positive.
\end{assumption}

Assumptions \ref{ass:derivative3} and \ref{ass:continuity} can be found in many other works, such as \cite{Glasserman2013Monte}. Under Assumption \ref{ass:derivative3}, according to the Taylor expansion, we have
\begin{align*}
\alpha(\theta_0 + h) = \alpha(\theta_0) + \alpha'(\theta_0) h+ \frac{\alpha^{(2)}(\theta_0)}{2} {h}^2 + \frac{\alpha^{(3)}(\theta_0)}{6} {h}^3 + o(h^3), \\
\alpha(\theta_0 - h) = \alpha(\theta_0) - \alpha'(\theta_0) h+ \frac{\alpha^{(2)}(\theta_0)}{2} h^2 - \frac{\alpha^{(3)}(\theta_0)}{6} h^3 + o(h^3).
\end{align*}
Subtraction eliminates $\alpha^{(2)}(\theta_0)$, leaving $\widetilde{\alpha}_h'(\theta_0) = \alpha'(\theta_0) + B h^2 + o(h^2)$, where $B = \alpha^{(3)}(\theta_0)/6$. Therefore, the bias of $\widehat{\Delta}_{n,h}$ is
\begin{align}\label{eq:estimate_bias}
	\Bias\left[\widehat{\Delta}_{n,h}\right] = \E\widehat{\Delta}_{n,h} - \alpha'(\theta_0) = \widetilde{\alpha}_h'(\theta_0) - \alpha'(\theta_0) = B h^2 + o(h^2).
\end{align}

From \eqref{eq:finite_difference_estimate}, the variance of $\widehat{\Delta}_{n,h}$ is
\begin{align*}
	\Var\left[\widehat{\Delta}_{n,h}\right] = \frac{1}{n}\Var\left[\frac{Y(\theta_0 + h) + Y(\theta_0 - h)}{2h}\right] = \frac{\sigma^2(\theta_0 + h) + \sigma^2(\theta_0 - h)}{4nh^2}.
\end{align*}
Under Assumption \ref{ass:continuity}, we have $\sigma^2(\theta_0 + h) = \sigma^2(\theta_0) + o(1)$ and $\sigma^2(\theta_0 - h) = \sigma^2(\theta_0) + o(1)$. Therefore,
\begin{align}\label{eq:estimate_variance}
	\Var\left[\widehat{\Delta}_{n,h}\right] = \frac{\sigma^2(\theta_0) + o(1)}{2nh^2}.
\end{align}
Then the MSE of $\widehat{\Delta}_{n,h}$ is 
\begin{align}\label{eq:estimate_MSE}
	\mbox{MSE}\left[\widehat{\Delta}_{n,h}\right] = \Bias^2\left[\widehat{\Delta}_{n,h}\right] + \Var\left[\widehat{\Delta}_{n,h}\right] = \left(B + o(1)\right)^2 h^4 + \frac{\sigma^2(\theta_0) + o(1)}{2nh^2}.
\end{align}

It follows from \eqref{eq:estimate_MSE} that $h$ controls a trade-off between the bias and variance. Minimizing the MSE w.r.t. $h$ leads to the optimal perturbation 
\begin{align}\label{eq:optimal_tune}
	h^* = \left(\frac{\sigma^2(\theta_{0})}{4nB^2}\right)^{1/6}.
\end{align}
Thus the optimal bias, variance and MSE are
\begin{align}\label{eq:optimal_bias_variance_MSE}
\begin{cases}
		\mbox{Bias}^* = \left(\dfrac{B\sigma^2(\theta_{0})}{4n}\right)^{1/3} + o\left(n^{-1/3}\right),\\
		\Var^* = \left(\dfrac{B^2 \sigma^4(\theta_0)}{2n^2}\right)^{1/3} + o\left(n^{-2/3}\right),\\
		\mbox{MSE}^* = 3\left(\dfrac{B^2 \sigma^4(\theta_0)}{16n^2}\right)^{1/3} + o\left(n^{-2/3}\right).
\end{cases}
\end{align}

Theoretically, $h^*$ in \eqref{eq:optimal_tune} yields an optimal $\widehat{\Delta}_{n,h^*}$. Although we know that $h^* = C n^{-1/6}$ from \eqref{eq:optimal_tune}, where $C=\left(\sigma^2(\theta_{0})/(4B^2)\right)^{1/6}$, the constant $C$ is typically unknown. The definition of $C$ indicates that its estimation depends on the model information, such as $\alpha^{(3)}(\theta_0)$ and $\sigma^2(\theta_0)$. However, estimating these values is more challenging than estimating the gradient itself, because it involves the higher-order derivatives of $\alpha(\theta)$.  Furthermore, experimental findings indicate that assigning a perturbation with a significant deviation from the true one leads to substantial discrepancies in the MSE of the CFD estimator. These motivate us to explore an approach to estimate $h^*$.

\section{A Sample-driven Method for Constant Estimation}\label{sec:bootstrap}

In this section, we propose a sample-driven method to estimate $C$, specifically $B$ and $\sigma^2(\theta_0)$. Our inspiration comes from \eqref{eq:estimate_bias} and \eqref{eq:estimate_variance}, which demonstrate that $\E\widehat{\Delta}_{n,h}$ and $\Var\left[\widehat{\Delta}_{n,h}\right]$ are linear w.r.t. $h^2$ and $1/h^2$, respectively. If we can accurately calculate $\E\widehat{\Delta}_{n,h}$ and $\Var\left[\widehat{\Delta}_{n,h}\right]$ for any given $h$, then choosing $h_1 \neq h_2$ and solving the system of linear equations 
\begin{align}\label{eq:linear}
\begin{cases}
	\E\widehat{\Delta}_{n,h_1} - \alpha'(\theta_0) = B h_1^2 + o(h_1^2),\\
	\E\widehat{\Delta}_{n,h_2} - \alpha'(\theta_0) = B h_2^2 + o(h_2^2),
\end{cases}
\end{align}
leads to an estimator of $B$. This estimator is noise-free and its deviation from the true value tends to 0 as $h_1, h_2 \to 0$ because the convergence rate of $o(h_1^2)$ and $o(h_2^2)$ is faster than that of $h_1^2$ and $h_2^2$, respectively. Likewise, $\sigma^2(\theta_0)$ can be estimated by $2nh_1^2\Var\left[\widehat{\Delta}_{n,h_1}\right]$.

However, $\E\widehat{\Delta}_{n,h}$ and $\Var\left[\widehat{\Delta}_{n,h}\right]$ are typically unknown, and estimating them may be as challenging as estimating $\alpha'(\theta_0)$. Therefore, we turn to seek their surrogates. A straightforward approach is to use sample mean and variance, respectively. That is, we generate several sample versions of $\widehat{\Delta}_{n,h}$, denoted by $\left\{\widehat{\Delta}_{n,h}(l), l = 1,2,...,p\right\}$, and then use their sample mean and variance as the surrogates of $\E\widehat{\Delta}_{n,h}$ and $\Var\left[\widehat{\Delta}_{n,h}\right]$, respectively. Nevertheless, for any perturbation $h$, we need $np$ sample pairs
to obtain these surrogates, which is unavailable due to the limitation of sample size. To address this limitation, we introduce a resampling technique, the bootstrap, to construct the sample mean and variance.

On the other hand, our primary objective is to estimate $\alpha'(\theta_0)$ rather than determining the optimal perturbation $h^*$. Consequently, the number of samples used for estimating $h^*$ is relatively small compared to the total sample size. As a result, the bootstrap sample means as surrogates of the expectations in \eqref{eq:linear} may not provide sufficient accuracy to yield a relatively precise estimator of $B$ by solving the two equations in \eqref{eq:linear}. To address this issue, we select $K$ different values of $h$, expanding \eqref{eq:linear} to encompass $K$ linear equations correspondingly. Then least-squares regression is applied to derive the slope of linear equations, i.e., the constant $B$. Similar procedure is also used to obtain the estimate of $\sigma^2(\theta_0)$.

In the following, we introduce the sample-driven method, including the bootstrap technique for estimating sample mean and variance and the least-squares regression procedure for estimating the constants, along with its theoretical guarantees.

\subsection{Bootstrap Sample Mean and Variance}

In this subsection, we apply the bootstrap technique to establish surrogates of $\E\widehat{\Delta}_{n,h}$ and $\Var\left[\widehat{\Delta}_{n,h}\right]$ based on some pilot samples, and we provide their corresponding theoretical analyses.

The bootstrap technique is described as follows. For a fixed $h$, assume that $n_b$ sample pairs have been generated, denoted by $\{(Y_i(\theta_0 +h),Y_i(\theta_0 - h)), i = 1,2,...,n_b\}$. Correspondingly, as defined in \eqref{def:Delta}, we have $n_b$ outputs $\boldsymbol{\Delta}(h) \triangleq  \{\Delta_i(h), i = 1,2,...,n_b\}$. We pick with replacement $n_b$ times from $\boldsymbol{\Delta}(h)$ independently and randomly and then get a group of bootstrap samples $\boldsymbol{\Delta}^{*}(h) = \{\Delta_{i}^{*}(h), i = 1,2,...,n_b\}$. Then, the estimator of $\alpha'(\theta_0)$ is expressed as
\begin{align*}
	\widehat{\Delta}^{b}_{n_b,h} = \frac{1}{n_b}\sum_{i=1}^{n_b}\Delta_{i}^{*}(h).
\end{align*}

It is worth mentioning that in this procedure, the bootstrap samples come from the same pilot samples (the $n_b$ sample pairs) and the generation of extra samples is not required. In essence, the bootstrap technique is based on the reuse of the pilot samples. 

It is natural to expect that the bootstrap mean and variance of $\widehat{\Delta}^{b}_{n_b,h}$ will exhibit similar asymptotic properties as described in \eqref{eq:estimate_bias} and \eqref{eq:estimate_variance}, respectively, when $n_b$ is sufficiently large. 
Let $\E_{*}$ and $\Var_{*}$ denote the expectation and variance under the bootstrap probability measure $\P_{*}(\cdot) \triangleq \P(\cdot|\boldsymbol{\Delta}(h))$. Note that $\P_{*}$ is a probability conditional on pilot samples. Also, $\P_{*}$ is a discrete probability distribution, and its probability mass function is shown in Table \ref{tab:pmf}.
\begin{table}[htbp!]
\fontsize{11pt}{13.6pt}\selectfont
\centering
\caption{The values and probabilities under $\P_{*}$.}
\smallskip
\label{tab:pmf}
\small
\renewcommand{\arraystretch}{0.85}
\begin{tabular}{lcccccccc}
\toprule
Value & &$\Delta_{1}(h)$ & &$\Delta_{2}(h)$ & &$\cdots$ & &$\Delta_{n_b}(h)$\\
\midrule
Probability & &$1/n_b$ & &$1/n_b$ & &$\cdots$ & &$1/n_b$\\
\bottomrule
\end{tabular}
\end{table}

Asymptotic properties of the bootstrap mean and variance of $\widehat{\Delta}^{b}_{n_b,h}$ are summarized in Theorem \ref{thm:bootstrap}. Note that in the context of simulations, obtaining samples is typically expensive, while bootstrap operations are relatively cheap. Therefore, we adopt the analysis of bootstrap method from the field of statistics \citep{shao2012jackknife}, i.e., theoretically studying $\E_{*}\widehat{\Delta}^{b}_{n_b,h}$ and $\Var_{*}\left[\widehat{\Delta}^{b}_{n_b,h}\right]$ and using their Monte Carlo versions in practical applications.

Before going further, we make the following Assumption \ref{ass:derivative5}, which helps us examine the term $o(h^2)$ in \eqref{eq:estimate_bias} and further assists us in analyzing the estimation error of $B$ and setting the initial perturbations for least-squares regression.
\begin{assumption}\label{ass:derivative5}
	$\alpha(\theta)$ is five times continuously differentiable in a neighborhood of $\theta_0$ and $\alpha^{(5)}(\theta_0) \neq 0$.
\end{assumption}
In many practical problems, the function $\alpha(\theta)$ is five times continuously differentiable. For example, in the $M/M/1$ queue, the average waiting time is a smooth function of the arrival and service rate \citep[see,][]{Kleinrock1996Queueing}. Now we provide the theorem and the proof is provided in Appendix \ref{app:prooftheorem1}.
\begin{theorem}\label{thm:bootstrap}
	Denote $\nu_4(h) = \E\left[Z_1(h)\right]^4$ and $\nu_4 = \underset{h\to 0}{\lim}\E\left[hZ_1(h)\right]^4$. If $\nu_4 < \infty$, then under Assumptions \ref{ass:continuity} and \ref{ass:derivative5}, as $n_b\to\infty$,
	\begin{align}
		&\E_{*}\widehat{\Delta}^{b}_{n_b,h} = \alpha'(\theta_0) + B h^2 + D h^4 + o(h^4) + \bar{Z}_{n_b,h}, \label{eq:bootstrap_bias}\\
		&\Var_{*}\left[\widehat{\Delta}^{b}_{n_b,h}\right] = \frac{(n_b - 1)(\sigma^2(\theta_0) + o(1))}{2n_b^2 h^2} + \phi(h),\label{eq:bootstrap_variance}
	\end{align}
	where $D = \alpha^{(5)}(\theta_0)/120$ and $\phi(h)$ is a zero-mean error with
	\begin{align}\label{eq:Var_phi_h}
		\Var[\phi(h)] = \frac{(n_b-1)^2}{n_b^4h^4} \left(\frac{\nu_4 + o(1)}{n_b} - \frac{n_b - 3}{n_b(n_b - 1)}\frac{\sigma^4(\theta_0) + o(1)}{4}\right).
	\end{align}
\end{theorem}

Despite the relatively small size of $n_b$ compared to the total sample size, Theorem \ref{thm:bootstrap} provides valuable insights into the properties of $\E_{*}\widehat{\Delta}^b_{n_b,h}$ and $\Var_{*}\left[\widehat{\Delta}^b_{n_b,h}\right]$. Specifically, it demonstrates that $\E_{*}\widehat{\Delta}^b_{n_b,h}$ and $\Var_{*}\left[\widehat{\Delta}^b_{n_b,h}\right]$ inherit the asymptotic properties of $\E\widehat{\Delta}_{n,h}$ and  $\Var\left[\widehat{\Delta}_{n,h}\right]$, respectively. That is, they are linear w.r.t. $h^2$ and $1/h^2$, respectively, and their slopes are respectively equal. As a result, we can estimate $B$ and $\sigma^2(\theta_0)$ by leveraging these linear relationships, which we will detail in the following subsection.

\subsection{Regression for Perturbation Selection}

As mentioned at the beginning of Section \ref{sec:bootstrap}, to obtain relatively precise estimators of $B$ and $\sigma^2(\theta_0)$, multiple equations like \eqref{eq:linear} have to be constructed, and least-squares regression is used to derive the slopes of these linear equations, i.e., the constants $B$ and $\sigma^2(\theta_0)$.

In our setting, we consider $K$ perturbation parameters $h_1,h_2,...,h_{K}$, where $K$ is finite. Using $K n_b$ pilot samples and \eqref{eq:bootstrap_bias} in Theorem \ref{thm:bootstrap}, we construct a system of $K$ linear equations like \eqref{eq:linear}, and we consider the vector version as follows.
\begin{align}\label{eq:regression_problem_bias}
	\boldsymbol{Y}_e &= \boldsymbol{X}_e \bbeta_e + \cE_e,
\end{align}
where
\begin{equation}\label{eq:regression_component_bias}
\begin{aligned}
&\quad \quad \quad \boldsymbol{Y}_e = \left[\E_{*}\widehat{\Delta}^b_{n_b,h_1},...,\E_{*}\widehat{\Delta}^b_{n_b,h_K}\right]^{\top}, 
\boldsymbol{X}_e = \left[\begin{array}{lll}
1 & ... & 1 \\
h_1^2 & ... & h_{K}^2
\end{array}\right]^{\top}, \\
&\bbeta_e = \left[\alpha'(\theta_0), B\right]^{\top}, 
\cE_e = \left[D h_1^4 + o(h_1^4) + \bar{Z}_{n_b,h_1},...,D h_{K}^4 + o(h_{K}^4) + \bar{Z}_{n_b,h_K}\right]^{\top}.
\end{aligned}
\end{equation}

Similarly, by \eqref{eq:bootstrap_variance} in Theorem \ref{thm:bootstrap}, the system of $K$ linear equations for estimating $\sigma^2(\theta_0)$ is
\begin{align}\label{eq:regression_problem_variance}
	\boldsymbol{Y}_v &= \boldsymbol{X}_v \bbeta_v + \cE_v,
\end{align}
where
\begin{equation}\label{eq:regression_component_variance}
\begin{aligned}
&\boldsymbol{Y}_v = \left[\Var_{*}\left[\widehat{\Delta}^b_{n_b,h_1}\right],...,\Var_{*}\left[\widehat{\Delta}^b_{n_b,h_K}\right]\right]^{\top}, 
\ \boldsymbol{X}_v = \left[\frac{n_b-1}{2n_b^2h_1^2},...,\frac{n_b-1}{2n_b^2h_{K}^2}\right]^{\top}, \\
&\bbeta_v = \sigma^2(\theta_0), \ 
\cE_v = \left[\frac{(n_b-1)o(1)}{2n_b^2h_1^2} + \phi(h_1),...,\frac{(n_b-1)o(1)}{2n_b^2h_{K}^2} + \phi(h_K)\right]^{\top}.
\end{aligned}
\end{equation}

Using the least-squares method, we derive the estimators of $\bbeta_e$ and $\bbeta_v$, denoted by $\widehat{\bbeta}_e$ and $\widehat{\bbeta}_v$, respectively, where
\begin{align*}
	\widehat{\bbeta}_e = \left(\boldsymbol{X}_e^{\top} \boldsymbol{X}_e\right)^{-1} \boldsymbol{X}_e^{\top} \boldsymbol{Y}_e, \quad \widehat{\bbeta}_v = \left(\boldsymbol{X}_v^{\top} \boldsymbol{X}_v\right)^{-1} \boldsymbol{X}_v^{\top} \boldsymbol{Y}_v.
\end{align*}

Let the first term of $\widehat{\bbeta}_e$ be $\widehat{\alpha}'(\theta_0)$ and the second term be $\widehat{B}$. In the following, we show that $\widehat{\bbeta}_e$ and $\widehat{\bbeta}_v$ are consistent estimators of $\bbeta_e$ and $\bbeta_v$, respectively.

\subsubsection{Consistency of $\widehat{B}$}

The following theorem presents the consistency of $\widehat{B}$, whose proof is provided in Appendix \ref{app:prooftheorem2}. The conditions about $h_k (k = 1,...,K)$ in Theorem \ref{thm:B_consistence} ensure that the regression error vanishes fast enough, which is necessary for the consistency of regression estimators. 

\begin{theorem}\label{thm:B_consistence}
	Suppose that Assumptions \ref{ass:derivative3} and \ref{ass:derivative5} hold. For any $k = 1,...,K (K \geq 2)$, denote $h_k = c_k n_b^{\gamma} (c_k \neq 0, \gamma<0)$ and for any $j \neq k$, let $c_j \neq c_k$. Then,
	\begin{align}\label{eq:B_bias_variance}
		\E\left[\widehat{B}\right] - B = H_Kn_b^{2\gamma} + o\left(n_b^{2\gamma}\right),\quad \Var\left[\widehat{B}\right] = V_K \frac{\sigma^2(\theta_0) + o(1)}{2n_b^{1+6\gamma}},
	\end{align}
where $H_K$ and $V_K$ are constants depending on $c_1,...,c_K$. Specifically,
	\begin{align*}
		H_K = \frac{K \sum_{k=1}^K Dc_k^6 - \sum_{k=1}^K c_k^2 \sum_{k=1}^K D c_k^4}{K \sum_{k=1}^K c_k^4-\left(\sum_{k=1}^K c_k^2\right)^2},\quad V_K  = \frac{- K^2\sum_{k=1}^K c_k^2 + \left(\sum_{k=1}^{K}c_k^2\right)^2\sum_{k=1}^{K}1/c_k^2}{\left(K \sum_{k=1}^K c_k^4-\left(\sum_{k=1}^K c_k^2\right)^2\right)^2}.
	\end{align*}
	In addition, if $-1/6 < \gamma < 0$, then $\widehat{B} \stackrel{p}{\longrightarrow} B$, where $\stackrel{p}{\longrightarrow}$ means convergence in probability.
\end{theorem}

Theorem \ref{thm:B_consistence} indicates that for all $h_k = c_k n_b^\gamma$, where $k = 1,...,K$ and $c_k \neq 0$, as long as $-1/6 < \gamma < 0$, $\widehat{B}$ is a consistent estimator of $B$. Furthermore, from \eqref{eq:B_bias_variance}, the convergence rate of the bias of $\widehat{B}$ is $n_b^{2\gamma}$, and that of the variance of $\widehat{B}$ is $n_b^{-1 - 6\gamma}$. In addition, Theorem \ref{thm:B_consistence} provides a valuable guidance for the selection of appropriate values of $h_k$ ($k=1,...,K$) during the pilot stage.

Except for fixed values of $c_1,...,c_K$, they can be generated from a proper distribution $\mathcal{P}_0$. In this case, $H_K$ and $V_K$ are both random. Since $c_1,...,c_K$ are coefficients in the perturbations, they must not equal 0, and are typically bounded. For these purposes, there are many optional distributions, such as the truncated normal distribution, denoted by $\psi\left(\mu_0, \sigma_0^2,L,U\right)$. Specifically, for any $k = 1,...,K$, let $c_k \stackrel{d}{=} X \1\{L \leq X \leq U\}$, where $\stackrel{d}{=}$ represents the same distribution, $X \sim \mathcal{N}\left(\mu_0, \sigma_0^2\right)$, and $0 < L < U$ denote the lower and upper bounds, respectively. Because $0 < L \leq c_k \leq U$, $H_K$ and $V_K$ are bounded.

When we generate $c_1,...,c_K$ from $\mathcal{P}_0$, the bias of $\widehat{B}$ is
\begin{align}\label{eq:B_bias_random}
	\mbox{Bias}\left[\widehat{B}\right] = \E\left[\E\left[\widehat{B} - B\big|\mathcal{P}_0\right]\right] = \E[H_K]n_b^{2\gamma} + o\left(n_b^{2\gamma}\right),
\end{align}
where the last equality is because $c_1,...,c_K$ are independent with the pilot samples and due to the bias result in Theorem \ref{thm:B_consistence}. If $c_k\sim \psi\left(\mu_0, \sigma_0^2,L,U\right)$ for $k=1,...,K$, then $c_k$'s are bounded. Therefore, $H_K$ and the constants in $o\left(n_b^{2\gamma}\right)$ are all bounded.

Likewise, the variance of $\widehat{B}$ is
\begin{align}\label{eq:B_variance_random}
	\Var\left[\widehat{B}\right] &= \Var\left[\E\left[\widehat{B}\big|\mathcal{P}_0\right]\right] + \E\left[\Var\left[\widehat{B}\big|\mathcal{P}_0\right]\right]\nonumber \\
	&= \Var\left[B + H_Kn_b^{2\gamma} + o\left(n_b^{2\gamma}\right)\right] + \E\left[V_K \frac{\sigma^2(\theta_0) + o(1)}{2n_b^{1+6\gamma}}\right]\nonumber\\
	&= \Var[H_K]n_b^{4\gamma} + o\left(n_b^{4\gamma}\right) + \E[V_K]\frac{\sigma^2(\theta_0) + o(1)}{2n_b^{1+6\gamma}}.
\end{align}

Comparing \eqref{eq:B_bias_variance} with \eqref{eq:B_bias_random} and \eqref{eq:B_variance_random}, it becomes evident that when $c_k$'s are generated from $\cP_0$, the bias of $\widehat{B}$ still converges at a rate of $n_b^{2\gamma}$, while the variance of $\widehat{B}$ is increased by $n_b^{4\gamma}$. However, it is important to note that the convergence rate of the MSE of $\widehat{B}$ remains unchanged. Therefore, during the pilot stage, the $h_k$'s selected based on the tradeoff between bias and variance in \eqref{eq:B_bias_variance} maintain the same order as those selected using the tradeoff between bias and variance in \eqref{eq:B_bias_random} and \eqref{eq:B_variance_random}.

Note that when estimating $B$, the first element of the vector $\widehat{\bbeta}_e$ is a rough estimator of $\alpha'(\theta_0)$, denoted by $\widehat{\alpha}'(\theta_0)$. In Appendix \ref{app:property_alpha}, Corollary \ref{cor:alpha_consistence} establishes the consistency of $\widehat{\alpha}'(\theta_0)$. However, directly using $\widehat{\alpha}'(\theta_0)$ as the FD estimator is not recommended, as it is based on a small sample size of $Kn_b$. Despite this, it is evident that $\widehat{\alpha}'(\theta_0)$ contains valuable gradient information, which can be leveraged to construct the correlation-induced FD estimator in Section \ref{sec:DSR}. Corollary \ref{cor:alpha_consistence} also plays a crucial role in proving the consistency of this new estimator.

\subsubsection{Consistency of \mbox{\boldmath $\widehat\beta$}$_v$}

To conduct a more precise analysis of the convergence of $\widehat{\bbeta}_v$, we provide the following Assumption \ref{ass:derivative1}.

\begin{assumption}\label{ass:derivative1}
	$\sigma(\theta)$ is continuously differentiable at $\theta_{0}$ and $\sigma'(\theta_0)$ is bounded.
\end{assumption}

Under Assumption \ref{ass:derivative1}, we are able to refine \eqref{eq:bootstrap_variance} as follows, 
\begin{align}\label{eq:bootstrap_variance_new}
	\Var_{*}\left[\widehat{\Delta}^{b}_{n_b,h}\right] = \frac{(n_b - 1)\left[\sigma^2(\theta_0) + \sigma'(\theta_0)^2h^2 + o(h^2)\right]}{2n_b^2 h^2} + \phi(h).
\end{align}
The proof of \eqref{eq:bootstrap_variance_new} is provided in Appendix \ref{app:eq:bootstrap_variance_new}. It follows from \eqref{eq:bootstrap_variance_new} that $\cE_v$ in \eqref{eq:regression_component_variance} is expressed as
\begin{align}\label{eq:regression_component_variance_new}
\cE_v = \left[\frac{(n_b-1)\left[\sigma'(\theta_0)^2h_1^2 + o(h_1^2)\right]}{2n_b^2h_1^2} + \phi(h_1),...,\frac{(n_b-1)\left[\sigma'(\theta_0)^2h_K^2 + o(h_K^2)\right]}{2n_b^2h_{K}^2} + \phi(h_K)\right]^{\top}.
\end{align}
Evidently, under Assumption \ref{ass:derivative1}, the terms $o(1)$'s in \eqref{eq:bootstrap_variance} and \eqref{eq:regression_component_variance} are explicitly expressed, which helps us analyze the bias and variance of $\widehat{\bbeta}_v$ more precisely.

Now we provide Theorem \ref{thm:sigma_consistence} which demonstrates the convergence rate of the bias and variance of $\widehat{\bbeta}_v$. The proof of Theorem \ref{thm:sigma_consistence} is provided in Appendix \ref{app:prooftheorem3}.
\begin{theorem}\label{thm:sigma_consistence}
	Suppose that Assumptions \ref{ass:continuity} and \ref{ass:derivative1} hold. For any $k = 1,...,K$ $(K \geq 1)$, denote $h_k = c_k n_b^{\gamma}$ $(c_k \neq 0, \gamma<0)$ and for any $j \neq k$, $c_j \neq c_k$. Then, as $n_b\to\infty$,
	\begin{align*}
		\E\left[\widehat{\bbeta}_v\right] - \bbeta_v = \widehat{H}_Kn_b^{2\gamma} + o\left(n_b^{2\gamma}\right), \quad \Var\left[\widehat{\bbeta}_v\right] = \widehat{V}_K \frac{4\nu_4(n_b-1) - \sigma^4(\theta_0)(n_b-3)}{n_b(n_b-1)} + o\left(\frac{1}{n_b}\right),
	\end{align*}
	where $\widehat{H}_K$ and $\widehat{V}_K$ are constants depending on $c_1,...,c_K$. Specifically,
	\begin{align*}
		\widehat{H}_K = \sigma'(\theta_0)^2\left(\sum_{k=1}^K\frac{1}{c_k^4}\right)^{-1}\sum_{k=1}^K\frac{1}{c_k^2}, \quad \widehat{V}_K = \left(\sum_{k=1}^K\frac{1}{c_k^4}\right)^{-2}\sum_{k=1}^{K}\frac{1}{c_k^8}.
	\end{align*}
	Consequently, we have $\widehat{\bbeta}_v \stackrel{p}{\longrightarrow} \bbeta_v$.
\end{theorem}

Theorem \ref{thm:sigma_consistence} indicates that for all $h_k = c_k n_b^\gamma$, where $k = 1,...,K$ and $c_k \neq 0$, as long as $\gamma < 0$, $\widehat{\bbeta}_v$ is a consistent estimator of $\bbeta_v$. The convergence rate of the variance of $\widehat{\bbeta}_v$ is $n_b^{-1}$, which is independent of $\gamma$ and faster than that of $\widehat{B}$. Additionally, the convergence rate of the bias of $\widehat{\bbeta}_v$ is $n_b^{2\gamma}$, which is equal to that of $\widehat{B}$.

\subsection{The Optimal Perturbation for Pilot Samples}

From the asymptotic properties of $\widehat{B}$ and $\widehat{\bbeta}_v$ as shown in Theorems \ref{thm:B_consistence} and \ref{thm:sigma_consistence}, respectively, we derive that the optimal perturbation the pilot samples is not the same as the optimal perturbation $h^*$ of CFD problem in \eqref{eq:optimal_tune}. In fact, since $\Var\left[\widehat{\bbeta}_v\right]$ is unrelated to $\gamma$, we only have to focus on \eqref{eq:B_bias_variance} and choose a suitable $\gamma$ to balance the bias and variance of $\widehat{B}$. Specifically, according to Theorem \ref{thm:B_consistence}, we consider the following optimization problem: 
\begin{align}\label{eq:B_optimizaion}
	&\underset{\gamma,c_1,...,c_K}{\mbox{minimize}} \quad \MSE\left[\widehat{B}\right] = H_K^2 n_b^{4\gamma} + V_K \frac{\sigma^2(\theta_0)}{2n_b^{1+6\gamma}}, \\
	&\mbox{subject to}\quad -1/6 < \gamma < 0.\nonumber
\end{align}
Solving \eqref{eq:B_optimizaion} gives $\gamma = -1/10$. 
On the other hand, if $c_1,...,c_K$ are generated from $\mathcal{P}_0$ randomly, the optimization problem comes from the combination of \eqref{eq:B_bias_random} and \eqref{eq:B_variance_random}:
\begin{align}\label{eq:B_optimizaion_random}
	&\underset{\gamma,\mathcal{P}_0}{\mbox{minimize}} \quad \E[H_K^2] n_b^{4\gamma} + \Var[H_K]n_b^{4\gamma} + \E[V_K]\frac{\sigma^2(\theta_0)}{2n_b^{1+6\gamma}}, \\
	&\mbox{subject to}\quad -1/6 < \gamma < 0.\nonumber
\end{align}
Solving \eqref{eq:B_optimizaion_random} also gives $\gamma = -1/10$. That is, whether $c_1,...,c_K$ are fixed or not, the optimal selection of $h_k$'s are of order $n_b^\gamma$ with $\gamma = -1/10$.

A more exact setting about $\{c_1,...,c_K\}$ or $\mathcal{P}_0$ necessitates knowledge of $D$, which is more challenging than the estimation of $\alpha'(\theta_0)$. In this paper, we treat $\{c_1,...,c_K\}$ and $\mathcal{P}_0$ as the hyperparameters whose selection is beyond the scope of our discussion.

\section{The Correlation-induced FD Estimator}\label{sec:DSR}

In this section, we use estimators $\widehat{B}$ and $\widehat{\bbeta}_v$ to estimate the optimal perturbation by \eqref{eq:optimal_tune} and construct the correlation-induced estimator for $\alpha'(\theta_0)$.

The straightforward method for estimating the optimal perturbation is neglecting the $Kn_b$ pilot sample pairs, i.e., setting
\begin{align}\label{eq:estimate_perturbation0}
	\widehat{h}_{n_2} = \left(\frac{\widehat{\bbeta}_v}{4n_2\widehat{B}^2}\right)^{1/6},
\end{align}
where $n_2 = n - Kn_b$. Then, $n_2$ sample pairs are used to construct $\widehat{\Delta}_{n_2,\widehat{h}_{n_2}}$ by \eqref{eq:finite_difference_estimate} as the estimator of $\alpha'(\theta_0)$. It is noteworthy that this estimator is similar to the one constructed in \cite{Li2020Optimally}, with the primary difference being the estimations of $\widehat{B}$ and $\widehat{\bbeta}_v$. From \eqref{eq:optimal_bias_variance_MSE}, given $\widehat{h}_{n_2}$, the expectation and variance of $\widehat{\Delta}_{n_2,\widehat{h}_{n_2}}$ are respectively 
\begin{align}\label{eq:bias_variance_conditional}
\begin{cases}
		\E\left[\widehat{\Delta}_{n_2,\widehat{h}_{n_2}}\left|\widehat{h}_{n_2}\right.\right] = \alpha'(\theta_0) + B\left(\dfrac{\widehat{\bbeta}_v}{4n_2\widehat{B}^2}\right)^{1/3} + o_p\left(n_2^{-1/3}\right),\\
		\Var\left[\widehat{\Delta}_{n_2,\widehat{h}_{n_2}}\left|\widehat{h}_{n_2}\right.\right] = \bbeta_v\left(\dfrac{\widehat{B}^2}{2n_2^2\widehat{\bbeta}_v}\right)^{1/3} + o_p\left(n_2^{-2/3}\right).
\end{cases}
\end{align}

According to $\E\left[\widehat{\Delta}_{n_2,\widehat{h}_{n_2}}\right] = \E\left[\E\left[\widehat{\Delta}_{n_2,\widehat{h}_{n_2}}\left|\widehat{h}_{n_2}\right.\right]\right]$ and $\Var\left[\widehat{\Delta}_{n_2,\widehat{h}_{n_2}}\right] = \Var\left[\E\left[\widehat{\Delta}_{n_2,\widehat{h}_{n_2}}\left|\widehat{h}_{n_2}\right.\right]\right] + \E\left[\Var\left[\widehat{\Delta}_{n_2,\widehat{h}_{n_2}}\left|\widehat{h}_{n_2}\right.\right]\right]$, we have
\begin{align}\label{eq:nonrecycling_bias_var_MSE}
\begin{cases}
		\mbox{Bias}\left[\widehat{\Delta}_{n_2,\widehat{h}_{n_2}}\right] = \left(\dfrac{B\bbeta_v}{4n_2}\right)^{1/3} + o\left(n_2^{-1/3}\right),\\
		\Var\left[\widehat{\Delta}_{n_2,\widehat{h}_{n_2}}\right] = \left(\dfrac{B^2\bbeta_v^2}{2n_2^2}\right)^{1/3} + o\left(n_2^{-2/3}\right),\\
		\mbox{MSE}\left[\widehat{\Delta}_{n_2,\widehat{h}_{n_2}}\right] = 3\left(\dfrac{B^2 \bbeta_v^2}{16n_2^2}\right)^{1/3} + o\left(n_2^{-2/3}\right).
\end{cases}
\end{align}
Because $\widehat{\Delta}_{n_2,\widehat{h}_{n_2}}$ is not the one we study in this paper, the discussion and proof of \eqref{eq:nonrecycling_bias_var_MSE} are provided in Appendix \ref{app:widetildeB}.

It is worth noting that $\widehat{\Delta}_{n_2,\widehat{h}_{n_2}}$ only uses $n_2$ $(< n)$ sample pairs. Although the MSE shown in \eqref{eq:nonrecycling_bias_var_MSE} achieves the theoretical optimality as shown in \eqref{eq:optimal_bias_variance_MSE} when $n_b = o(n)$, neglecting the pilot samples results in reduced efficiency when the sample size is small.

\subsection{Constructing the Estimator by Inducing Correlated Samples}

To avoid neglecting pilot samples in the construction of CFD estimators, we attempt to combine the pilot samples with the additional $n_2$ sample pairs. In fact, as will be indicated in Section \ref{AnalysesCor-CFD}, theoretical analyses suggest that $n_2=0$ is still feasible.  Next, we must address two questions: (1) how to generate the $n_2$ sample pairs; and (2) how to reuse the pilot samples.

To answer the first question, because we hope to utilize the whole $n$ sample pairs, the desired CFD estimator should be constructed based on the perturbation $\widehat{h}_n$, where 
\begin{align}\label{eq:estimate_perturbation}
	\widehat{h}_n = \left(\frac{\widehat{\bbeta}_v}{4n\widehat{B}^2}\right)^{1/6}.
\end{align}
Note that $\widehat{h}_n$ is derived from the entire sample-pair set of size $n$, whereas $\widehat{h}_{n_2}$, as defined in \eqref{eq:estimate_perturbation0}, is based on the subset of size $n_2$. Next, the additional $n_2$ sample pairs are generated at $\widehat{h}_n$, and we can calculate $\left\{\Delta_i\left(\widehat{h}_n\right), i = 1,...,n_2\right\}$ based on \eqref{def:Delta}. When no confusion arises, the subscript $i$ of $\Delta_i\left(\widehat{h}_n\right)$ is omitted to denote a generic $\Delta\left(\widehat{h}_n\right)$.

To answer the second question, recall that our objective is to estimate $\alpha'(\theta_0)$ under the criteria of minimizing the MSE of its estimator. To reuse the pilot samples, we transform $\Delta(h_k)$ which are constructed based on the pilot samples, by adjusting their location and scale, so that after the transformation, they have the same expectation and variance as $\Delta\left(\widehat{h}_n\right)$. 

Specifically, for any $j = 1,...,n_b$ and $k = 1,...,K$, we transform $\Delta_j(h_k)$ to
\begin{align}\label{eq:transform}
	\left(\dfrac{\Var\left[\Delta\left(\widehat{h}_n\right)\right]}{\Var[\Delta(h_k)]}\right)^{1/2} [\Delta_j(h_k) - \E\Delta(h_k)] + \E\Delta\left(\widehat{h}_n\right),
\end{align}
where $\Var\left[\Delta\left(\widehat{h}_n\right)\right]$ and $\Var[\Delta(h_k)]$ are the variances at $\widehat{h}_n$ and $h_k$, respectively, and $\E\Delta\left(\widehat{h}_n\right)$ and $\E\Delta(h_k)$ are the expectations at $\widehat{h}_n$ and $h_k$, respectively. Evidently, $[\Delta_j(h_k) - \E\Delta(h_k)]/\left(\Var[\Delta(h_k)]\right)^{1/2}$ in \eqref{eq:transform} represents a standardization with mean 0 and variance 1. Then \eqref{eq:transform} possesses the same expectation and variance as $\Delta\left(\widehat{h}_n\right)$. Therefore, a desired estimator for $\alpha'(\theta_0)$ can be formulated as:
\begin{align}\label{eq:desired_estimator}
	 \frac{n_2}{n} \frac{1}{n_2}\sum_{i=1}^{n_2}\Delta_i\left(\widehat{h}_n\right) + \frac{K n_b}{n}\frac{1}{Kn_b}\sum_{k=1}^{K}\sum_{j=1}^{n_b} \eqref{eq:transform}.
\end{align}
This is a weighted average of an estimator at $\widehat{h}_n$ constructed through the $n_2$ sample pairs and the one by the transformation from the pilot samples. It is reasonable to expect that under mild conditions, such as ensuring that the covariance of two terms in \eqref{eq:desired_estimator} converges faster than the variance of either term, the bias, variance and MSE of \eqref{eq:desired_estimator} all achieve their optimal values as in \eqref{eq:optimal_bias_variance_MSE}.

However, expectations and variances of $\Delta(h_k)$ and $\Delta\left(\widehat{h}_n\right)$ in \eqref{eq:transform} are both unknown. Therefore, we have to substitute them in \eqref{eq:desired_estimator} by their estimators. In light of results in Section \ref{sec:bootstrap}, we choose estimators as shown in Table \ref{tab:options}.
\begin{table}[htbp!]
\centering
\caption{Estimators of expectations and variances of $\Delta(h_k)$ and $\Delta\left(\widehat{h}_n\right)$.}
\label{tab:options}
\begin{tabular}{m{3cm} R{3cm}}
\hline
Statistics & Estimators\\
\hline
$\E\Delta(h_k)$ & $\left(1,h_k^2\right)\widehat{\beta}_e$ \\
$\E\Delta\left(\widehat{h}_n\right)$ & $\left(1,{\widehat{h}_n}^{2}\right)\widehat{\beta}_e$ \\
$\Var[\Delta(h_k)]$ & ${\widehat{\bbeta}_v^2}\big/{\left(2h_k^2\right)}$\\
$\Var\left[\Delta\left(\widehat h_n\right)\right]$ & ${\widehat{\bbeta}_v^2}\big/{\left(2{\widehat{h}_n}^{2}\right)}$\\[1.5ex]
\hline
\end{tabular}
\end{table}
Specifically, the estimators are all constructed through solutions to regression problems \eqref{eq:regression_problem_bias} and \eqref{eq:regression_problem_variance}. 

For any $j = 1,...,n_b$ and $k = 1,...,K$, we substitute these estimators into \eqref{eq:transform},  and \eqref{eq:transform} is approximated by
\begin{align}\label{eq:sample_transform}
		\Delta_j^{cor}(h_k) = \frac{|h_k|}{\left|\widehat{h}_n\right|}\left(\Delta_j(h_k) - \left(1,h_k^2\right)\widehat{\bbeta}_e\right) + \left(1,{\widehat{h}_n}^2\right)\widehat{\bbeta}_e.
\end{align}
Compared with \eqref{eq:transform}, the transformation induces correlation to $\Delta_j^{cor}(h_k)$ for $j = 1,...,n_b$, and so we use the superscript ``$cor$" to highlight the correlation. Then, the {\it correlation-induced CFD (Cor-CFD) estimator}, i.e., the estimated version of \eqref{eq:desired_estimator}, is established as follows.
\begin{align}\label{eq:DSR}
		\widehat{\Delta}_{n,\widehat{h}_n} = \frac{1}{n}\left(\sum_{i=1}^{n_2}\Delta_i\left(\widehat{h}_n\right) + \sum_{k=1}^{K}\sum_{j=1}^{n_b}\Delta_j^{cor}(h_k)\right).
\end{align}

\subsection{Asymptotic Analyses of the Cor-CFD Estimator}\label{AnalysesCor-CFD}

The results for the Cor-CFD estimator $\widehat{\Delta}_{n,\widehat{h}_n}$ are summarized in the following theorem and proposition. To simplify the proof, we consider the non-random hyperparameters $c_1, ..., c_K$ such that when $n_b$ is given, $h_1, ..., h_K$ are also given. The proof is similar but more tedious when considering $\mathcal{P}_0$ as the hyperparameter, i.e., $c_1, ..., c_K$ are random.
\begin{theorem}\label{thm:DSR}
	Suppose that Assumptions \ref{ass:derivative3}, \ref{ass:continuity} and \ref{ass:derivative5} hold. For any $k = 1,...,K$ $(K \geq 2)$, let $h_k = c_k n_b^{-1/10}$ $(c_k \neq 0)$ and for any $j \neq k$, $c_j \neq c_k$. If $n_b,n \to \infty$, then we have
	\begin{align}
	\E\left[\widehat{\Delta}_{n,\widehat{h}_n}\right] =& \alpha'(\theta_0) + \left(\frac{B\sigma^2(\theta_{0})}{4n}\right)^{1/3} + \left(\frac{4B^2}{\sigma^2(\theta_0)}\right)^{1/6}\sqrt{\frac{n_b}{n}}D \Lambda n^{-1/3} + o\left(n^{-1/3}\right),\label{eq:DSR_bias}\\
	\Var\left[\widehat{\Delta}_{n,\widehat{h}_n}\right] =& \left(\frac{B^2 \sigma^4(\theta_0)}{2n^2}\right)^{1/3} + \left(\frac{B^2 \sigma^4(\theta_0)}{2}\right)^{1/3}\frac{n_b}{n} (q - K) n^{-2/3} + o\left(n^{-2/3}\right),\label{eq:DSR_variance}\\
	\MSE\left[\widehat{\Delta}_{n,\widehat{h}_n}\right] =& 3\left(\dfrac{B^2 \sigma^4(\theta_0)}{16n^2}\right)^{1/3} + \left\{\left(\frac{4B^2}{\sigma^2(\theta_0)}\right)^{1/3}\sqrt{\frac{n_b}{n}}D \Lambda \left(\sqrt{\frac{n_b}{n}}D \Lambda + {\rm{sign}}(B)\sigma(\theta_0)\right)\right.\nonumber\\
	&+ \left.\left(\frac{B^2 \sigma^4(\theta_0)}{2}\right)^{1/3}\frac{n_b}{n} (q - K) \right\}n^{-2/3} + o\left(n^{-2/3}\right),\label{eq:DSR_MSE}
\end{align}
 where ${\boldsymbol{c}} = [|c_1|,...,|c_K|]^{\top}$, ${\boldsymbol{c}^4} = \left[c_1^4,...,c_K^4\right]^{\top}$, $\Lambda = \boldsymbol{c^{\top}Pc^4}$, ${\boldsymbol{P}}=\boldsymbol{I} - \boldsymbol{X}_e(\boldsymbol{X}_e^{\top}\boldsymbol{X}_e)^{-1}\boldsymbol{X}_e^{\top}$, $\boldsymbol{X}_e$ is defined in \eqref{eq:regression_component_bias}, $q = \left\|{\rm{Diag}}(\boldsymbol{c^{-1}})\boldsymbol{Pc}\right\|_2^2$, ${\rm{Diag}}(\boldsymbol{c^{-1}}) = {\rm{Diag}}\left(1/|c_1|,...,1/|c_K| \right)$, ${\rm{sign}}(B)$ is the sign of $B$, and $||{\boldsymbol{v}}||_2 = \left(\sum_{k=1}^{K}v_k^2\right)^{1/2}$ for any ${\boldsymbol{v}} \in \R^K$.
\end{theorem}

The proof of Theorem \ref{thm:DSR} is provided in Appendix \ref{app:prooftheorem4}. By comparing \eqref{eq:DSR_bias}, \eqref{eq:DSR_variance} and \eqref{eq:DSR_MSE} with \eqref{eq:optimal_bias_variance_MSE}, it becomes attractive that if $B$ and $D \Lambda$ have opposite signs, the squared bias of $\widehat{\Delta}_{n,\widehat{h}_n}$ will be smaller than that of the traditional optimal CFD estimator (see \eqref{eq:optimal_bias_variance_MSE}). In addition, when $c_1,...,c_K$ are chosen such that $q \leq K$, the variance will be reduced. Proposition \ref{prop:DSR}, a corollary of Theorem \ref{thm:DSR}, further elaborates on these aspects.

\begin{proposition}\label{prop:DSR}
	Under the same conditions as in Theorem \ref{thm:DSR}, we have
	\begin{itemize}
		\item[1.] {If $n_b = o(n)$, then} 
		\begin{align}\label{eq:thmDSR_case1}
			\begin{cases}
				\E\widehat{\Delta}_{n,\widehat{h}_n} = \alpha'(\theta_0) + \left(\dfrac{B\sigma^2(\theta_{0})}{4n}\right)^{1/3} + o\left(n^{-1/3}\right),\\
				\Var\left[\widehat{\Delta}_{n,\widehat{h}_n}\right] = \left(\dfrac{B^2 \sigma^4(\theta_0)}{2n^2}\right)^{1/3} + o\left(n^{-2/3}\right),\\
				\MSE \left[\widehat{\Delta}_{n,\widehat{h}_n}\right] = 3\left(\dfrac{B^2 \sigma^4(\theta_0)}{16n^2}\right)^{1/3} + o\left(n^{-2/3}\right).
			\end{cases}
		\end{align}
		\item[2.] If $n_b = O(n)$, specifically, $\lim\limits_{n_b,n\to\infty} Kn_b/n = r$, where $0 < r \leq 1$, then
		\begin{itemize}
			\item[i.] {If $-1 - {\rm{sign}}(B) \leq \sqrt{\dfrac{r}{K}}\dfrac{2}{\sigma(\theta_0)}D \Lambda \leq 1 - {\rm{sign}}(B)$, then}
			\begin{align*}
				\left|\E\widehat{\Delta}_{n,\widehat{h}_n} - \alpha'(\theta_0)\right| \leq \left(\dfrac{B\sigma^2(\theta_{0})}{4n}\right)^{1/3} + o\left(n^{-1/3}\right).
			\end{align*}

			\item[ii.] {Else, we have}
			\begin{align*}
				\left|\E\widehat{\Delta}_{n,\widehat{h}_n} - \alpha'(\theta_0)\right| \leq \left(1 + \sqrt{\dfrac{r}{K}}\dfrac{2}{\sigma(\theta_0)}\left|D \Lambda\right|\right) \left(\dfrac{B\sigma^2(\theta_{0})}{4n}\right)^{1/3} + o\left(n^{-1/3}\right).
			\end{align*}
		\end{itemize}
		\item[3.] If $n_b = O(n)$, specifically, $\lim\limits_{n_b,n\to\infty} Kn_b/n = r$, where $0 < r \leq 1$, and $\max_{k\in \{1,...,K\}}c_k^2 \leq 2\min_{k\in \{1,...,K\}}c_k^2$, then
		\begin{align*}
			\Var\left[\widehat{\Delta}_{n,\widehat{h}_n}\right] \leq \left(\dfrac{B^2 \sigma^4(\theta_0)}{2n^2}\right)^{1/3} + o\left(n^{-2/3}\right).
		\end{align*}
	\end{itemize}
\end{proposition}

These results in Proposition \ref{prop:DSR} can be easily derived by the same technique as in Theorem \ref{thm:DSR} and the proof is provided in Appendix \ref{app:proofcorDSR}. 
Proposition \ref{prop:DSR} guides the following conclusions:

\begin{enumerate}

   \item The first result indicates that the Cor-CFD estimator regains the traditional optimal CFD estimator when $n_b = o(n)$ (refer to \eqref{eq:optimal_bias_variance_MSE} and \eqref{eq:thmDSR_case1}). This aligns with intuition since we allocate a majority of the samples to gradient estimation rather than constant estimation.
   
   \item The second result suggests that if $n_b = O(n)$, then the absolute value of the bias of the Cor-CFD estimator converges at the same rate as that of the traditional optimal CFD estimator. In particular, {\it 2.i} in Proposition \ref{prop:DSR} demonstrates that the former is smaller than the latter. On the other hand, in Appendix \ref{app:projection_matrix}, we show that $\Lambda$ is typically controllable. Therefore, {\it 2.ii} in Proposition \ref{prop:DSR} illustrates that even if the absolute value of the bias of the Cor-CFD estimator is not reduced, its increment is moderate and will not blow up.
   
   \item The third result demonstrates that if $n_b = O(n)$ and $\max_{k\in \{1,...,K\}}c_k^2 \leq 2\min_{k\in \{1,...,K\}}c_k^2$, the Cor-CFD estimator has a reduced variance compared to the traditional optimal CFD estimator. It is worth mentioning that the condition $\max_{k\in \{1,...,K\}}c_k^2 \leq 2\min_{k\in \{1,...,K\}}c_k^2$ is a highly sufficient condition of $q \leq K$, and the latter yields a reduced variance (see Theorem \ref{thm:DSR}).
   
   \item An appealing observation from Proposition \ref{prop:DSR} is the robustness of the Cor-CFD estimator w.r.t. $r$, in the sense that it exhibits favorable theoretical performance for $0\leq r \leq 1$. In particular, when $r = 1$, i.e., the entire simulation budget is used as the pilot samples, the Cor-CFD estimator remains efficient. This theoretical insight serves as a guide, suggesting that when faced with limited simulation resources, allocating the entire budget to generate pilot samples, and the correlation induced samples can lead to an efficient FD estimator.

\end{enumerate}

The variance reduction and possible bias reduction stems from the correlated property of $\Delta_j^c(h_k)$ ($j = 1,...,n_b$) induced by the substitution of $\E[\Delta(h_k)],k=1,...,K$, with their regression estimators. To illustrate this, let us impractically assume that $\E\Delta(h_k), k = 1,...,K$ are all known precisely and remain them in \eqref{eq:transform} but substituting other unknown variables with their estimators in Table \ref{tab:options}. This results in a new transformed estimator of \eqref{eq:transform}, denoted by
	\begin{align*}
		\Delta_j^t(h_k) = \frac{|h_k|}{\left|\widehat{h}_n\right|}\left(\Delta_j(h_k) - \E\Delta(h_k)\right) + \left(1,{\widehat{h}_n}^2\right)\widehat{\bbeta}_e.
	\end{align*}
Replacing $\Delta_j^{cor}(h_k)$ in \eqref{eq:DSR} by $\Delta_j^t(h_k)$, leads to a new Cor-CFD estimator:
\begin{align}
		\widehat{\Delta}_{n,\widehat{h}_n}^{new} =& \frac{1}{n}\left(\sum_{i=1}^{n_2}\Delta_i\left(\widehat{h}_n\right) + n_b\sum_{k=1}^{K}\frac{|h_k|}{\left|\widehat{h}_n\right|}\left(\widehat{\Delta}_{n_b,h_k} - \E\Delta(h_k)\right) + Kn_b\left(1,{\widehat{h}_n}^2\right)\widehat{\bbeta}_e\right). \label{eq:remark4_1}
	\end{align}
Based on this estimator, $\widehat{\Delta}_{n,\widehat{h}_n}$ can be rewritten as 
\begin{align}
		\widehat{\Delta}_{n,\widehat{h}_n} = \widehat{\Delta}_{n,\widehat{h}_n}^{new} - \frac{n_b}{n}\sum_{k=1}^{K}\frac{|h_k|}{\left|\widehat{h}_n\right|}\left(\left(1,h_k^2\right)\widehat{\bbeta}_e - \E\Delta(h_k)\right).\label{eq:remark4_3}
	\end{align}

In Appendix \ref{app:corn_MSE_reduction}, we show that $\widehat{\Delta}_{n,\widehat{h}_n}^{new}$ has the same asymptotic bias and variance as the traditional optimal CFD estimator. Therefore, by comparing results in Theorem \ref{thm:DSR} and Appendix \ref{app:corn_MSE_reduction}, the reasons for variance reduction and potential bias reduction become evident:

	\begin{itemize}
	
		\item {\it{Variance reduction:}}  The second term of \eqref{eq:remark4_3} is highly positively correlated with the second term of \eqref{eq:remark4_1} because $\left(1,h_k^2\right)\widehat{\bbeta}_e$ approximates $\widehat{\Delta}_{n_b,h_k}$. Although the former is not mean 0, it can be viewed as a control variate of the latter. Therefore, the second term of \eqref{eq:remark4_3} reduces the variance of $\widehat{\Delta}_{n,\widehat{h}_n}^{new}$ to that of $\widehat{\Delta}_{n,\widehat{h}_n}$.
		
		\item {\it{Possible bias reduction:}} The expectation of the second term of \eqref{eq:remark4_3} depends on $D$ (see \eqref{eq:regression_component_bias}) while the bias of $\widehat{\Delta}_{n,\widehat{h}_n}^{new}$ depends on $B$ (see \eqref{eq:optimal_bias_variance_MSE}). If these two terms tend to cancel each other out, the overall bias will be reduced.
		
	\end{itemize}
	
In Section \ref{sec:experiments}, we will present several numerical examples to validate the aforementioned properties of our proposed Cor-CFD estimator.

\section{Implementation}\label{sec:implementation}

This section provides the implementations for estimating $\bbeta_e$ and $\bbeta_v$ and constructing the Cor-CFD estimator.

\subsection{Estimating $B$ and $\sigma^2(\theta_0)$} \label{subsec:EstimateBsigma}

In this subsection, we present the specific steps for estimating unknown constants using the bootstrap method as described in Section \ref{sec:bootstrap}.  We generate i.i.d. $c_1,...,c_K$ from $\mathcal{P}_0$ and for $k = 1,...,K$, let $h_k = c_k n_b^{-1/10}$.  At each perturbation $h_k$, we generate $n_b$ samples $\boldsymbol{\Delta}(h_k) = \{\Delta_1(h_k),...,\Delta_{n_b}(h_k)\}$. Repeat the bootstrap procedure in Section \ref{sec:bootstrap} independently for $I$ times, and then we obtain $I$ versions of $\widehat{\Delta}^{b}_{n_b,h_k}$, denoted by $\left\{\widehat{\Delta}^{b}_{n_b,h_k}(1),...,\widehat{\Delta}^{b}_{n_b,h_k}(I)\right\}$. Then $\E_{*}\widehat{\Delta}_{n_b,h_k}$ is approximated by
\begin{align}\label{eq:bootstrap_sample_mean}
	\bar{\Delta}^{b}_{n_b,h_k} = \frac{1}{I}\sum_{q=1}^{I}\widehat{\Delta}^{b}_{n_b,h_k}(q), 
\end{align}
and $\Var_{*}\left[\widehat{\Delta}_{n_b,h_k}\right]$ is approximated by
\begin{align}\label{eq:bootstrap_sample_variance}
	s^2_{n_b,h_k} = \frac{1}{I}\sum_{q=1}^{I}\left[\widehat{\Delta}^{b}_{n_b,h_k}(q) - \bar{\Delta}^{b}_{n_b,h_k}\right]^2.
\end{align}
According to the law of large numbers, as $I$ increases, $\bar{\Delta}^{b}_{n_b,h_k}$ and $s^2_{n_b,h_k}$ are respectively consistent estimators of $\E_{*}\widehat{\Delta}^{b}_{n_b,h_k}$ and $\Var_{*}\left[\widehat{\Delta}^{b}_{n_b,h_k}\right]$.

It is worth noting that both of the regression problems \eqref{eq:regression_problem_bias} and \eqref{eq:regression_problem_variance} exhibit heteroscedasticity. This arises from the fact that elements of $\cE_e$ in \eqref{eq:regression_component_bias} and $\cE_v$ in \eqref{eq:regression_component_variance} possess varying variances, due to different values of $h_k$'s.
The primary approach to addressing heteroscedasticity is the weighted least squares method. This involves determining appropriate weights to eliminate or mitigate the heteroscedasticity in the regression models. Because the estimator obtained through the weighted least squares method exhibits the same convergence rate as that derived from the ordinary least squares method \citep{Hayashi2011}, we omit an in-depth analysis of the weighted least squares method. 

While estimating the parameter $B$, we can effectively tackle heteroscedasticity by estimating the variance at each perturbation using the $n_b$ samples, and the reciprocals of the standard deviations are appropriate weights. That is, we regress $\widetilde{\boldsymbol{Y}}_e$ on $\widetilde{\boldsymbol{X}}_e$, where
\begin{align}\label{RegressYXe}
	\widetilde{\boldsymbol{Y}}_e &= \left[\bar{\Delta}^{b}_{n_b,h_1}\big/s_{n_b,h_1},...,\bar{\Delta}^{b}_{n_b,h_{K}}\big/s_{n_b,h_K}\right]^{\top}, 
\widetilde{\boldsymbol{X}}_e = \left[\begin{array}{lll}
1\big/s_{n_b,h_1} & ... & 1\big/s_{n_b,h_K} \\
h_1^2\big/s_{n_b,h_1} & ... & h_{K}^2\big/s_{n_b,h_K}
\end{array}\right]^{\top},
\end{align}
and the slope is the estimator $\widehat{B}$.

However, when estimating $\sigma^2(\theta_0)$, we lack such standard deviations. From $\cE_v$ in \eqref{eq:regression_component_variance}, it can be seen that the heteroscedasticity of the regression model \eqref{eq:regression_problem_variance} arises from the different variances of $\phi(\cdot)$ for different values of $h_k$'s. It follows from \eqref{eq:Var_phi_h} that $h_k^2 \phi(h_k)$ ($k=1,...,K$) share the same variance. Therefore, we can handle the heteroscedasticity resulting from differences in perturbations by multiplying both sides of \eqref{eq:regression_problem_variance} by $h_k^2$ ($k=1,...,K$) simultaneously. That is, we regress $\widetilde{\boldsymbol{Y}}_v $ on $\widetilde{\boldsymbol{X}}_v$, where
\begin{align}\label{RegressYXv}
\widetilde{\boldsymbol{Y}}_v = \left[h_1^2 s^2_{n_b,h_1},...,h_K^2 s^2_{n_b,h_{K}}\right]^{\top}, 
\widetilde{\boldsymbol{X}}_v = \left[\frac{n_b-1}{2n_b^2},...,\frac{n_b-1}{2n_b^2}\right]^{\top},
\end{align}
and the slope is the estimator $\widehat{\bbeta}_v$.

\subsection{Estimating $\alpha'(\theta_0)$}

Once we have generated $\Delta_j(h_k)$, where $j=1,2,...,n_b$ and $k=1,2,...,K$, and estimated the unknown constants, we can transform these samples using \eqref{eq:sample_transform} and construct the Cor-CFD estimator using \eqref{eq:DSR}. We summarize the above procedure as Algorithm \ref{alg:algorithm_estimate}:

\begin{enumerate}
	
	\item Usually, we choose $\mathcal{P}_0$ as a truncated normal distribution $\mathcal{P}_0 = \psi\left(\mu_0,\sigma_0^2,L,U\right)$, where $0<L<U<\infty$. This setting ensures that $\E[H_K]$, $\Var[H_K]$, $\E[V_K]$, $\E\left[\widetilde{H}_K\right]$, $\Var\left[\widetilde{H}_K\right]$, $\E\left[\widetilde{V}_K\right]$, $\E\left[\widehat{H}_K\right]$,$\Var\left[\widehat{H}_K\right]$ and $\E\left[\widehat{V}_K\right]$ are all bounded. For $c_1,...,c_K \sim \mathcal{P}_0$, the condition $\max_{k\in \{1,...,K\}}c_k^2 \leq 2\min_{k\in \{1,...,K\}}c_k^2$ holds only when $U \leq \sqrt{2}L$, which is a stringent condition. However, we have mentioned that $\max_{k\in \{1,...,K\}}c_k^2 \leq 2\min_{k\in \{1,...,K\}}c_k^2$ is a sufficient condition of $q \leq K$ which is satisfied more easily for $\mathcal{P}_0$ with different parameters.
	
	\item In step 1, the pilot samples are generated at the perturbations $h_k = c_k n_b^{-1/10}$, rather than $c_k n_b^{-1/6}$. This is because \eqref{eq:B_optimizaion} and \eqref{eq:B_optimizaion_random} indicate that the optimal perturbation for the pilot samples is of order $n_b^{-1/10}$.
	
	\item In step 1, when estimating $B$ and $\sigma^2(\theta_0)$, we apply the weighted least squares regression proposed in Section \ref{subsec:EstimateBsigma}.
		
	\item In step 2, we generate the remaining samples at $\widehat{h}_n$, instead of $\widehat{h}_{n_2}$ defined as in \eqref{eq:estimate_perturbation0}, because we will utilize the total $n$ samples to construct the CFD estimator.
		
\end{enumerate}

\begin{algorithm2e}[t!]
    \fontsize{11pt}{13.6pt}\selectfont
	\caption{Cor-CFD Algorithm.}
	\label{alg:algorithm_estimate}
	\BlankLine
	
	 \textit{\textbf{Input: }} The number of total sample pairs $n$, the number of perturbation parameters $K$, the number of pilot sample pairs at each perturbation $n_b$, the number of bootstrap $I$ and initial perturbation generator $\mathcal{P}_0$.
	
	\noindent\textit{\textbf{Step 1.} Estimate the unknown constants $B$, $\sigma^2(\theta_0)$ and the optimal perturbation.}
	
	\begin{enumerate}

	\item Generate $c_1,c_2,...,c_{K} \stackrel{i.i.d.}{\sim} \mathcal{P}_0$ and let $h_k = c_k n_b^{-1/10}$ for any $k=1,2,...,K$. Generate pilot samples and calculate $\Delta_{j}(h_k)$, where $j=1,2,...,n_b$ and $k=1,2,...,K$.
	
	\item Perform bootstrap resampling $I$ times for each set of samples at $h_1, h_2, ..., h_K$ and calculate $\bar{\Delta}^{b}_{n_b,h_k}$ and $s^2_{n_b,h_k}$ for $k=1,...,K$, according to \eqref{eq:bootstrap_sample_mean} and \eqref{eq:bootstrap_sample_variance}.
	
	\item Regress $\widetilde{\boldsymbol{Y}}_e$ on $\widetilde{\boldsymbol{X}}_e$, where $\widetilde{\boldsymbol{Y}}_e$ and $\widetilde{\boldsymbol{X}}_e$ are defined in \eqref{RegressYXe}, and the slope is $\widehat{B}$. 

	Regress $\widetilde{\boldsymbol{Y}}_v $ on $\widetilde{\boldsymbol{X}}_v$, where $\widetilde{\boldsymbol{Y}}_v $ and $\widetilde{\boldsymbol{X}}_v$ are defined in \eqref{RegressYXv},
	and the slope is $\widehat{\bbeta}_v$.
	
	Set the perturbation by \eqref{eq:estimate_perturbation}, i.e.,
        \begin{align*}
        	\widehat{h}_n = \left(\frac{\widehat{\bbeta}_v}{4n\widehat{B}^2} \right)^{1/6}.
        \end{align*}
        
        \end{enumerate}

	\noindent\textit{\textbf{Step 2.} Construct the Cor-CFD estimator of $\alpha'(\theta_0)$.}		
	\begin{enumerate}
        \item Reuse $\Delta_{j}(h_k)$, where $j=1,2,...,n_b$ and $k=1,2,...,K$. That is, calculate $\Delta_j^{cor}(h_k) = \dfrac{|h_k|}{\left|\widehat{h}_n\right|}\left(\Delta_j(h_k) - \left(1,h_k^2\right)\widehat{\bbeta}_e\right) + \left(1,{\widehat{h}}_n^2\right)\widehat{\bbeta}_e$ for $j = 1,2,...,n_b$, and $k = 1,2,...,K$.
		
	\item Generate $n_2 = n - K n_b$ sample pairs at $\widehat{h}_n$ and calculate $\left\{\Delta_1(\widehat{h}_n),...,\Delta_{n_2}(\widehat{h}_n)\right\}$.
	
	\item The Cor-CFD estimator 
		\begin{align*}
       \widehat{\Delta}_{n,\widehat{h}_n} = \frac{1}{n}\left(\sum_{i=1}^{n_2}\Delta_i\left(\widehat{h}_n\right) + \sum_{k=1}^{K}\sum_{j=1}^{n_b}\Delta_j^{cor}(h_k)\right).
        \end{align*}
        \end{enumerate}

        \textit{\textbf{Output: }} $\widehat{\Delta}_{n,\widehat{h}_n}$.
        
\end{algorithm2e}

\section{Numerical Experiments}\label{sec:experiments}

In this section, we apply Algorithm \ref{alg:algorithm_estimate} to three examples to verify theoretical results, including the constants estimation and the asymptotic variance reduction and possible bias reduction of our proposed Cor-CFD estimator. All numerical results are based on 1000 replications and additional results are provided in Appendix \ref{app:addition_experiments}.

Across all examples, when applying Algorithm \ref{alg:algorithm_estimate}, we set the initial perturbation generator $\mathcal{P}_0 = \psi\left(0,1,0.1,\infty\right)$, the number of bootstrap $I=1000$, and the number of regression points $K = 10$ or $20$.

\begin{example}\label{exa:ksinx}
Consider estimating the first order derivative of the function $\alpha(\theta) = \kappa\cdot {\rm sin}(\theta)$ at the point $\theta_{0} = 0$, where $\kappa$ is a priori unknown parameter.
\begin{itemize}
	\item {\textit{Case 1:} The observed variables at $\theta$ are assumed to obey a normal distribution with mean $\alpha(\theta)$ and variance $1$, i.e., $Y(\theta) \sim \mathcal{N}(\alpha(\theta),1)$.}
	\item {\textit{Case 2:} The observed variables at $\theta$ are assumed to obey a normal distribution with mean $\alpha(\theta)$ and variance ${\rm e}^{-3\theta}$, i.e., $Y(\theta) \sim \mathcal{N}(\alpha(\theta),{\rm e}^{-3\theta})$.}
\end{itemize}
Cases 1 and 2 represent homoscedasticity and heteroscedasticity, respectively. From a simple calculation we obtain
\begin{align*}
	\alpha'(\theta_{0}) = \kappa,\quad B = -\kappa/6,\quad  \sigma^{2}(\theta_{0}) = 1,
\end{align*}
as the real parameters to measure the performance of our proposed method. During the execution of each algorithm, we assume that the parameters $\alpha'(\theta_0)$, $B$ and $\sigma^{2}(\theta_{0})$ are unknown.
\end{example}

\begin{figure}[t]
	\centering
	\caption{Comparison of the estimates of the unknown constants and perturbation for case 1 in Example \ref{exa:ksinx}.}
	\vspace{7pt}
	\hspace*{-2.0cm}
	\includegraphics[scale = 0.45]{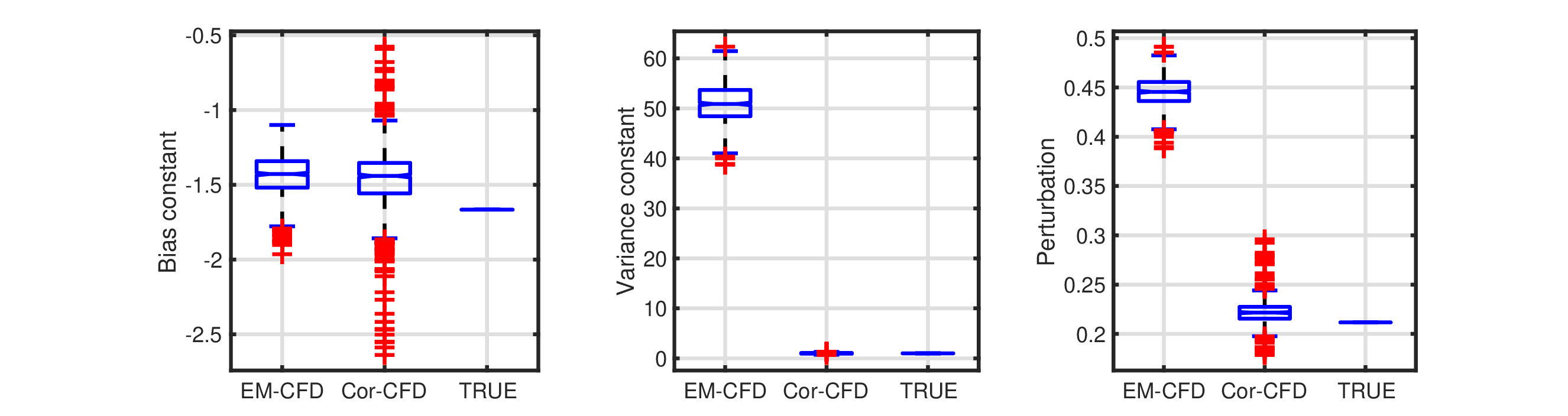}
	\label{fig:example1_case1}
	\vspace{-15pt}
\end{figure}

\begin{figure}[t]
	\centering
	\caption{Comparison of the estimates of the unknown constants and perturbation for case 2 in Example \ref{exa:ksinx}.}
	\vspace{7pt}
	\hspace*{-2.0cm}
	\includegraphics[scale = 0.45]{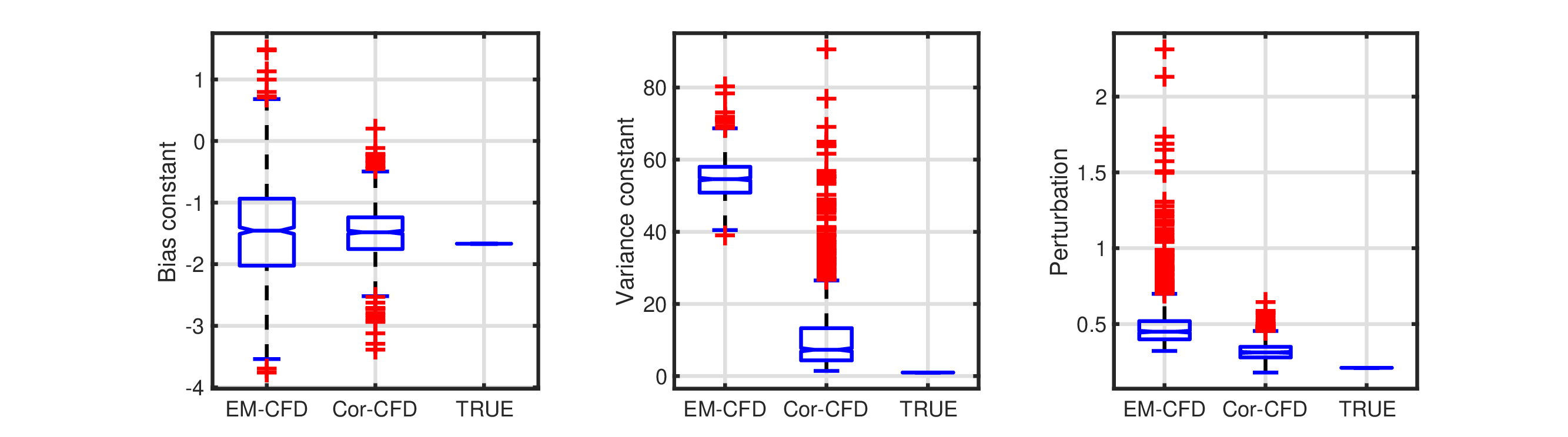}
	\label{fig:example1_case2}
	\vspace{-15pt}
\end{figure}

Example \ref{exa:ksinx}, from \cite{Li2020Optimally}, is used to assess the validity and robustness of the sample-driven method for estimating unknown constants and perturbations. We compare the performance of our proposed method and the estimation-minimization CFD (EM-CFD) method proposed by \cite{Li2020Optimally}. In the implementation, we set $\kappa=10$, $K=10$ and utilize 200 pilot sample pairs. The results, presented in Figures \ref{fig:example1_case1} and \ref{fig:example1_case2}, display three subfigures from left to right, representing the estimators of $B$, $\sigma^2(\theta_0)$, and $h$ across 1000 replications. We compare the performance of our proposed estimator and the estimation-minimization CFD (EM-CFD) estimator proposed by \cite{Li2020Optimally}. When executing the EM-CFD algorithm, to align with the setting in \cite{Li2020Optimally}, we generate $h_1, \ldots, h_{200} \stackrel{i.i.d.}{\sim} \mathcal{N}\left(0,1\times 200^{-1/5}\right)$.

In the left subfigure of Figure \ref{fig:example1_case1}, we observe that our method's robustness in estimating the bias constant is slightly lower than that of the EM-CFD algorithm in case 1, as indicated by the wider box for the Cor-CFD method. This could be due to the larger number of regression points used in the EM-CFD method; the Cor-CFD method employs only $K=10$ regression points, whereas EM-CFD uses 200 sample pairs as regression points. Nonetheless, the two estimators of the bias constant are still comparable overall, likely due to the reduced error at each regression point in our proposed sample-driven method. The middle subfigure of Figure \ref{fig:example1_case1} indicates that our method effectively estimates the variance constant, a challenging task for the EM-CFD algorithm as it incorporates the variance of the perturbation generator. The right subfigure provides a comprehensive view of perturbation estimation. Across all 1000 experiments, the estimated perturbations through our method are much closer to the true value.

In Figure \ref{fig:example1_case2}, we illustrate the impact of different methods on the estimation of $B$, $\sigma^2(\theta_0)$, and $h$ in case 2 of Example \ref{exa:ksinx}, which deals with scenarios of heteroskedasticity. Analyzing the pair of subfigures presented on the left, it is apparent that our method captures the variance information effectively and provides a more robust estimator of the bias constant. From the right subfigure, we observe that our method yields a more accurate and reliable estimator of the perturbation. 

Furthermore, in Appendix \ref{app:addition_experiments1}, we verify the sensitivity of different methods w.r.t. $r$, demonstrating the importance of inducing correlated samples in our method.

\begin{example}\label{exa:poly}
	We consider estimating the first order derivatives of the polynomial function $\alpha(\theta) = 1 - 6\theta + 6\theta^2 - 2.5\theta^3 + 0.1\theta^5$ at different $\theta_0$'s. Assume that the observed variables at $\theta$ obey a normal distribution with mean $\alpha(\theta)$ and variance 1, i.e., $Y(\theta) \sim \mathcal{N}(\alpha(\theta),1)$. From a simple calculation we obtain
\begin{align*}
\alpha'(\theta_0) = -6 + 12\theta_0 - 7.5\theta_0^2 + 0.5\theta_0^4, \quad B = -2.5 + \theta_{0}^2, \quad \sigma^{2}(\theta_0) = 1,
\end{align*}
as the real parameters to measure the performance of our proposed method. During the implementation of each algorithm, we assume that the parameters $\alpha'(\theta_0)$, $B$ and $\sigma^{2}(\theta_0)$ are unknown.
\end{example} 

In this example, we attempt to verify the asymptotic variance reduction and possible bias reduction of our proposed Cor-CFD estimator. Table \ref{tab:TestTheorem} presents a comparison of bias, variance, and MSE between the Cor-CFD estimator (denoted by Bias-Cor, Var-Cor and MSE-Cor, respectively) and the Opt-CFD estimator (denoted by Bias-Opt, Var-Opt and MSE-Opt, respectively) across various sample-pair sizes and $\theta_0$'s in Example \ref{exa:poly}. Additional experiment results are provided in Appendix \ref{app:addition_experiments2}, comparing the estimated MSEs of different methods. In Table \ref{tab:TestTheorem}, $B$ and $D$ are respectively the true values and Opt-CFD is meant to set $B$ and $\sigma^2(\theta_0)$ as their true values, i.e., $B = -2.5 + \theta_0^2$ and $\sigma^2(\theta_0) = 1$. In addition, $\Lambda = \boldsymbol{c^{\top}Pc^4} \approx -2.2450$. We take the total samples to estimate the unknown constants (i.e., $r=1$) and set $K = 10$ for the Cor-CFD estimator.

\begin{table}[h!]
\caption{Comparison of the bias, variance and MSE of the Cor-CFD and Opt-CFD estimators in Example \ref{exa:poly}.}
\label{tab:TestTheorem}
\smallskip
\centering
\small
\renewcommand{\arraystretch}{0.85}
\begin{tabular}{|c|c|c|c|c|c|c|c|c|c|}
\hline
Budget & $x$ & $B$ & $D$ & Bias-Cor & Bias-Opt & Var-Cor & Var-Opt & MSE-Cor & MSE-Opt \\ \hline
\multirow{4}{*}{$10^2$} & 0 & $-2.5$ & 0.1 & $-0.3034$ & $-0.1851$ & 0.0711 & 0.0655 & 0.1631  & 0.0997 \\ \cline{2-10} 
                                 & 1 & $-1.5$ & 0.1 & $-0.2616$ & $-0.1581$ & 0.0679  & 0.0489 & 0.1362 & 0.0739 \\ \cline{2-10} 
                                 & 2 & 1.5 & 0.1 & 0.0351 & 0.1501 & 0.0658  & 0.0474 & 0.0670 & 0.0699 \\ \cline{2-10}  
                                 & 3 & 6.5 & 0.1 & 0.1673 & 0.2340 & 0.0750 & 0.1313 & 0.0134 & 0.1029 \\ \hline
\multirow{4}{*}{$10^3$} & 0 & $-2.5$ & 0.1 & $-0.1256$ & $-0.0865$ & 0.0111 & 0.0131 & 0.0269  & 0.0205 \\ \cline{2-10} 
                                 & 1 & $-1.5$ & 0.1 & $-0.1169$ & $-0.0670$ & 0.0099  & 0.0100 & 0.0235 & 0.0145 \\ \cline{2-10} 
                                 & 2 & 1.5 & 0.1 & 0.0345 & 0.0757 & 0.0099  & 0.0098 & 0.0111 & 0.0155 \\ \cline{2-10} 
                                 & 3 & 6.5 & 0.1 & 0.0889 & 0.1150 & 0.0117 & 0.0291 & 0.0196 & 0.0423 \\ \hline
\multirow{4}{*}{$10^4$} & 0 & $-2.5$ & 0.1 & $-0.0559$ & $-0.0380$ & 0.0018 & 0.0034 & 0.0049  & 0.0048 \\ \cline{2-10} 
                                 & 1 & $-1.5$ & 0.1 & $-0.0487$ & $-0.0352$ & 0.0016  & 0.0023 & 0.0040 & 0.0035 \\ \cline{2-10} 
                                 & 2 & 1.5 & 0.1 & 0.0157 & 0.0309 & 0.0017  & 0.0023 & 0.0019 & 0.0033 \\ \cline{2-10} 
                                 & 3 & 6.5 & 0.1 & 0.0414 & 0.0521 & 0.0021 & 0.0061 & 0.0038 & 0.0088 \\ \hline
\end{tabular}
\end{table}

From Table \ref{tab:TestTheorem}, we have the following observations:

\begin{itemize}

	\item {Bias:} The bias is influenced by the signs of $B$ and $D$. If $B$ and $D$ have the same (opposite) signs, the absolute bias of the Cor-CFD method is smaller (larger) than that of the Opt-CFD method. This aligns with Theorem \ref{thm:DSR} because $\Lambda$ is negative. 

	\item {Variance:} It is observed that as the sample size increases, the variance of the Cor-CFD estimator consistently outperforms that of the Opt-CFD estimator, aligning with Proposition \ref{prop:DSR}. For instance, at $x = 3$ with a sample budget of $10^4$, the variances for the Cor-CFD and Opt-CFD estimators are $0.0021$ and $0.0061$, respectively.
	
	\item {MSE:} It is important to note that the bias and variance of the Cor-CFD estimator consistently approximate and are smaller than those of the Opt-CFD estimator, respectively. Consequently, when employing MSE as the error criterion, the Cor-CFD estimator consistently performs nearly as well as (or even outperforms) the Opt-CFD estimator.
	
\end{itemize}

\begin{example}[Queuing system]\label{exa:exa1}
In this example, we consider two cases.
	\begin{itemize}
		\item \textit{Case 1:} We consider a critically loaded system, where we set $\lambda = \mu = 4$ and $N = 10$. In this case, the true derivative is $-0.2501$, which is calculated using the likelihood ratio/score function method \citep[see, e.g.,][]{Glynn1990Likelihood} with $10^6$ simulation repetitions \citep{Lam2023Enhanced}.
		\item \textit{Case 2:} We consider a non-critically loaded queue, where we set $\lambda = 3$, $\mu = 5$ and $N = 10$. In this case, the true derivative is $-0.1136$.
	\end{itemize}
\end{example}
 
\begin{figure}[h!]
	\centering
	\caption{Comparison of the MSE among different estimators for cases 1 and 2 in Example \ref{exa:exa1}.}
	\vspace{7pt}
	\hspace*{-0.5cm}
	\includegraphics[scale = 0.55]{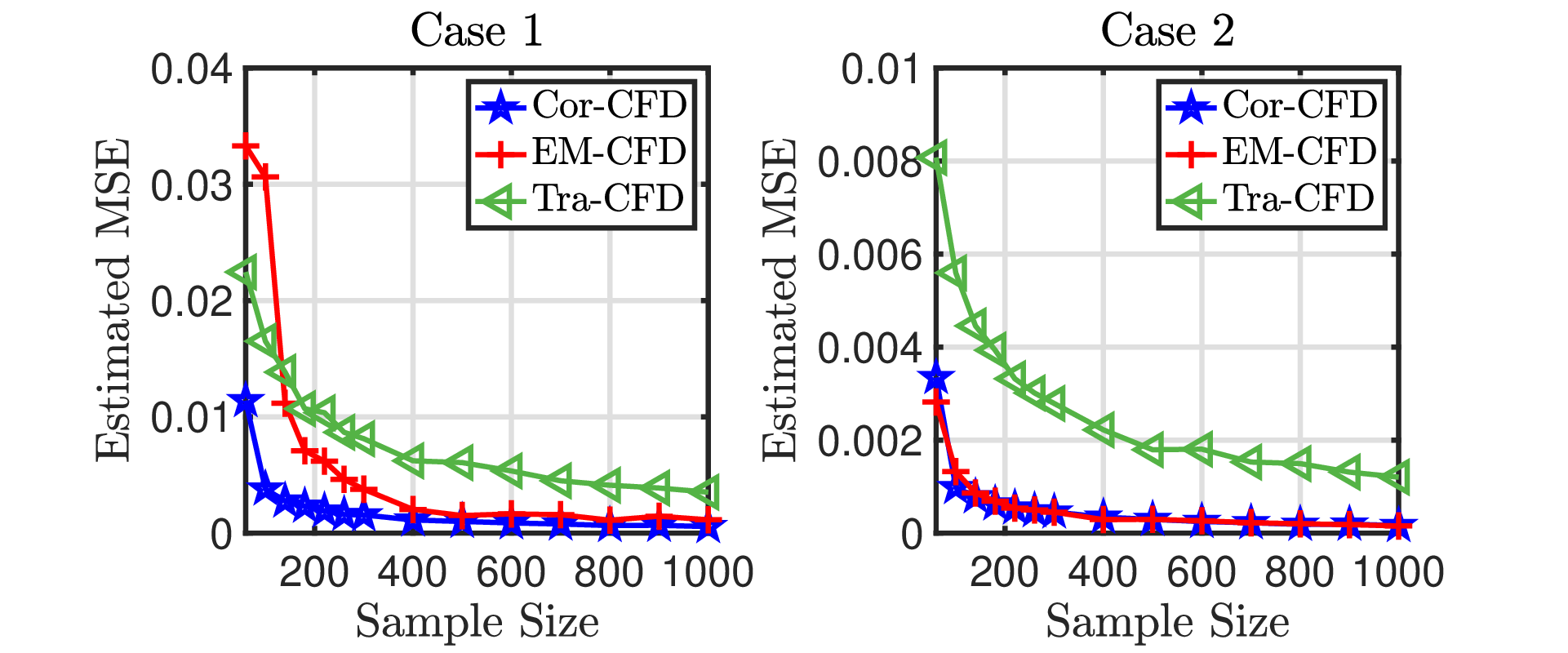}
	\label{fig:example3}
\end{figure}

Figure \ref{fig:example3} illustrates the numerical results for different methods in Example \ref{exa:exa1} when the sample-pair size changes from $n = 60$ to $n=1000$. We compare the following three methods: the traditional CFD (Tra-CFD) method, EM-CFD method and Cor-CFD method. For the Tra-CFD method, we do not know any information about the model in advance and set $B = 5$ and $\sigma^2(\theta_0) = 1$. When considering the EM-CFD method, we set $r = 0.1$ and generate $h_1,...,h_{rn} \stackrel{i.i.d.}{\sim} \mathcal{N}\left(0,1\times (rn)^{-1/5}\right)$. When applying the Cor-CFD method, we set $K = 20$ and $r = 1$ ($r = 1$ means the total samples are pilot samples).

Figure \ref{fig:example3} illustrates that the performance of the Tra-CFD estimator is inferior to the other two estimators, underscoring the importance of estimating unknown constants in finite difference methods. In case 1 of Example \ref{exa:exa1}, the Cor-CFD estimator consistently yields the most favorable results. For instance, with a sample-pair size of 60, the MSEs for the Cor-CFD and EM-CFD estimators are $0.011$ and $0.033$, respectively. In case 2, the performances of these two estimators are comparable.

\section{Derivative-Free Optimization Using Cor-CFD}\label{sec:DFO}

In this section, we consider the DFO problem, i.e., a stochastic optimization problem where gradient information is unavailable and only the stochastic function values can be obtained. Our goal is to incorporate the Cor-CFD estimator into the well-known L-BFGS algorithm \citep{Nocedal1980} and numerically verify its efficiency in solving the DFO problem. We leave the corresponding theoretical studies for future research.
 
Consider an optimization problem 
\begin{align}\label{eq:SO}
	\underset{\theta \in \R^d}{\min}\ \alpha(\theta) = \E[Y(\theta)].
\end{align}
Suppose that Assumptions \ref{ass:derivative3} and \ref{ass:derivative5} hold, ensuring the applicability of our Cor-CFD method. Similarly to the operations in \cite{shi2023numerical}, we replace the gradient estimator in the L-BFGS algorithm with our proposed Cor-CFD estimator, constructing a finite difference L-BFGS series:
\begin{align*}
	\theta_{k+1} = \theta_k - a_k H_k \widehat{\Delta}_{T_k,\widehat{h}_{T_k}},
\end{align*}
where $k$ represents the iterative process, $a_k > 0$ is the step length, $H_k$ is an approximation to the inverse Hessian matrix and $\widehat{\Delta}_{T_k,\widehat{h}_{T_k}}$ is a Cor-CFD estimator based on $T_k$ sample pairs.

To ensure the seamless progress of iterations, two issues must be considered: how to determine the appropriate sample size $T_k$ for each iteration, and how to choose the step length $a_k$. For the first issue, as iterations proceed and $\theta_k$ approaches the optimal point, the accuracy of the finite difference estimator becomes increasingly important. Intuitively, $T_k$ should therefore increase as $k$ increases. The theoretical determination of the optimal $T_k$ could be left as a topic for future research.
 
For the second issue, we apply the stochastic Armijo condition \cite[see, e.g.,][]{berahas2019derivative,shi2023numerical},
	\begin{align}\label{Armijo}
		Y\left(\theta_k - a_k H_k \widehat{\Delta}_{T_k,\widehat{h}_{T_k}}\right) \leq Y(\theta_k) + l_1 a_k \widehat{\Delta}_{T_k,\widehat{h}_{T_k}}^{\top} H_k \widehat{\Delta}_{T_k,\widehat{h}_{T_k}} + 2\sigma.
	\end{align}
Specifically, we shrink $a_k \leftarrow l_2 a_k$ until the above condition holds, where $0 < l_1 < l_2 < 1$. Note that $\sigma$ need not to be chosen too precisely and it suffices to select it as the upper bound of $\sigma(\theta_k)$ \citep{berahas2019derivative}. For example, when $\sigma$ is relatively large, the stochastic Armijo condition is naturally satisfied and $a_k$ degenerates a constant step length. 

The Cor-CFD-based L-BFGS (CorCFD-L-BFGS) algorithm is shown in Algorithm \ref{alg:algorithm_SO}. This algorithm is a quasi-Newton algorithm for DFO problems. We will compare our proposed Algorithm \ref{alg:algorithm_SO} with the traditional CFD-based L-BFGS (TraFD-L-BFGS) algorithm \citep{shi2023numerical} and the new unconstrained optimization algorithm (NEWUOA) \citep{powell2006newuoa}. The TraCFD-L-BFGS algorithm replaces the gradient estimator in L-BFGS algorithm by the finite difference estimator constructed from only a sample pair. Note that the NEWUOA is used as a benchmark in \cite{shi2023numerical}. The detailed settings for each algorithm are provided Appendix \ref{app:setting}.

\begin{algorithm2e}[t!]
	\fontsize{11pt}{13.6pt}\selectfont
	\caption{Cor-CFD-based L-BFGS algorithm.}
	\label{alg:algorithm_SO}
	\BlankLine
	\textit{\textbf{Input: }} The number of total sample pairs $T$, the number of perturbation parameters $K$, the number of initial batch size $T_0$, the number of function evluations $t = 0$, the number of bootstrap $I$, initial perturbation generator $\mathcal{P}_0$, the line search parameters $(l_1,l_2)$, starting point $\theta_0$ and initial step length for line search $a_0 > 0$.
	
	\textit{\textbf{Initialization: }} Set $k = 0, H_0 = I_{d\times d}$, where $I_{d\times d}$ is an identity matrix, and $d$ is the problem dimension. Obtain the estimator $\widehat{\Delta}_{T_0,\widehat{h}_{T_0}}$ using Algorithm \ref{alg:algorithm_estimate}.
	
	\textit{\textbf{While $t < 2T$ do}} 
	\begin{enumerate}
		\item Use stochastic Armijo line search \eqref{Armijo} to find an appropriate step length $0 < a_k < a_0$ and set $t = t + t_{ls}$, where $t_{ls}$ denotes the function evaluation counter during the line search procedure.
		\item Update $\theta_{k+1} = \theta_k - a_k H_k \widehat{\Delta}_{T_k,\widehat{h}_{T_k}}$.
		\item Set $T_{k+1} = \lfloor(T_k + k + 1)/K \rfloor \times K$ and obtain the estimator $\widehat{\Delta}_{T_{k+1},\widehat{h}_{T_{k+1}}}$ using Algorithm \ref{alg:algorithm_estimate}.
		\item Update $H_{k+1}$ via the L-BFGS algorithm.
		\item Set $t = t + 2dT_k$ and $k = k + 1$.
	\end{enumerate}
	
	\textit{\textbf{Output: }} The ultimate estimate $\theta_{k+1}$.
\end{algorithm2e}

We consider two stochastic optimization problems constructed by artificially imposing a standard normal noise $\mathcal{N}(0,1)$ on two classical test functions in optimization area -- Rosenbrock function and Zakharov function \citep[see,e.g.,][]{schittkowski2012more,jamil2013literature} and examine the efficiency of Algorithm \ref{alg:algorithm_SO}. The definitions of these two functions are shown as follows:

\begin{itemize}

\item The Rosenbrock function $f: \boldsymbol{x} \mapsto f(\boldsymbol{x})$ where $\boldsymbol{x} \in \R^2$ is defined as
	\begin{align*}
		f(\boldsymbol{x}) = 100(x_2 - x_1^2)^2 + (x_1 - 1)^2.
	\end{align*}

\item A $d$-dimensional Zakharov function $f: \boldsymbol{x} \mapsto f(\boldsymbol{x})$ where $\boldsymbol{x} \in \R^d$ is defined as
	\begin{align*}
		f(\boldsymbol{x}) = \sum_{i=1}^{d}x_i^2 + \left(\sum_{i=1}^{d}0.5ix_i\right)^2 + \left(\sum_{i=1}^{d}0.5ix_i\right)^4.
	\end{align*}

\end{itemize}
That is, in \eqref{eq:SO}, $Y(\theta) = f(\theta) + \varepsilon$ with $\varepsilon\sim\mathcal{N}(0,1)$.

\begin{figure}[t!]
	\centering
	\caption{Performances of the optimality gaps by different algorithms for Rosenbrock function w.r.t. different sample sizes.}
	\vspace{7pt}
	\hspace*{-0.2cm}
	\includegraphics[scale = 0.5]{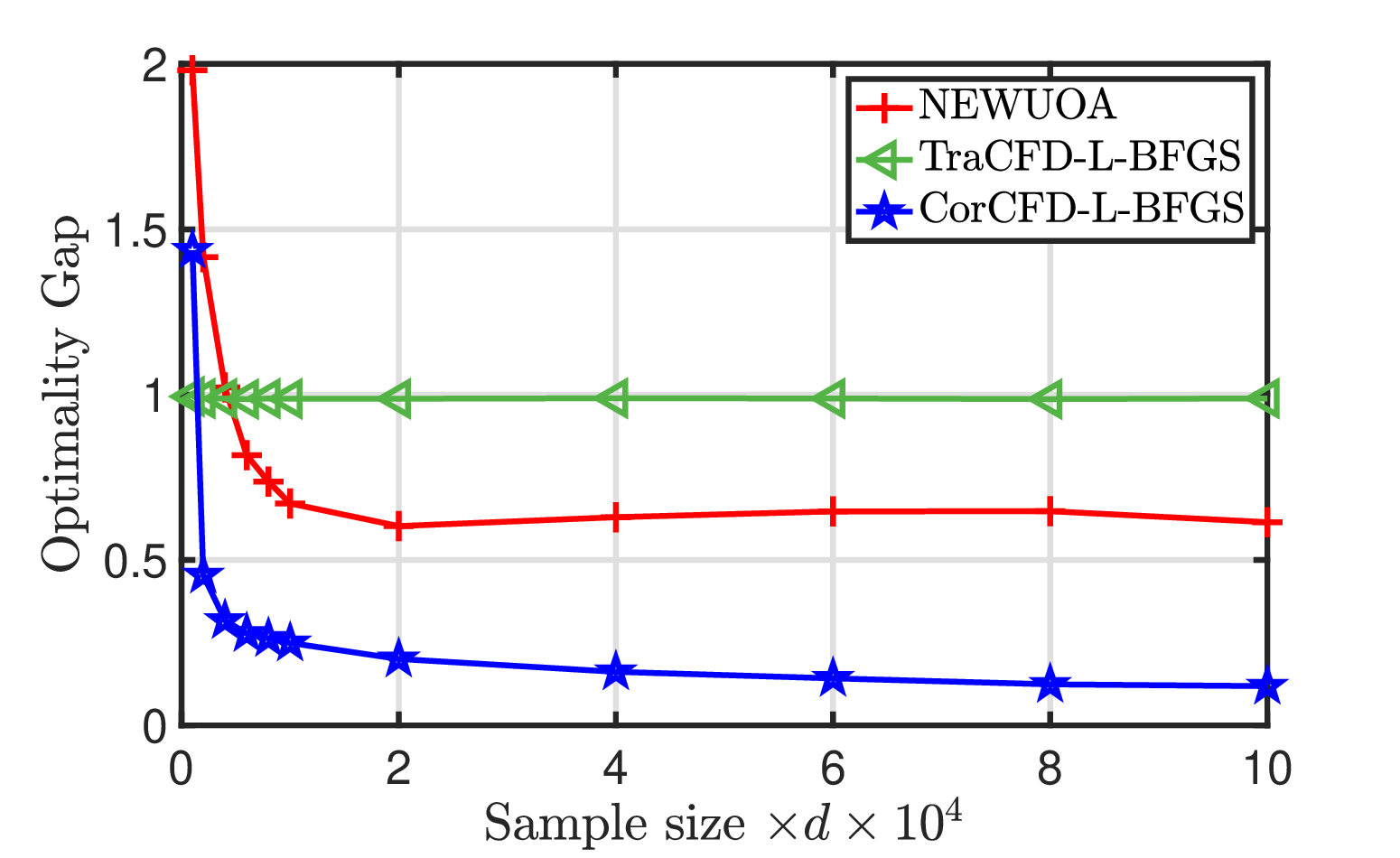}
	\label{fig:Banana}
\end{figure}

We show the optimality gap (OG, i,e., $\alpha(\theta_k) - \alpha(\theta^*)$, where $\theta^*$ is the global minimum point) for different algorithms on the Rosenbrock function in Figure \ref{fig:Banana}. Figure \ref{fig:Banana} reveals that the TraCFD-L-BFGS algorithm fails to converge due to the large error in gradient estimation when using only one sample pair, which prevents the algorithm from identifying the correct descent direction. In contrast, the CorCFD-L-BFGS algorithm employs a batch of samples to estimate the gradient, facilitating the algorithm's gradual convergence to the minimum point. In addition, compared to NEWUOA, the CorCFD-L-BFGS algorithm also produces better outputs. For example, when the sample-pair size is $10^5\times d$, the optimal gaps for the CorCFD-L-BFGS algorithm and NEWUOA are 0.12 and 0.61, respectively. 

Note that the TraCFD-L-BFGS algorithm is slightly better when the sample size is small. This may be attributed to two factors. Firstly, the TraCFD-L-BFGS algorithm undergoes more iterations than the CorCFD-L-BFGS algorithm because the finite difference estimator in the former is constructed by only one sample pair. Secondly, in the initial a few iterations, the gradient descent through the rough gradient estimation may be more important than the accuracy of the gradient descent direction. However, when the sample-pair size is relatively larger, i.e., $2\times10^3$, the CorCFD-L-BFGS algorithm outperforms the TraCFD-L-BFGS algorithm, and so the former is more recommended.

Next, we study the comparison among these algorithms on the Zakharov function. Table \ref{tab:Zakharov} exhibits the solution gap (SG, denoted by $||\theta_k - \theta^*||_2$) and OG related to these algorithms. It is worth mentioning that in this example, the starting point $\theta_0 = (\underbrace{1,...,1}_{d})^\top$, and the corresponding function values are about $f(\theta_0) = 1.3125$, $5.7191\times10^5$ and $4.0649\times10^{13}$ for $d=1, 10$ and 100, respectively. The optimal point and the optimal function value are $\theta^*=0$ and $f(\theta^*)=0$, respectively. In the table, ``$\approx 0$'' indicates a value less than $10^{-3}$. 

\begin{table}[t!]
\caption{Performances of the solution gaps and optimality gaps by different algorithms for Zakharov function w.r.t. different dimensions and different sample sizes.}
\centering
\vspace{7pt}
\label{tab:Zakharov}
\hspace*{-0.5cm}
\renewcommand{\arraystretch}{0.85}
\begin{tabular}{ccccccccccccc}
\hline
Dimension                           && \multicolumn{3}{c}{$d=1$}                      && \multicolumn{3}{c}{$d=10$}   && \multicolumn{3}{c}{$d=100$}   \\ 
 \cline{1-1}  \cline{3-5}  \cline{7-9}  \cline{11-13}
Sample Size ($\times d$)    && $10^3$ & $10^4$ & $10^5$                   && $10^3$ & $10^4$ & $10^5$  && $10^3$& $10^4$ & $10^5$ \\ 
 \cline{1-1}  \cline{3-5}  \cline{7-9}  \cline{11-13}
NEWUOA (SG)                   && 0.059    & $\approx 0$   & $\approx 0$  && 5.654    & 7.542    & 7.532    && 8.254    & 32.03    & 54.09    \\ 
TraCFD-L-BFGS (SG)          && 0.087    & 0.081    & 0.079                    && 3.591    & 3.684    & 3.697    && 7.718    & 14.92    & 17.20    \\ 
CorCFD-L-BFGS (SG)             && 0.002    & 0.002    & $\approx 0$                    && 0.330    & 0.099    & 0.056    && 4.930    & 1.825    & 0.694   \\ 
 \cline{1-1}  \cline{3-5}  \cline{7-9}  \cline{11-13}
NEWUOA (OG)                  && 0.018    & $\approx 0$    & $\approx 0$   && 30.93    & 57.38    & 57.25   && 68.51    & 1027     & 2927     \\ 
TraCFD-L-BFGS (OG)         && 0.015    & 0.013             & 0.012             && 37.33    & 34.74    & 36.23     && 688.4    & 1469     & 1015     \\ 
CorCFD-L-BFGS (OG)             && 0.004    & 0.002             & $\approx 0$             && 0.176    & 0.037    & 0.022     && 24.42    & 3.527    & 0.569    \\ 
\hline
\end{tabular}
\end{table}

From Table \ref{tab:Zakharov}, it can be seen that in the case of low dimension, e.g., $d=1$, the results of three algorithms are comparable in the sense of both SG and OG. All of them are close to 0. When the dimension is higher, e.g., $d=10$ and 100, the results of our proposed CorCFD-L-BFGS algorithm exhibit convergence as the sample-pair size increases, while those of the other algorithms do not. This is because the TraCFD-L-BFGS algorithm cannot find the correct gradient descent direction due to rough gradient estimation with only one sample pair, and the performance of NEWUOA may be improved through finding better algorithm parameter values tailored to the higher dimension cases. The detailed discussion on this topic will be provided in Appendix \ref{app:AdditionalResults1}. In particular, when $d=100$ and the sample-pair size is $10^3\times d$, our proposed CorCFD-L-BFGS algorithm reduces the estimated optimal function value to 24.42, down from the initial value of $4.0649\times10^{13}$. Furthermore, with an increased sample-pair size of $10^5\times d = 10^7$, the estimated optimal function value decreases to 0.569. It is worth mentioning that this sample-pair size, i.e., $10^7$ is considered acceptable for DFO problems with dimensions as large as $d=100$ \cite[see, e.g.,][]{bollapragada2024derivative}.

In addition, KW and SPSA algorithms are classical approaches in simulation optimization, a direct comparison with our proposed CorCFD-L-BFGS algorithm may not be entirely equitable. This is because KW and SPSA correspond to the first-order method, whereas CorCFD-L-BFGS corresponds to the second-order method. Nevertheless, for the sake of completeness, we provide a brief comparison and discussion of these methods in Appendix \ref{app:AdditionalResultSPSA}. A more comprehensive and in-depth comparative analysis of these algorithms could be left as a topic for future research.

\section{Conclusions}\label{sec:conclusions}

The choice of the perturbation significantly influences the accuracy of the FD estimator in stochastic gradient estimation. Although the order of the optimal perturbation is known, determining the unknown constants remains challenging. To address this issue, we have proposed a sample-driven method. Furthermore, we reuse samples, thereby inducing correlated samples to construct an efficient FD estimator. This estimator has an asymptotic variance reduction and possible bias reduction compared to the traditional optimal FD estimator. Numerical results demonstrate that our proposed correlation-induced FD estimator performs well even with small sample sizes. In addition, we propose an algorithm that combines our estimator with the renowned L-BFGS algorithm to solve DFO problems. Numerical results show that the proposed algorithm performs effectively when solving DFO problems with dimensions as large as 100.

For future work, bias reduction and variance reduction techniques for our proposed finite difference estimator could be explored. Section 7.1.2 in \citep{Glasserman2013Monte} reviews an extrapolation method for reducing the bias of conventional finite difference estimators, which could potentially be applied to our proposed estimator to achieve similar bias reduction.

On the other hand, as systems become increasingly complex, it is often easier to experiment with them rather than to fully understand their intricacies \citep{Golovin2017GoogleVizier}. Consequently, DFO is gaining importance, and FD approximation emerges as a powerful technique for solving DFO problems \citep{shi2023numerical}. However, this research area remains underdeveloped. In this paper, we establish an efficient FD estimator for practical implementation and propose Algorithm \ref{alg:algorithm_SO} to solve DFO problems. In this research direction, the first topic is to prove the convergence and convergence rate of this algorithm, which intuitively, can be derived from Theorem 1 in \cite{hu2024convergence}. The second one involves determining the optimal sample size $T_k$ in each iteration (e.g., $T_k$ in the step 3 of Algorithm \ref{alg:algorithm_SO}). The third one is how to apply the proposed FD estimator to DFO problems with large dimensions, such as $10^4$ or larger. A potential approach could be the combination of the FD estimator with coordinate descent methods \citep[see e.g.,][]{Nesterov2012CoordinateDescent}.

\bibliographystyle{plainnat}
\bibliography{mybibfile}

\newpage

\appendix
\renewcommand{\appendixname}{Appendix~\Alph{section}}
\section{Proofs in Section \ref{sec:bootstrap}}

\subsection{Proof of Theorem \ref{thm:bootstrap}} \label{app:prooftheorem1}
We prove \eqref{eq:bootstrap_bias} and \eqref{eq:bootstrap_variance} separately.

$\boldsymbol{\mbox{Proof of \eqref{eq:bootstrap_bias}}:}$
\begin{align}\label{eq:proof_bootstrap_bias}
	\E_{*}\widehat{\Delta}^{b}_{n_b,h} = \E_{*}[\Delta_1^{*}(h)] = \frac{1}{n_b}\sum_{i=1}^{n_b}\Delta_i(h) = \widetilde{\alpha}_h'(\theta_0) + \bar{Z}_{n_b,h},
\end{align}
where the first equality is due to the linearity of expectation, the second one is due to Table \ref{tab:pmf} and the last one is from \eqref{eq:estimate_expression}. Under Assumption \ref{ass:derivative5}, using the Taylor expansion, we get
\begin{align*}
	\alpha(\theta_0 + h) = \alpha(\theta_0) + \alpha'(\theta_0) h + \frac{\alpha^{(2)}(\theta_0)}{2} {h}^2 + B {h}^3 + \frac{\alpha^{(4)}(\theta_0)}{24}h^4 + \frac{\alpha^{(5)}(\theta_0)}{120}h^5 + o(h^5), \\
\alpha(\theta_0 - h) = \alpha(\theta_0) - \alpha'(\theta_0) h + \frac{\alpha^{(2)}(\theta_0)}{2} h^2 - B h^3 + \frac{\alpha^{(4)}(\theta_0)}{24}h^4 - \frac{\alpha^{(5)}(\theta_0)}{120}h^5 + o(h^5).
\end{align*}
Subtraction eliminates $\alpha^{(2)}(\theta_0)$ and $\alpha^{(4)}(\theta_0)$, leaving
\begin{align*}
	\widetilde{\alpha}_h'(\theta_0) = \alpha'(\theta_0) + B h^2 + D h^4 + o(h^4),
\end{align*}
where $D = \alpha^{(5)}(\theta_0)/120$. Combining with \eqref{eq:proof_bootstrap_bias} gives
\begin{align*}
	\E_{*}\widehat{\Delta}^{b}_{n_b,h} = \alpha'(\theta_0) + B h^2 + D h^4 + o(h^4) + \bar{Z}_{n_b,h}.
\end{align*}

$\boldsymbol{\mbox{Proof of \eqref{eq:bootstrap_variance}}:}$

Under $\P_{*}$, $\Delta_1^{*}(h),\Delta_2^{*}(h),...,\Delta_{n_b}^{*}(h)$ are i.i.d., so the variance of $\widehat{\Delta}^{b}_{n_b,h}$ is
\begin{align*}
	\Var_{*}\left[\widehat{\Delta}^{b}_{n_b,h}\right] &= \frac{1}{n_b}\Var_{*}[\Delta_1^{*}(h)] = \frac{1}{n_b}\left(\E_{*}\left[\Delta_1^{*}(h)\right]^2 - \left[\E_{*}\Delta_1^{*}(h)\right]^2\right) \\
	&= \frac{1}{n_b}\left(\frac{1}{n_b}\sum_{i=1}^{n_b}\Delta_i^2(h) - \left(\frac{1}{n_b}\sum_{i=1}^{n_b}\Delta_i(h)\right)^2\right)\\
	&= \frac{1}{n_b^2}\sum_{i=1}^{n_b}\left(\Delta_i(h) - \frac{1}{n_b}\sum_{j=1}^{n_b}\Delta_j(h)\right)^2.
\end{align*}
Denote
\begin{align}\label{eq:proof_sample_variance}
	S^2 = \frac{1}{n_b - 1}\sum_{i=1}^{n_b}\left(\Delta_i(h) - \frac{1}{n_b}\sum_{j=1}^{n_b}\Delta_j(h)\right)^2 = \frac{1}{n_b - 1}\sum_{i=1}^{n_b}\left(Z_i(h) - \bar{Z}_{n_b,h}\right)^2,
\end{align}
and then we have 
\begin{align}\label{eq:proof_bootstrap_variance}
	\Var_{*}\left[\widehat{\Delta}^{b}_{n_b,h}\right] = \frac{n_b-1}{n_b^2} S^2.
\end{align}

In the following we study $S^2$. Since $Z_1(h),Z_2(h),...,Z_{n_b}(h)$ are i.i.d., $S^2$ is an unbiased estimate of $\Var[Z_1(h)]$, i.e., 
\begin{align*}
	\E[S^2] = \Var[Z_1(h)] = \Var[\Delta_1(h)] = \frac{\sigma^2(\theta_0 + h) + \sigma^2(\theta_0 - h)}{4h^2}.
\end{align*}
Under Assumption \ref{ass:continuity}, we have $\sigma^2(\theta_0 + h) = \sigma^2(\theta_0) + o(1)$ and $\sigma^2(\theta_0 - h) = \sigma^2(\theta_0) + o(1)$. Therefore,
\begin{align}\label{eq:proof_mean_sample_variance}
	\E[S^2] = \Var[Z_1(h)] = \frac{\sigma^2(\theta_0) + o(1)}{2h^2}.
\end{align}
Let $\xi(h)$ is the zero-mean error associated with $S^2$ and we have
\begin{align}\label{eq:proof_sample_variance_rewrite}
	S^2 = \frac{\sigma^2(\theta_0) + o(1)}{2h^2} + \xi(h).
\end{align}
The variance of $\xi(h)$ is
\begin{align}\label{eq:proof_variance_sample_variance}
	\Var[\xi(h)] = \E[\xi(h)]^2 = \E\left[S^2 - \frac{\sigma^2(\theta_0) + o(1)}{2 h^2}\right]^2 = \E[S^2]^2 - \frac{\sigma^4(\theta_0) + o(1)}{4h^4}.
\end{align}
We calculate $\E[S^2]^2$ as follows. It follows from \eqref{eq:proof_sample_variance} that
\begin{align}
	&\quad \ (n_b - 1)^2 \E[S^2]^2\nonumber \\
	&= \E\left[\sum_{i=1}^{n_b}\left(Z_i(h) - \bar{Z}_{n_b,h}\right)^2 \right]^2 = \E\left[\sum_{i=1}^{n_b}Z_i^2(h) - n_b \bar{Z}_{n_b}^2(h)\right] ^2 \nonumber \\
	&= \E\left[\sum_{i=1}^{n_b}Z_i^2(h)\right]^2 - 2n_b\E\left[\sum_{i=1}^{n_b}Z_i^2(h)\bar{Z}_{n_b}^2(h)\right] + n_b^2\E\left[\bar{Z}_{n_b,h}^4\right]\nonumber \\
	&= \E\left[\sum_{i=1}^{n_b}Z_i^2(h)\right]^2 - 2n_b^2\E\left[Z_1^2(h)\bar{Z}_{n_b}^2(h)\right] + n_b^2\E\left[Z_1(h)\bar{Z}_{n_b}^3(h)\right]\nonumber\\
	&= \E\left[\sum_{i=1}^{n_b}Z_i^2(h)\right]^2 - 2\E\left[Z_1^2(h)\left(\sum_{i=1}^{n_b}Z_i(h)\right)^2\right] + \frac{1}{n_b}\E\left[Z_1(h)\left(\sum_{i=1}^{n_b}Z_i(h)\right)^3\right],\label{eq:proof_second_moment_sample_variance}
\end{align}
where the fourth equality holds because for any $1 \leq i \leq n_b$, we have 
\begin{align*}
	\E\left[Z_i^2(h)\bar{Z}_{n_b}^2(h)\right] = \E\left[Z_1^2(h)\bar{Z}_{n_b}^2(h)\right],\quad \E\left[Z_i(h)\bar{Z}_{n_b}^3(h)\right] = \E\left[Z_1(h)\bar{Z}_{n_b}^3(h)\right].
\end{align*}

In the following, we study the three terms on RHS of \eqref{eq:proof_second_moment_sample_variance} separately.
\begin{align}
	\mbox{The first term on the RHS of \eqref{eq:proof_second_moment_sample_variance}} &= \E\left[\sum_{i=1}^{n_b}Z_i^4(h) + 2\sum_{i<j}Z_i^2(h)Z_j^2(h)\right]\nonumber\\
	&=n_b\E[Z_1(h)]^4 + n_b(n_b - 1)\E[Z_1(h)]^2 \E[Z_2(h)]^2\nonumber\\
	 &= n_b\nu_4(h) + n_b(n_b - 1)\frac{\sigma^4(\theta_0) + o(1)}{4h^4},\label{eq:proof_second_moment_sample_variance_1}
\end{align}
where $\nu_4(h) = \E[Z_1(h)]^4$ and $\E[Z_1(h)]^2$ and $\E[Z_2(h)]^2$ are equal to $\Var[Z_1(h)]$ which can be found in \eqref{eq:proof_mean_sample_variance}. Then, we calculate the second term on the RHS of \eqref{eq:proof_second_moment_sample_variance} as
\begin{align}
	&\mbox{The second term on the RHS of \eqref{eq:proof_second_moment_sample_variance}}\nonumber \\
	 =& -2\E\left[Z_1^2(h)\left(\sum_{i=1}^{n_b}Z_i^2(h) + 2\sum_{i<j}Z_i(h)Z_j(h)\right)\right]\nonumber \\
	 =& -2\E[Z_1(h)]^4 - 2\E\left[Z_1^2(h)\sum_{i > 1}Z_i^2(h)\right] - 4\E\left[Z_1^2(h)\sum_{i<j}Z_i(h)Z_j(h)\right]\nonumber\\
	 =& -2\nu_4(h) -2(n_b - 1)\E[Z_1^2(h)] \E[Z_2^2(h)] - 4\E\left[Z_1^3(h)\sum_{j > 1}Z_j(h) + Z_1^2(h)\sum_{1<i<j}Z_i(h)Z_j(h)\right]\nonumber\\
	 =& -2\nu_4(h) -2(n_b - 1)\frac{\sigma^4(\theta_0) + o(1)}{4h^4},\label{eq:proof_second_moment_sample_variance_2}
\end{align}
where the last equality is because for any $j > 1$, $\E[Z_1^3(h)Z_j(h)] = \E[Z_1^3(h)]\E[Z_j(h)] = 0$, and for any $1 < i < j$, $\E[Z_1^2(h)Z_i(h)Z_j(h)] = \E[Z_1^2(h)]\E[Z_i(h)]\E[Z_j(h)] = 0$.
\begin{align}
	&\mbox{The third term on the RHS of \eqref{eq:proof_second_moment_sample_variance}}\nonumber\\
	=& \frac{1}{n_b}\E\left[Z_1(h)\left(\sum_{i=1}^{n_b}Z_i^3(h) + \sum_{i \neq j} 3Z_i^2(h)Z_j(h) + \sum_{i<j<k}6Z_i(h)Z_j(h)Z_k(h)\right)\right]\nonumber\\
	=& \frac{1}{n_b}\E\left[Z_1^4(h) + Z_1(h)\sum_{i>1}Z_i^3(h) + \sum_{j>1}3Z_1^3(h)Z_j(h) + \sum_{i>1}3Z_1^2(h)Z_i^2(h) + \sum_{i,j>1, i\neq j}3Z_1(h)Z_i^2(h)Z_j(h)\right.\nonumber\\
	 &+ \left. \sum_{1<j<k}6Z_1^2(h)Z_j(h)Z_k(h) + \sum_{1<i<j<k}6Z_1(h)Z_i(h)Z_j(h)Z_k(h)\right]\nonumber\\
	=& \frac{1}{n_b}\E\left[Z_1^4(h) + \sum_{i > 1}3Z_1^2(h)Z_i^2(h)\right] =  \frac{\nu_4(h)}{n_b} + \frac{3(n_b-1)}{n_b}\frac{\sigma^4(\theta_0) + o(1)}{4h^4},\label{eq:proof_second_moment_sample_variance_3}
\end{align}
where the third equality is because for any $i\neq j$, $\E[Z_i(h)Z_j^3(h)] = \E[Z_i(h)]\E[Z_j^3(h)] = 0$, and for any $i\neq j, i\neq k, j\neq k$, $\E[Z_i^2(h)Z_j(h)Z_k(h)] = \E[Z_i^2(h)]\E[Z_j(h)]\E[Z_k(h)] = 0, \E[Z_1(h)Z_i(h)Z_j(h)Z_k(h)] = \E[Z_1(h)]\E[Z_i(h)]\E[Z_j(h)]\E[Z_k(h)] = 0$.
Combine \eqref{eq:proof_second_moment_sample_variance}, \eqref{eq:proof_second_moment_sample_variance_1}, \eqref{eq:proof_second_moment_sample_variance_2} and \eqref{eq:proof_second_moment_sample_variance_3}, leading to
\begin{align*}
	(n_b - 1)^2 \E[S^2]^2 = \frac{(n_b-1)^2}{n_b}\nu_4(h) + \frac{(n_b^2 - 2n_b + 3)(n_b - 1)}{n_b}\frac{\sigma^4(\theta_0) + o(1)}{4h^4}.
\end{align*}
Therefore,
\begin{align*}
	\E[S^2]^2 = \frac{\nu_4(h)}{n_b} + \frac{n_b^2 - 2n_b + 3}{n_b(n_b - 1)}\frac{\sigma^4(\theta_0) + o(1)}{4h^4},
\end{align*}
and \eqref{eq:proof_variance_sample_variance} is
\begin{align}\label{eq:proof_variance_sample_variance_rewrite}
	\Var[\xi(h)] = \frac{\nu_4(h)}{n_b} - \frac{n_b - 3}{n_b(n_b - 1)}\frac{\sigma^4(\theta_0) + o(1)}{4h^4}.
\end{align}
According to \eqref{eq:proof_bootstrap_variance} and \eqref{eq:proof_sample_variance_rewrite}, we have
\begin{align*}
	\Var_{*}\left[\widehat{\Delta}^{b}_{n_b,h}\right] = \frac{(n_b - 1)(\sigma^2(\theta_0) + o(1))}{2n_b^2 h^2} + \phi(h), 
\end{align*}
where $\phi(h) = (n_b - 1)\xi(h)/n_b^2$ and 
	\begin{align*}
		\Var[\phi(h)] &= \frac{(n_b-1)^2}{n_b^4} \left(\frac{\nu_4(h)}{n_b} - \frac{n_b - 3}{n_b(n_b - 1)}\frac{\sigma^4(\theta_0) + o(1)}{4h^4}\right)\\
		&= \frac{(n_b-1)^2}{n_b^4h^4} \left(\frac{\nu_4 + o(1)}{n_b} - \frac{n_b - 3}{n_b(n_b - 1)}\frac{\sigma^4(\theta_0) + o(1)}{4}\right),
	\end{align*}
where the second equality is because
\begin{align*}
	\nu_4(h) = \E[Z_1(h)]^4 = \frac{\E[hZ_1(h)]^4}{h^4} = \frac{\nu_4 + o(1)}{h^4}.
\end{align*}
The proof is complete.

\subsection{Proof of Theorem \ref{thm:B_consistence}} \label{app:prooftheorem2}
Using the least squares method, we get
	\begin{align*}
	\widehat{\bbeta}_e &= \left(\boldsymbol{X}_e^{\top} \boldsymbol{X}_e\right)^{-1} \boldsymbol{X}_e^{\top} \boldsymbol{Y}_e.
	\end{align*}
    Note that
    \begin{align*}
	\bbeta_e &= \left(\boldsymbol{X}_e^{\top} \boldsymbol{X}_e\right)^{-1} \boldsymbol{X}_e^{\top} \boldsymbol{X}_e \bbeta_e.
	\end{align*}
	Then, according to \eqref{eq:regression_component_bias}, we have
	\begin{align*}
		\widehat{\bbeta}_e - \bbeta_e &= \left(\boldsymbol{X}_e^{\top} \boldsymbol{X}_e\right)^{-1} \boldsymbol{X}_e^{\top}(\boldsymbol{Y}_e - \boldsymbol{X}_e \bbeta_e) = \left(\boldsymbol{X}_e^{\top} \boldsymbol{X}_e\right)^{-1} \boldsymbol{X}_e^{\top}\cE_e\\
		&= \left(\left[\begin{array}{lll}
	1 & ... & 1 \\
	h_1^2 & ... & h_{K}^2
	\end{array}\right] \left[\begin{array}{lll}
	1 & ... & 1 \\
	h_1^2 & ... & h_{K}^2
	\end{array}\right]^{\top}\right)^{-1}\left[\begin{array}{lll}
	1 & ... & 1 \\
	h_1^2 & ... & h_{K}^2
	\end{array}\right]\left[\begin{array}{c}
	D h_1^4 + o(h_1^4) + \bar{Z}_{n_b,h_1} \\
	\vdots \\
	D h_K^4 + o(h_K^4) + \bar{Z}_{n_b,h_K}
	\end{array}\right]\\
	&=\left[\begin{array}{cc}
	K & \sum_{k=1}^{K}h_k^2 \\
	\sum_{k=1}^{K}h_k^2& \sum_{k=1}^{K}h_k^4
	\end{array}\right]^{-1}\left[\begin{array}{c}
	\sum_{k=1}^{K}\left(D h_k^4 + o(h_k^4) + \bar{Z}_{n_b,h_k}\right) \\
	\sum_{k=1}^{K}\left(D h_k^6 + o(h_k^4) + h_k^2\bar{Z}_{n_b,h_k}\right)
	\end{array}\right]\\
	&=\frac{\left[\begin{array}{cc}
	\sum_{k=1}^K h_k^4 & -\sum_{k=1}^K h_k^2 \\
	-\sum_{k=1}^K h_k^2 & K
	\end{array}\right]\left[\begin{array}{c}
	\sum_{k=1}^{K}\left(D h_k^4 + o(h_k^4) + \bar{Z}_{n_b,h_k}\right) \\
	\sum_{k=1}^{K}\left(D h_k^6 + o(h_k^6) + h_k^2\bar{Z}_{n_b,h_k}\right)
	\end{array}\right]}{K \sum_{k=1}^K h_k^4-\left(\sum_{k=1}^K h_k^2\right)^2}.
	\end{align*}
	Therefore,
	\begin{align*}
	\widehat{B} - B = \frac{K \sum_{k=1}^K\left(D h_k^6+o(h_k^6)+h_k^2 \bar{Z}_{n_b,h_k}\right)-\sum_{k=1}^K h_k^2 \sum_{k=1}^K\left(D h_k^4+o(h_k^4)+\bar{Z}_{n_b,h_k}\right)}{K \sum_{k=1}^K h_k^4-\left(\sum_{k=1}^K h_k^2\right)^2}.
	\end{align*}
	
	We divide $\widehat{B} - B$ into two parts, i.e.,
	\begin{align}
		\widehat{B} - B = &\frac{K \sum_{k=1}^K\left(D h_k^6+o(h_k^6)\right) - \sum_{k=1}^K h_k^2 \sum_{k=1}^K\left(D h_k^4+o(h_k^4)\right)}{K \sum_{k=1}^K h_k^4-\left(\sum_{k=1}^K h_k^2\right)^2}\label{eq:proof_B_nonrandom}\\
	 	&+ \frac{K \sum_{k=1}^Kh_k^2 \bar{Z}_{n_b,h_k}-\sum_{k=1}^K h_k^2 \sum_{k=1}^K\bar{Z}_{n_b,h_k}}{K \sum_{k=1}^K h_k^4-\left(\sum_{k=1}^K h_k^2\right)^2}.\label{eq:proof_B_random}
	\end{align}
	Note that \eqref{eq:proof_B_nonrandom} is the non-random term, while \eqref{eq:proof_B_random} is the random term whose variance is the same as that of $\widehat{B}$. We now separately calculate the bias, variance, and then the MSE of $\widehat{B}$. 
	
	For any $h_k$, $\E\left[\bar{Z}_{n_b,h_k}\right] = 0$, so the expectation of \eqref{eq:proof_B_random} is 0. Furthermore, the bias of $\widehat{B}$ is
	\begin{align}\label{eq:proof_B_bias}
		\mbox{Bias}\left[\widehat{B}\right] = \E\left[\widehat{B} - B\right] &= \frac{K \sum_{k=1}^K\left(D h_k^6+o(h_k^6)\right) - \sum_{k=1}^K h_k^2 \sum_{k=1}^K\left(D h_k^4+o(h_k^4)\right)}{K \sum_{k=1}^K h_k^4-\left(\sum_{k=1}^K h_k^2\right)^2}\nonumber\\
		&= \frac{K \sum_{k=1}^K Dc_k^6 - \sum_{k=1}^K c_k^2 \sum_{k=1}^K D c_k^4}{K \sum_{k=1}^K c_k^4-\left(\sum_{k=1}^K c_k^2\right)^2} n_b^{2\gamma} + o\left(n_b^{2\gamma}\right)\nonumber\\
		&= H_Kn_b^{2\gamma} + o\left(n_b^{2\gamma}\right),
	\end{align}
	where
	\begin{align*}
		H_K = \frac{K \sum_{k=1}^K Dc_k^6 - \sum_{k=1}^K c_k^2 \sum_{k=1}^K D c_k^4}{K \sum_{k=1}^K c_k^4-\left(\sum_{k=1}^K c_k^2\right)^2}.
	\end{align*}
	
	For the variance of $\widehat{B}$, by \eqref{eq:proof_B_random}, 
	\begin{align}
		&\Var\left[\widehat{B}\right]\left(K \sum_{k=1}^K h_k^4-\left(\sum_{k=1}^K h_k^2\right)^2\right)^2 = \Var\left[K \sum_{k=1}^Kh_k^2 \bar{Z}_{n_b,h_k}-\sum_{k=1}^K h_k^2 \sum_{k=1}^K\bar{Z}_{n_b,h_k}\right]\nonumber\\
		=& K^2 \sum_{k=1}^Kh_k^4 \Var\left[\bar{Z}_{n_b,h_k}\right] - 2K\sum_{k=1}^K h_k^2 \sum_{k=1}^K h_k^2\Var\left[\bar{Z}_{n_b,h_k}\right] + \left(\sum_{k=1}^{K}h_k^2\right)^2\sum_{k=1}^{K}\Var\left[\bar{Z}_{n_b,h_k}\right]\nonumber\\
		=& \frac{\sigma^2(\theta_0) + o(1)}{2n_b} \left(K^2 \sum_{k=1}^Kh_k^2 - 2K^2\sum_{k=1}^K h_k^2 + \left(\sum_{k=1}^{K}h_k^2\right)^2\sum_{k=1}^{K}\frac{1}{h_k^2}\right)\nonumber\\
		=& \frac{\sigma^2(\theta_0) + o(1)}{2n_b^{1+6\gamma}} \left(- K^2\sum_{k=1}^K c_k^2 + \left(\sum_{k=1}^{K}c_k^2\right)^2\sum_{k=1}^{K}\frac{1}{c_k^2}\right).\nonumber
	\end{align}
	Furthermore, the variance of $\widehat{B}$ is
	\begin{align}\label{eq:proof_B_variance}
		\Var\left[\widehat{B}\right] = V_K \frac{\sigma^2(\theta_0) + o(1)}{2n_b^{1+6\gamma}},
	\end{align}
	where
	\begin{align*}
		V_K  = \frac{- K^2\sum_{k=1}^K c_k^2 + \left(\sum_{k=1}^{K}c_k^2\right)^2\sum_{k=1}^{K}1/c_k^2}{\left(K \sum_{k=1}^K c_k^4-\left(\sum_{k=1}^K c_k^2\right)^2\right)^2}.
	\end{align*}
	
	Combining \eqref{eq:proof_B_bias} and \eqref{eq:proof_B_variance} gives the MSE of $\widehat{B}$.
	\begin{align}\label{eq:proof_B_MSE}
		\mbox{MSE}\left[\widehat{B}\right] = \mbox{Bias}^2\left[\widehat{B}\right] + \Var\left[\widehat{B}\right] = H_K^2n_b^{4\gamma} + o\left(n_b^{4\gamma}\right) + V_K \frac{\sigma^2(\theta_0) + o(1)}{2n_b^{1+6\gamma}}.
	\end{align}
	From \eqref{eq:proof_B_MSE}, when $-1/6 < \gamma < 0$, we have $\mbox{MSE}\left[\widehat{B}\right] \to 0$ as $n_b \to \infty$. Therefore, according to Chebyshev's inequality, for any $\epsilon > 0$, as $n_b \to \infty$
	\begin{align*}
	\P\left(\left|\widehat{B} - B\right| \geq \epsilon \right) \leq \frac{\E\left[\widehat{B} - B\right]^2}{\epsilon^2} = \frac{\mbox{MSE}\left[\widehat{B}\right]}{\epsilon^2}\to 0.
	\end{align*}
	So $\widehat{B} \stackrel{p}{\longrightarrow} B$.
	The proof is complete.

\subsection{Asymptotic Property of $\widehat{\alpha}'(\theta_0)$} \label{app:property_alpha}
\begin{corollary}\label{cor:alpha_consistence}
	Under the same conditions as in Theorem \ref{thm:B_consistence}, we have
	\begin{align}\label{eq:alpha_bias_variance}
	\E\left[\widehat{\alpha}'(\theta_0) - \alpha'(\theta_0)\right] = \widetilde{H}_{K}n_b^{4\gamma} + o\left(n_b^{4\gamma}\right), \quad \Var\left[\widehat{\alpha}'(\theta_0)\right] = \widetilde{V}_{K} \frac{\sigma^2(\theta_0) + o(1)}{2n_b^{1+2\gamma}},
	\end{align}
	where $\widetilde{H}_K$ and $\widetilde{V}_K$ are constants depending on $c_1,...,c_K$. Specifically,
	\begin{align*}
	\widetilde{H}_{K} &= \frac{\sum_{k=1}^{K}c_k^4 \sum_{k=1}^K D c_k^4 - \sum_{k=1}^K c_k^2 \sum_{k=1}^K D c_k^6}{K \sum_{k=1}^K c_k^4-\left(\sum_{k=1}^K c_k^2\right)^2},\\
		\widetilde{V}_{K} &= \frac{\left(\sum_{k=1}^{K}c_k^2\right)^3 - 2K\sum_{k=1}^{K}c_k^4\sum_{k=1}^{K}c_k^2 + \left(\sum_{k=1}^{K}c_k^4\right)^2 \sum_{k=1}^{K}1/c_k^2}{\left({K \sum_{k=1}^K c_k^4-\left(\sum_{k=1}^K c_k^2\right)^2}\right)^2}.
	\end{align*}
	In addition, if $-1/6 < \gamma < 0$, then $\widehat{\alpha}'(\theta_0) \stackrel{p}{\longrightarrow} \alpha'(\theta_0)$.
\end{corollary}

The proof of Corollary \ref{cor:alpha_consistence} is identical to that of Theorem \ref{thm:B_consistence}. \eqref{eq:alpha_bias_variance} provides the convergence rates for the bias and the variance of $\widehat{\alpha}'(\theta_0)$, which are $O\left(n_b^{4\gamma}\right)$ and $O\left(n_b^{-1 - 2\gamma}\right)$, respectively. Additionally, when $c_1,...,c_K$ are generated from a proper distribution $\mathcal{P}_0$, the convergence rates of the bias and the variance of $\widehat{\alpha}'(\theta_0)$ are $O\left(n_b^{4\gamma}\right)$ and $O\left(n_b^{8\gamma}\right) + O\left(n_b^{-1 - 2\gamma}\right)$, respectively.

\subsection{Analysis of \eqref{eq:bootstrap_variance_new}} \label{app:eq:bootstrap_variance_new}

Under Assumption \ref{ass:derivative1}, the left hand side (LHS) of \eqref{eq:proof_mean_sample_variance} is
\begin{align*}
	&\E[S^2] = \Var[Z_1(h)] = \Var[\Delta_1(h)]
	 = \frac{\sigma^2(\theta_0 + h) + \sigma^2(\theta_0 - h)}{4h^2}\\
	 =& \frac{\left(\sigma(\theta_0) + \sigma'(\theta_0)h + o(h)\right)^2 + \left(\sigma(\theta_0) - \sigma'(\theta_0)h + o(h)\right)^2}{4h^2}
	 = \frac{\sigma^2(\theta_0) + \sigma'(\theta_0)^2h^2 + o(h^2)}{2h^2},
\end{align*}
where the fourth equality is due to the Taylor expansion. Then, following the proof of Theorem \ref{thm:bootstrap}, \eqref{eq:bootstrap_variance_new} is obtained.

\subsection{Proof of Theorem \ref{thm:sigma_consistence}} \label{app:prooftheorem3}

Using the least squares method, we get
\begin{align*}
\widehat{\bbeta}_v &= \left(\boldsymbol{X}_v^{\top} \boldsymbol{X}_v\right)^{-1} \boldsymbol{X}_v^{\top} \boldsymbol{Y}_v. 
\end{align*}
Note that
\begin{align*}
\bbeta_v &= \left(\boldsymbol{X}_v^{\top} \boldsymbol{X}_v\right)^{-1} \boldsymbol{X}_v^{\top} \boldsymbol{X}_v \bbeta_v.
\end{align*}
Then, according to \eqref{eq:regression_component_variance_new}, we have
\begin{align*}
	\widehat{\bbeta}_v - \bbeta_v &= \left(\boldsymbol{X}_v^{\top} \boldsymbol{X}_v\right)^{-1} \boldsymbol{X}_v^{\top}(\boldsymbol{Y}_v - \boldsymbol{X}_v \bbeta_v) = \left(\boldsymbol{X}_v^{\top} \boldsymbol{X}_v\right)^{-1} \boldsymbol{X}_v^{\top}\cE_v\\
	&= \left(\sum_{k=1}^K\frac{(n_b -1)^2}{4n_b^4h_k^4}\right)^{-1}\sum_{k=1}^{K}\left(\frac{(n_b - 1)^2 \left[\sigma'(\theta_0)^2h_k^2 + o(h_k^2)\right]}{4n_b^4h_k^4} + \frac{n_b - 1}{2n_b^2h_k^2}\phi(h_k)\right).
\end{align*}

We divide $\widehat{\bbeta}_v - \bbeta_v$ into two parts, i.e., 
\begin{align}
	\widehat{\bbeta}_v - \bbeta_v = &\left(\sum_{k=1}^K\frac{(n_b -1)^2}{4n_b^4h_k^4}\right)^{-1}\sum_{k=1}^{K}\frac{(n_b - 1)^2 \left[\sigma'(\theta_0)^2h_k^2 + o(h_k^2)\right]}{4n_b^4h_k^4}\label{eq:proof_sigma_nonrandom}\\
	&+ \left(\sum_{k=1}^K\frac{(n_b -1)^2}{4n_b^4h_k^4}\right)^{-1}\sum_{k=1}^{K}\frac{n_b - 1}{2n_b^2h_k^2}\phi(h_k).\label{eq:proof_sigma_random}
\end{align}
Note that \eqref{eq:proof_sigma_nonrandom} is the non-random term, while \eqref{eq:proof_sigma_random} is the random term whose variance is the same as that of $\widehat{\bbeta}_v$. We now separately calculate the bias, variance, and then the MSE of $\widehat{\bbeta}_v$.

For any $h_k$, $\E\left[\phi(h_k)\right] = 0$, so the expectation of \eqref{eq:proof_sigma_random} is 0. Furthermore, the bias of $\widehat{\bbeta}_v$ is
\begin{align}\label{eq:proof_sigma_bias}
	\mbox{Bias}\left[\widehat{\bbeta}_v\right] = \E\left[\widehat{\bbeta}_v - \bbeta_v\right] &= \left(\sum_{k=1}^K\frac{(n_b -1)^2}{4n_b^4h_k^4}\right)^{-1}\sum_{k=1}^{K}\frac{(n_b - 1)^2 \left[\sigma'(\theta_0)^2h_k^2 + o(h_k^2)\right]}{4n_b^4h_k^4}\nonumber\\
	&= \left(\sum_{k=1}^K\frac{(n_b -1)^2}{4n_b^4c_k^4n_b^{4\gamma}}\right)^{-1}\sum_{k=1}^{K}\frac{(n_b - 1)^2 \left[\sigma'(\theta_0)^2c_k^2n_b^{2\gamma} + o\left(n_b^{2\gamma}\right)\right]}{4n_b^4c_k^4n_b^{4\gamma}}\nonumber\\
	&= \frac{4n_b^{4+4\gamma}}{(n_b -1)^2}\left(\sum_{k=1}^K\frac{1}{c_k^4}\right)^{-1}\frac{\sigma'(\theta_0)^2(n_b -1)^2n_b^{2\gamma}}{4n_b^{4+4\gamma}}\sum_{k=1}^K\frac{1 + o(1)}{c_k^2}\nonumber\\
	&= \widehat{H}_Kn_b^{2\gamma} + o\left(n_b^{2\gamma}\right),
\end{align}
where
\begin{align*}
	\widehat{H}_K = \sigma'(\theta_0)^2\left(\sum_{k=1}^K\frac{1}{c_k^4}\right)^{-1}\sum_{k=1}^K\frac{1}{c_k^2}.
\end{align*}

The variance of $\widehat{\bbeta}_v$ is 
\begin{align}\label{eq:proof_sigma_variance}
	\Var\left[\widehat{\bbeta}_v\right] &= \left(\sum_{k=1}^K\frac{(n_b -1)^2}{4n_b^4h_k^4}\right)^{-2}\sum_{k=1}^{K}\frac{(n_b - 1)^2}{4n_b^4h_k^4}\Var[\phi(h_k)]\nonumber\\
	&= \left(\sum_{k=1}^K\frac{(n_b -1)^2}{4n_b^4c_k^4n_b^{4\gamma}}\right)^{-2}\sum_{k=1}^{K}\frac{(n_b - 1)^2}{4n_b^4c_k^4n_b^{4\gamma}}\Var[\phi(h_k)]\nonumber\\
	&= \frac{16n_b^{8+8\gamma}}{(n_b-1)^4}\left(\sum_{k=1}^K\frac{1}{c_k^4}\right)^{-2}\frac{(n_b - 1)^2}{4n_b^{4+4\gamma}}\sum_{k=1}^{K}\frac{\Var[\phi(h_k)]}{c_k^4}\nonumber\\
	&= \frac{4n_b^{4+4\gamma}}{(n_b-1)^2}\left(\sum_{k=1}^K\frac{1}{c_k^4}\right)^{-2}\sum_{k=1}^{K}\frac{\Var[\phi(h_k)]}{c_k^4}\nonumber\\
	&= \frac{4n_b^{4+4\gamma}}{(n_b-1)^2}\left(\sum_{k=1}^K\frac{1}{c_k^4}\right)^{-2}\sum_{k=1}^{K}\frac{1}{c_k^4}\frac{(n_b-1)^2}{n_b^4h_k^4} \left(\frac{\nu_4 + o(1)}{n_b} - \frac{n_b - 3}{n_b(n_b - 1)}\frac{\sigma^4(\theta_0) + o(1)}{4}\right)\nonumber\\
	&= \frac{4n_b^{4+4\gamma}}{(n_b-1)^2}\left(\sum_{k=1}^K\frac{1}{c_k^4}\right)^{-2}\sum_{k=1}^{K}\frac{1}{c_k^4}\frac{(n_b-1)^2}{n_b^4c_k^4n_b^{4\gamma}} \left(\frac{\nu_4 + o(1)}{n_b} - \frac{n_b - 3}{n_b(n_b - 1)}\frac{\sigma^4(\theta_0) + o(1)}{4}\right)\nonumber\\
	&= \widehat{V}_K \frac{4\nu_4(n_b-1) - \sigma^4(\theta_0)(n_b-3)}{n_b(n_b-1)} + o\left(\frac{1}{n_b}\right),
\end{align}
where the fifth equality is from Theorem \ref{thm:bootstrap} and 
\begin{align*}
	\widehat{V}_K = \left(\sum_{k=1}^K\frac{1}{c_k^4}\right)^{-2}\sum_{k=1}^{K}\frac{1}{c_k^8}.
\end{align*}

Combining \eqref{eq:proof_sigma_bias} and \eqref{eq:proof_sigma_variance} gives the MSE of $\widehat{\bbeta}_v$.
	\begin{align}\label{eq:proof_sigma_MSE}
		\mbox{MSE}\left[\widehat{\bbeta}_v\right] &= \mbox{Bias}^2\left[\widehat{\bbeta}_v\right] + \Var\left[\widehat{\bbeta}_v\right]\nonumber\\
		&= \widehat{H}_K^2n_b^{4\gamma} + o\left(n_b^{4\gamma}\right) + \widehat{V}_K \frac{4\nu_4(n_b-1) - \sigma^4(\theta_0)(n_b-3)}{n_b(n_b-1)} + o\left(\frac{1}{n_b}\right)
	\end{align}
	From \eqref{eq:proof_sigma_MSE}, when $\gamma < 0$, we have $\mbox{MSE}\left[\widehat{\bbeta}_v\right] \to 0$ as $n_b \to \infty$. Therefore, according to Chebyshev's inequality, for any $\epsilon > 0$, as $n_b \to \infty$
	\begin{align*}
	\P\left(\left|\widehat{\bbeta}_v - \bbeta_v\right| \geq \epsilon \right) \leq \frac{\E\left[\widehat{\bbeta}_v - \bbeta_v\right]^2}{\epsilon^2} = \frac{\mbox{MSE}\left[\widehat{\bbeta}_v\right]}{\epsilon^2}\to 0.
	\end{align*}
The proof is complete.

\section{Discussions and Proofs in Section \ref{sec:DSR}}

\subsection{Further Explanations on \eqref{eq:nonrecycling_bias_var_MSE}} \label{app:widetildeB}

We need to demonstrate that the leading terms and $o_p(\cdot)$ terms in \eqref{eq:bias_variance_conditional} converge to their corresponding terms in \eqref{eq:nonrecycling_bias_var_MSE}. To prevent excessive technical complexity of proving the convergence of the $o_p(\cdot)$ terms in \eqref{eq:bias_variance_conditional}, we adopt the operation in \cite{hong2017kernel} and thus directly assume that their uniform integrability holds, which ensures their convergence. Next, we prove the convergence of the leading terms in \eqref{eq:bias_variance_conditional}.

The continuous mapping theorem implies that $\left({\widehat{\bbeta}_v}/{\widehat{B}^2}\right)^{1/3}$ and $\left({\widehat{B}^2}/{\widehat{\bbeta}_v}\right)^{1/3}$ converge to $\left({\bbeta_v}/{B^2}\right)^{1/3}$ and $\left({B^2}/{\bbeta_v}\right)^{1/3}$ in probability, respectively. To prove the convergence of the leading terms in \eqref{eq:bias_variance_conditional}, we aim to apply the dominated convergence theorem. This theorem requires that $\left({\widehat{\bbeta}_v}/{\widehat{B}^2}\right)^{1/3}$ and $\left({\widehat{B}^2}/{\widehat{\bbeta}_v}\right)^{1/3}$ are bounded by some integrable random variable. This requirement is not apparently satisfied because 0 cannot be excluded from the support sets of the regression estimators $\widehat{\bbeta}_v$ and $\widehat{B}$. In the following, we provide a technique, inspired by \cite{hong2017kernel}, to prove the convergence of the leading terms in \eqref{eq:bias_variance_conditional}.

Take the convergence of the bias term in \eqref{eq:bias_variance_conditional} for example. Note that $B \neq 0$. Specifically, we can assume that there exists a positive number $\epsilon$ such that $B > \epsilon > 0$. Then, we can solve \eqref{eq:regression_problem_bias} under the constraint $\widehat{B} > \epsilon$. In others words, define a new estimator $\widetilde{B} = \epsilon + \max{\left\{\widehat{B} - \epsilon,0\right\}}$ and replace $\widehat{B}$ with $\widetilde{B}$ in the bias term. The following Lemma \ref{lem:ContinuousOfB} and Proposition \ref{pro:widetildeB} demonstrate that the asymptotic properties of $\widetilde{B}$ are identical to those of $\widehat{B}$. As a result, the dominated convergence theorem ensures the convergence of the leading terms in \eqref{eq:bias_variance_conditional} to their corresponding terms in \eqref{eq:nonrecycling_bias_var_MSE}.

\begin{lemma}\label{lem:ContinuousOfB}
	Suppose that Assumption \ref{ass:derivative5} holds and assume that $f(\cdot)$ is third continuously differentiable and has a uniformly bounded third derivative. For any $k = 1,...,K (K \geq 2)$, denote $h_k = c_k n_b^{\gamma} (c_k \neq 0, \gamma<0)$ and for any $j \neq k$, let $c_j \neq c_k$. Then,
	\begin{align}
		\E\left[f\left(\widehat{B}\right)\right] &= f(B) + f'(B)H_Kn_b^{2\gamma} + o\left(n_b^{2\gamma}\right),\label{eq:applemmaexpectation}\\
		\Var\left[f\left(\widehat{B}\right)\right] = f(B)f''(B)H_K^2n_b^{4\gamma} + &\left(f'(B)^2 + f(B)f''(B)\right)V_K\frac{\sigma^2(\theta_0)}{2n_b^{1+6\gamma}} + o\left(n_b^{4\gamma}\right) + o\left(\frac{1}{n_b^{1+6\gamma}}\right),\label{eq:applemmavariance}
	\end{align}
	where $H_K$ and $V_K$ are defined as in Theorem \ref{thm:B_consistence}.
\end{lemma}

\begin{proof}

We prove \eqref{eq:applemmaexpectation} and \eqref{eq:applemmavariance}, respectively.

{\bf{Proof of \eqref{eq:applemmaexpectation}: }}
By Taylor expansion, we have
\begin{align*}
	f\left(\widehat{B}\right) = f(B) + f'(B)\left(\widehat{B} - B\right) + \frac{f''\left(\breve{B}\right)}{2}\left(\widehat{B} - B\right)^2,
\end{align*}
where $\breve{B}$ is a random variable between $B$ and $\widehat{B}$. Therefore,
\begin{align}\label{eq:applemmaexpectationproof}
	\E\left[f\left(\widehat{B}\right)\right] = f(B) + f'(B)\E\left[\widehat{B} - B\right] + o\left(n_b^{2\gamma}\right) = f(B) + f'(B)H_Kn_b^{2\gamma} + o\left(n_b^{2\gamma}\right),
\end{align}
where $H_K$ is defined as in Theorem \ref{thm:B_consistence}.

{\bf{Proof of \eqref{eq:applemmavariance}: }}
Note that 
\begin{align}\label{eq:applemmavarianceproof1}
	\Var\left[f\left(\widehat{B}\right)\right] = \E\left[f^2\left(\widehat{B}\right)\right] - \E^2\left[f\left(\widehat{B}\right)\right].
\end{align}
Our task now is to calculate $\E\left[f^2\left(\widehat{B}\right)\right]$. According to Taylor expansion, 
\begin{align*}
	f^2\left(\widehat{B}\right) =& f^2(B) + 2f(B)f'(B)\left(\widehat{B} - B\right) + \left(f'(B)^2 + f(B)f''(B)\right)\left(\widehat{B} - B\right)^2\\
	&+\frac{1}{3}\left(3f'\left(\mathring{B}\right)f''\left(\mathring{B}\right) + f\left(\mathring{B}\right)f'''\left(\mathring{B}\right)\right)\left(\widehat{B} - B\right)^3,
\end{align*}
where $\mathring{B}$ is a random variable between $B$ and $\widehat{B}$. Therefore,
\begin{align}\label{eq:applemmavarianceproof2}
	\E\left[f^2\left(\widehat{B}\right)\right] =& f^2(B) + 2f(B)f'(B)\E\left[\widehat{B} - B\right] + \left(f'(B)^2 + f(B)f''(B)\right)\E\left[\widehat{B} - B\right]^2\nonumber\\
	&+ o\left(n_b^{4\gamma}\right) + o\left(\frac{1}{n_b^{1+6\gamma}}\right),\nonumber\\
	=& f^2(B) + 2f(B)f'(B)H_Kn_b^{2\gamma} + \left(f'(B)^2 + f(B)f''(B)\right)\left(H_K^2n_b^{4\gamma} + V_K\frac{\sigma^2(\theta_0)}{2n_b^{1+6\gamma}}\right)\nonumber\\
	&+ o\left(n_b^{4\gamma}\right) + o\left(\frac{1}{n_b^{1+6\gamma}}\right),
\end{align}
where $V_K$ is defined as in Theorem \ref{thm:B_consistence}.

Combining \eqref{eq:applemmaexpectationproof}, \eqref{eq:applemmavarianceproof1} and \eqref{eq:applemmavarianceproof2} gives that
\begin{align*}
	\Var\left[f\left(\widehat{B}\right)\right] = f(B)f''(B)H_K^2n_b^{4\gamma} + &\left(f'(B)^2 + f(B)f''(B)\right)V_K\frac{\sigma^2(\theta_0)}{2n_b^{1+6\gamma}} + o\left(n_b^{4\gamma}\right) + o\left(\frac{1}{n_b^{1+6\gamma}}\right).
\end{align*}

The proof is complete.
\end{proof}

\begin{proposition}\label{pro:widetildeB}
	Suppose that Assumption \ref{ass:derivative5} holds and assume that $B$ is bounded below away from 0. For any $k = 1,...,K (K \geq 2)$, denote $h_k = c_k n_b^{\gamma} (c_k \neq 0, \gamma<0)$ and for any $j \neq k$, let $c_j \neq c_k$. For $0 < \epsilon < B$, define $\widetilde{B} = \epsilon + \max{\left\{\widehat{B} - \epsilon, 0\right\}}$. Then,
	\begin{align}\label{eq:widetildeB_bias_variance}
		\E\left[\widetilde{B}\right] - B = H_Kn_b^{2\gamma} + o\left(n_b^{2\gamma}\right),\quad \Var\left[\widetilde{B}\right] = V_K\frac{\sigma^2(\theta_0) + o(1)}{2n_b^{1+6\gamma}},
	\end{align}
	where $H_K$ and $V_K$ are defined as in Theorem \ref{thm:B_consistence}.
\end{proposition}

\begin{proof}

For a given $\delta > 0$, we construct a smooth approximation function of $f(x) = \epsilon + \max{\{x-\epsilon,0\}}$, denoted by $f_\delta(x)$, as follows:
\begin{align*}
	f_\delta(x) = \epsilon + \left(\frac{1}{2}(x+\delta-\epsilon) - \frac{\delta}{\pi}\cos\left(\frac{x-\epsilon}{2\delta}\pi\right)\right)\1_{\{\epsilon - \delta \leq x \leq \epsilon + \delta\}} + (x - \epsilon)\1_{\{x > \epsilon + \delta\}}.
\end{align*}
It is easily verified that for a fixed $\delta > 0$, $f_\delta(x)$ is third continuously differentiable and has a uniformly bounded third derivative. Specifically,
\begin{align*}
	f'_\delta(x) =& \left(\frac{1}{2} + \frac{1}{2}\sin\left(\frac{x-\epsilon}{2\delta}\pi\right)\right)\1_{\{\epsilon - \delta \leq x \leq \epsilon + \delta\}} + \1_{\{x > \epsilon + \delta\}},\\
	f''_\delta(x) =& \frac{\pi}{4\delta}\cos\left(\frac{x-\epsilon}{2\delta}\pi\right)\1_{\{\epsilon - \delta \leq x \leq \epsilon + \delta\}},\\
	f'''_\delta(x) =& -\frac{\pi^2}{8\delta^2}\sin\left(\frac{x-\epsilon}{2\delta}\pi\right)\1_{\{\epsilon - \delta \leq x \leq \epsilon + \delta\}}.
\end{align*}

Therefore, according Lemma \ref{lem:ContinuousOfB}, for a fixed $\delta > 0$, we have
\begin{align*}
		\E\left[f_\delta\left(\widehat{B}\right)\right] &= f_\delta(B) + f_\delta'(B)H_Kn_b^{2\gamma} + o\left(n_b^{2\gamma}\right),\\
		\Var\left[f_\delta\left(\widehat{B}\right)\right] = f_\delta(B)f_\delta''(B)H_K^2n_b^{4\gamma} + &\left(f_\delta'(B)^2 + f_\delta(B)f_\delta''(B)\right)V_K\frac{\sigma^2(\theta_0)}{2n_b^{1+6\gamma}} + o\left(n_b^{4\gamma}\right) + o\left(\frac{1}{n_b^{1+6\gamma}}\right),
	\end{align*}
where $H_K$ and $V_K$ are defined as in Theorem \ref{thm:B_consistence}. As $\delta \to 0$, by the dominated convergence theorem we get
\begin{align*}
	&\E\left[\widetilde{B}\right] = \E\left[\epsilon + \max{\left\{\widehat{B} - \epsilon, 0\right\}}\right] = \E\left[\underset{\delta \to 0}{\lim} f_\delta\left(\widehat{B}\right)\right] = \underset{\delta \to 0}{\lim}\E\left[f_\delta\left(\widehat{B}\right)\right]\\
	=& \underset{\delta \to 0}{\lim}f_\delta(B) + \underset{\delta \to 0}{\lim}f_\delta'(B)H_Kn_b^{2\gamma} + o\left(n_b^{2\gamma}\right)
	= \epsilon + (B - \epsilon)\1_{\{B > \epsilon\}} + \1_{\{B > \epsilon\}}H_Kn_b^{2\gamma} + o\left(n_b^{2\gamma}\right)\\
	=& B + H_Kn_b^{2\gamma} + o\left(n_b^{2\gamma}\right),
\end{align*}
and 
\begin{align}\label{eq:variancetildeBo}
	&\Var\left[\widetilde{B}\right] = \Var\left[\epsilon + \max{\left\{\widehat{B} - \epsilon, 0\right\}}\right] = \Var\left[\underset{\delta \to 0}{\lim} f_\delta\left(\widehat{B}\right)\right] = \underset{\delta \to 0}{\lim}\Var\left[f_\delta\left(\widehat{B}\right)\right]\nonumber\\
	=& \underset{\delta \to 0}{\lim}f_\delta'(B)^2V_K\frac{\sigma^2(\theta_0)}{2n_b^{1+6\gamma}} + o\left(n_b^{4\gamma}\right) + o\left(\frac{1}{n_b^{1+6\gamma}}\right)
	= \1_{\{B > \epsilon\}}V_K\frac{\sigma^2(\theta_0)}{2n_b^{1+6\gamma}} + o\left(n_b^{4\gamma}\right) + o\left(\frac{1}{n_b^{1+6\gamma}}\right)\nonumber\\
	=& V_K\frac{\sigma^2(\theta_0)}{2n_b^{1+6\gamma}} + o\left(n_b^{4\gamma}\right) + o\left(\frac{1}{n_b^{1+6\gamma}}\right),
\end{align}
where the fourth equality is since that $B > \epsilon$, there exists $\delta' > 0$ such that $B > \epsilon + \delta'$. Therefore, for any $\delta < \delta'$, $f_\delta''(B) = \dfrac{\pi}{4\delta}\cos\left(\dfrac{B-\epsilon}{2\delta}\pi\right)\1_{\{\epsilon - \delta \leq B \leq \epsilon + \delta\}} = 0$, i.e., $\underset{\delta \to 0}{\lim}f_\delta''(B) = 0$. Similarly, $\underset{\delta \to 0}{\lim}f_\delta'''(B) = 0$. As a result, it is obvious that $o\left(n_b^{4\gamma}\right)$ in \eqref{eq:variancetildeBo} is 0. Therefore,
\begin{align*}
	\Var\left[\widetilde{B}\right] = V_K\frac{\sigma^2(\theta_0)}{2n_b^{1+6\gamma}} + o\left(\frac{1}{n_b^{1+6\gamma}}\right).
\end{align*}

The proof is complete.
\end{proof}

\subsection{Proof of Theorem \ref{thm:DSR}} \label{app:prooftheorem4}

From \eqref{eq:finite_difference_estimate}, \eqref{eq:sample_transform} and \eqref{eq:DSR}, it follows that
	\begin{align*}
		\widehat{\Delta}_{n,\widehat{h}_n} = \frac{1}{n}\left(\sum_{i=1}^{n_2}\Delta_i\left(\widehat{h}_n\right) + n_b\left(\sum_{k=1}^{K}\frac{|h_k|}{\left|\widehat{h}_n\right|}\left(\widehat{\Delta}_{n_b,h_k} - \left(1,h_k^2\right)\widehat{\bbeta}_e\right) + K\left(1,{\widehat{h}_n}^2\right)\widehat{\bbeta}_e\right)\right),
	\end{align*}
	where $\widehat{\bbeta}_e = \left(\boldsymbol{X}_e^{\top} \boldsymbol{X}_e\right)^{-1} \boldsymbol{X}_e^{\top} \boldsymbol{Y}_e$. From \eqref{eq:proof_bootstrap_bias}, 
	\begin{align*}
		\E_{*}\widehat{\Delta}^{b}_{n_b,h} = \frac{1}{n_b}\sum_{i=1}^{n_b}\Delta_i(h) = \widehat{\Delta}_{n_b,h},
	\end{align*} 
	so $\widehat{\bbeta}_e = \left(\boldsymbol{X}_e^{\top} \boldsymbol{X}_e\right)^{-1} \boldsymbol{X}_e^{\top} \left[\widehat{\Delta}_{n_b,h_1},...,\widehat{\Delta}_{n_b,h_K}\right]^{\top}$ and we can represent $\widehat{\Delta}_{n,\widehat{h}_n}$ in matrix form
	\begin{align*}
	\widehat{\Delta}_{n,\widehat{h}_n} = \frac{1}{n}\left(\sum_{i=1}^{n_2}\Delta_i\left(\widehat{h}_n\right) + n_b\boldsymbol{h}^{\top}\boldsymbol{P}\boldsymbol{\Delta_h} +Kn_b\left(1,{\widehat{h}_n}^2\right)\widehat{\bbeta}_e\right),
	\end{align*}
	where ${\boldsymbol{h}} = [|h_1|,...,|h_K|]^{\top}$, ${\boldsymbol{\Delta_h}} = \left[\widehat{\Delta}_{n_b,h_1}\big/\left|\widehat{h}_n\right|,...,\widehat{\Delta}_{n_b,h_K}\big/\left|\widehat{h}_n\right|\right]^{\top}$ and ${\boldsymbol{P}}=\boldsymbol{I} - \boldsymbol{X}_e(\boldsymbol{X}_e^{\top}\boldsymbol{X}_e)^{-1}\boldsymbol{X}_e^{\top}$.
	Note that \eqref{eq:DSR_MSE} is a straightforward corollary of \eqref{eq:DSR_bias} and \eqref{eq:DSR_variance}, so we only have to prove \eqref{eq:DSR_bias} and \eqref{eq:DSR_variance} separately.

	${\boldsymbol{\mbox{Proof of \eqref{eq:DSR_bias}:}}}$
	According to the linearity of the expectation,  
	\begin{align}\label{eq:proof_DSR_bias}
		n \E\widehat{\Delta}_{n,\widehat{h}_n} = \E\left[\sum_{i=1}^{n_2}\Delta_i\left(\widehat{h}_n\right)\right] + Kn_b\E\left[\left(1,{\widehat{h}_n}^2\right)\widehat{\bbeta}_e\right] + n_b\E\left[\boldsymbol{h}^{\top}\boldsymbol{P}\boldsymbol{\Delta_h}\right].
	\end{align}
	We consider the three terms on the RHS of \eqref{eq:proof_DSR_bias} separately. Firstly, we study the first term on the RHS of \eqref{eq:proof_DSR_bias}.
	\begin{align}\label{eq:proof_DSR_bias_first}
		\mbox{The first term on the RHS of \eqref{eq:proof_DSR_bias}} =& \E\left[\E\left[\sum_{i=1}^{n_2}\Delta_i\left(\widehat{h}_n\right)\left|\widehat{h}_n\right.\right]\right]\nonumber\\
		=& \E\left[n_2\left(\alpha'(\theta_0) + B\widehat{h}_n^2 + o\left(\widehat{h}_n^2\right)\right)\right]\nonumber\\
		 =& n_2\left(\E\left[\alpha'(\theta_0) + B\widehat{h}_n^2\right] + o\left(n^{-1/3}\right)\right),
	\end{align}
	where the second equality is due to \eqref{eq:estimate_bias}. As for the third equality, we omit the lengthy discussion on the technical assumptions needed to ensure $\E\left[o\left(\widehat{h}_n^2\right)\right] = o\left(n^{-1/3}\right)$\textsuperscript{\ref{fn:uniformaly_integrable}}. In the following, we keep assuming that the operators $\E[\cdot]$ (or $\Var[\cdot]$) and $o(\cdot)$ (or $o_p(\cdot)$) can be exchanged.

	Secondly, we consider the second term on the RHS of \eqref{eq:proof_DSR_bias}.
	\begin{align}\label{eq:proof_DSR_bias_second}
		&\mbox{The second term on the RHS of \eqref{eq:proof_DSR_bias}}\nonumber\\
		=& Kn_b\E\left[\left(1,{\widehat{h}_n}^2\right){\bbeta}_e + \left(1,{\widehat{h}_n}^2\right)\left(\widehat{\bbeta}_e-{\bbeta}_e\right)\right]\nonumber\\
		=& Kn_b\E\left[\alpha'(\theta_0) + B\widehat{h}_n^2\right] + Kn_b\E\left[\widehat{\alpha}'(\theta_0) - \alpha'(\theta_0)\right] + Kn_b\E\left[\left(\widehat{B} - B\right)\widehat{h}_n^2\right]\nonumber\\
		=& Kn_b\E\left[\alpha'(\theta_0) + B\widehat{h}_n^2\right] + Kn_b\widetilde{H}_K n_b^{-2/5} + Kn_b\E\left[\left(\widehat{B} - B\right)\widehat{h}_n^2\right]\nonumber\\
		\leq & Kn_b\E\left[\alpha'(\theta_0) + B\widehat{h}_n^2\right] + Kn_b O\left(n_b^{-2/5}\right) + Kn_b\left(\E\left[\widehat{B} - B\right]^2\E\left[{\widehat{h}_n^4}\right]\right)^{1/2}\nonumber\\
		=& Kn_b\E\left[\alpha'(\theta_0) + B\widehat{h}_n^2\right] + Kn_b O\left(n_b^{-2/5}\right) + Kn_bO\left(n_b^{-1/5}\right)O\left(n^{-1/3}\right),
	\end{align}
	where the last two equalities are due to Corollary \ref{cor:alpha_consistence} and Theorem \ref{thm:B_consistence}, respectively, and the inequality is because of Cauchy-Schwarz inequality.

	Lastly, we study the third term on the RHS of \eqref{eq:proof_DSR_bias}.
	\begin{align}\label{eq:proof_DSR_bias_third}
		&\mbox{The third term on the RHS of \eqref{eq:proof_DSR_bias}} = n_b\boldsymbol{h}^{\top}\boldsymbol{P}\E\left[\boldsymbol{\Delta_h}\right]\nonumber\\
	= & n_bD\left(\frac{4nB^2}{\sigma^2(\theta_0)}\right)^{1/6}{\boldsymbol{h^{\top}Ph^4}} + o\left(n_b^{1/2}\right)O\left(n^{1/6}\right)\nonumber\\
	= & D\left(\frac{4B^2}{\sigma^2(\theta_0)}\right)^{1/6}{\boldsymbol{c^{\top}Pc^4}}n_b^{1/2} n^{1/6} + o\left(n_b^{1/2}\right)O\left(n^{1/6}\right),
	\end{align}
	where ${\boldsymbol{h}^4} = [h_1^4,...,h_K^4]^{\top}$, ${\boldsymbol{c}} = [|c_1|,...,|c_K|]^{\top}$ and ${\boldsymbol{c}^4} = [c_1^4,...,c_K^4]^{\top}$ and the second equality is because $\boldsymbol{P}$ is a projection matrix.\footnote{Specifically, for any $\boldsymbol{v} \in \mathcal{C}(\boldsymbol{X}_e)$, $\boldsymbol{P v = 0}$.} Note that here we omit the lengthy discussion to ensure that the remainder terms have the form $o\left(n_b^{1/2}\right)O\left(n^{1/6}\right)$.
	
	Combining \eqref{eq:proof_DSR_bias}, \eqref{eq:proof_DSR_bias_first}, \eqref{eq:proof_DSR_bias_second} and \eqref{eq:proof_DSR_bias_third} gives
	\begin{align*}
		n \E\widehat{\Delta}_{n,\widehat{h}_n} =& (n_2 + Kn_b)\E\left[\alpha'(\theta_0) + B\widehat{h}_n^2\right] + D\left(\frac{4B^2}{\sigma^2(\theta_0)}\right)^{1/6}{\boldsymbol{c^{\top}Pc^4}}n_b^{1/2} n^{1/6}\\
		&+ n_2o\left(n^{-1/3}\right) + Kn_b O\left(n_b^{-2/5}\right) + Kn_bO\left(n_b^{-1/5}\right)O\left(n^{-1/3}\right) + o\left(n_b^{1/2}\right)O\left(n^{1/6}\right).
	\end{align*}
	Therefore,
	\begin{align*}
		\E\widehat{\Delta}_{n,\widehat{h}_n} =& \E\left[\alpha'(\theta_0) + B\widehat{h}_n^2\right] + D\left(\frac{4B^2}{\sigma^2(\theta_0)}\right)^{1/6}{\boldsymbol{c^{\top}Pc^4}}n_b^{1/2} n^{-5/6} + o\left(n^{-1/3}\right)\\
		=&\alpha'(\theta_0) + \left(\frac{B\sigma^2(\theta_{0})}{4n}\right)^{1/3} + \left(\frac{4B^2}{\sigma^2(\theta_0)}\right)^{1/6}\sqrt{\frac{n_b}{n}}D{\boldsymbol{c^{\top}Pc^4}}n^{-1/3} + o\left(n^{-1/3}\right).
	\end{align*}

	${\boldsymbol{\mbox{Proof of \eqref{eq:DSR_variance}:}}}$
	The variance of $\widehat{\Delta}_{n_b,\widehat{h}_n}$ is written as
	\begin{align}\label{eq:proof_variance_form}
		\Var\left[\widehat{\Delta}_{n_b,\widehat{h}_n}\right] =& \frac{1}{n^2}\left(\Var\left[\sum_{i=1}^{n_2}\Delta_i\left(\widehat{h}_n\right)\right] + \Var\left[n_b\boldsymbol{h}^{\top}\boldsymbol{P}\boldsymbol{\Delta_h}\right] + \Var\left[Kn_b\left(1,{\widehat{h}_n}^2\right)\widehat{\bbeta}_e\right]\right.\nonumber\\
		& +2\Cov\left[\sum_{i=1}^{n_2}\Delta_i\left(\widehat{h}_n\right), n_b\boldsymbol{h}^{\top}\boldsymbol{P}\boldsymbol{\Delta_h}\right] + 2\Cov\left[\sum_{i=1}^{n_2}\Delta_i\left(\widehat{h}_n\right), Kn_b\left(1,{\widehat{h}_n}^2\right)\widehat{\bbeta}_e\right]\nonumber\\
		&\left.+2\Cov\left[n_b\boldsymbol{h}^{\top}\boldsymbol{P}\boldsymbol{\Delta_h}, Kn_b\left(1,{\widehat{h}_n}^2\right)\widehat{\bbeta}_e\right] \right).
	\end{align}
	
	We look at the six terms on the RHS of \eqref{eq:proof_variance_form} one by one.
	
	Firstly, we consider the first term on the RHS of \eqref{eq:proof_variance_form}.
	\begin{align}\label{eq:proof_variance_form_1}
	&\mbox{The first term on the RHS of \eqref{eq:proof_variance_form}}\nonumber\\ 
	=& \frac{1}{n^2}\Var\left[\E\left[\sum_{i=1}^{n_2}\Delta_i\left(\widehat{h}_n\right)\left|\widehat{h}_n\right.\right]\right] + \frac{1}{n^2}\E\left[\Var\left[\sum_{i=1}^{n_2}\Delta_i\left(\widehat{h}_n\right)\left|\widehat{h}_n\right.\right]\right]\nonumber \\
		=& \frac{1}{n^2}\Var\left[n_2\left(\alpha'(\theta_0) + B\widehat{h}_n^2 + o\left(\widehat{h}_n^2\right)\right)\right] + \frac{1}{n^2}\E\left[n_2\frac{\sigma^2(\theta_0) + o(1)}{2\widehat{h}_n^2}\right]\nonumber\\
		=& \frac{n_2^2}{n^2}\Var\left[B\widehat{h}_n^2 + o\left(\widehat{h}_n^2\right)\right] + \frac{1}{n^2}\E\left[n_2\frac{\sigma^2(\theta_0) + o(1)}{2\widehat{h}_n^2}\right]\nonumber\\
		=& \frac{n_2^2 o\left(n^{-2/3}\right)}{n^2} + \E\left[n_2\frac{\sigma^2(\theta_0) + o(1)}{2n^2\widehat{h}_n^2}\right]\nonumber\\
		=& \E\left[\frac{n_2\sigma^2(\theta_0)}{2n^2\widehat{h}_n ^2}\right] + o\left(n^{-2/3}\right),
\end{align}
where the second equality is due to \eqref{eq:estimate_bias} and \eqref{eq:estimate_variance} and the fourth equality is because
\begin{align*}
	n^{2/3}\Var\left[\widehat{h}_n^2\right] = \Var\left[\left(\frac{\widehat{\bbeta}_v}{4\widehat{B}^2}\right)^{1/3}\right] = \E\left[\frac{\widehat{\bbeta}_v}{4\widehat{B}^2}\right]^{2/3} - \E^2\left[\frac{\widehat{\bbeta}_v}{4\widehat{B}^2}\right]^{1/3} \to \frac{\bbeta_v^{2/3}}{(4B^2)^{2/3}} - \left(\frac{\bbeta_v^{1/3}}{(4B^2)^{1/3}}\right)^2 = 0
\end{align*}
implies $\Var\left[\widehat{h}_n^2\right] = o\left(n^{-2/3}\right)$.

Secondly, we study the second term on the RHS of \eqref{eq:proof_variance_form}.
\begin{align*}
	&\mbox{The second term on the RHS of \eqref{eq:proof_variance_form}}
	= \frac{n_b^2}{n^2}\boldsymbol{h}^{\top}\boldsymbol{P}\Var[\boldsymbol{\Delta_h}]\boldsymbol{Ph}\\
	=& \frac{n_b^2}{n^2}\boldsymbol{h}^{\top}\boldsymbol{P}\left[\begin{array}{ccc}
\Var\left[\frac{\widehat{\Delta}_{n_b,h_1}}{\left|\widehat{h}_n\right|}\right] & ... & \Cov\left[\frac{\widehat{\Delta}_{n_b,h_1}}{\left|\widehat{h}_n\right|},\frac{\widehat{\Delta}_{n_b,h_K}}{\left|\widehat{h}_n\right|}\right]\\
\vdots & & \vdots \\
\Cov\left[\frac{\widehat{\Delta}_{n_b,h_K}}{\left|\widehat{h}_n\right|},\frac{\widehat{\Delta}_{n_b,h_1}}{\left|\widehat{h}_n\right|}\right] & ... & \Var\left[\frac{\widehat{\Delta}_{n_b,h_K}}{\left|\widehat{h}_n\right|}\right]
\end{array}\right]\boldsymbol{Ph}.
 \end{align*}
 Note that
 \begin{align*}
	&\Cov\left[\frac{\widehat{\Delta}_{n_b,h_1}}{\left|\widehat{h}_n\right|},\frac{\widehat{\Delta}_{n_b,h_K}}{\left|\widehat{h}_n\right|}\right] = \Cov\left[\frac{\widehat{\Delta}_{n_b,h_1}}{h^*} + \widehat{\Delta}_{n_b,h_1}\left(\frac{1}{\left|\widehat{h}_n\right|} - \frac{1}{h^*}\right),\frac{\widehat{\Delta}_{n_b,h_K}}{h^*} + \widehat{\Delta}_{n_b,h_K}\left(\frac{1}{\left|\widehat{h}_n\right|} - \frac{1}{h^*}\right)\right]\\
	=& \Cov\left[\frac{\widehat{\Delta}_{n_b,h_1}}{h^*}, \frac{\widehat{\Delta}_{n_b,h_K}}{h^*}\right] + \Cov\left[\frac{\widehat{\Delta}_{n_b,h_1}}{h^*}, \widehat{\Delta}_{n_b,h_K}\left(\frac{1}{\left|\widehat{h}_n\right|} - \frac{1}{h^*}\right)\right] + \Cov\left[\widehat{\Delta}_{n_b,h_1}\left(\frac{1}{\left|\widehat{h}_n\right|} - \frac{1}{h^*}\right), \frac{\widehat{\Delta}_{n_b,h_K}}{h^*}\right]\\
	&+ \Cov\left[\widehat{\Delta}_{n_b,h_1}\left(\frac{1}{\left|\widehat{h}_n\right|} - \frac{1}{h^*}\right), \widehat{\Delta}_{n_b,h_K}\left(\frac{1}{\left|\widehat{h}_n\right|} - \frac{1}{h^*}\right)\right]\\
	=& \Cov\left[\frac{\widehat{\Delta}_{n_b,h_1}}{h^*}, \widehat{\Delta}_{n_b,h_K}\left(\frac{1}{\left|\widehat{h}_n\right|} - \frac{1}{h^*}\right)\right] + \Cov\left[\widehat{\Delta}_{n_b,h_1}\left(\frac{1}{\left|\widehat{h}_n\right|} - \frac{1}{h^*}\right), \frac{\widehat{\Delta}_{n_b,h_K}}{h^*}\right]\\
	&+ \Cov\left[\widehat{\Delta}_{n_b,h_1}\left(\frac{1}{\left|\widehat{h}_n\right|} - \frac{1}{h^*}\right), \widehat{\Delta}_{n_b,h_K}\left(\frac{1}{\left|\widehat{h}_n\right|} - \frac{1}{h^*}\right)\right]\\
	\leq & \sqrt{\Var\left[\frac{\widehat{\Delta}_{n_b,h_1}}{h^*}\right] \Var\left[\widehat{\Delta}_{n_b,h_K}\left(\frac{1}{\left|\widehat{h}_n\right|} - \frac{1}{h^*}\right)\right]} + \sqrt{\Var\left[\widehat{\Delta}_{n_b,h_1}\left(\frac{1}{\left|\widehat{h}_n\right|} - \frac{1}{h^*}\right)\right] \Var\left[\frac{\widehat{\Delta}_{n_b,h_K}}{h^*}\right]}\\
	&+ \sqrt{\Var\left[\widehat{\Delta}_{n_b,h_1}\left(\frac{1}{\left|\widehat{h}_n\right|} - \frac{1}{h^*}\right)\right] \Var\left[\widehat{\Delta}_{n_b,h_K}\left(\frac{1}{\left|\widehat{h}_n\right|} - \frac{1}{h^*}\right)\right]}.
\end{align*}

Because $n^{-1/6}\widehat{\Delta}_{n_b,h_k}\left(\frac{1}{\left|\widehat{h}_n\right|} - \frac{1}{h^*}\right) = o_p\left(\widehat{\Delta}_{n_b,h_k}\right)$, for any $k = 1,2,...,K$, $\Var\left[\widehat{\Delta}_{n_b,h_k}\left(\frac{1}{\left|\widehat{h}_n\right|} - \frac{1}{h^*}\right)\right] = n^{1/3}\Var\left[o_p\left(\widehat{\Delta}_{n_b,h_k}\right)\right] = o\left(n_b^{-4/5}\right)O\left(n^{1/3}\right)$. As a result, $\Cov\left[\frac{\widehat{\Delta}_{n_b,h_1}}{\left|\widehat{h}_n\right|},\frac{\widehat{\Delta}_{n_b,h_K}}{\left|\widehat{h}_n\right|}\right] = o\left(n_b^{-4/5}\right)O\left(n^{1/3}\right)$. Similarly, for any $k = 1,2,...,K$, \begin{align*}
	\Var\left[\frac{\widehat{\Delta}_{n_b,h_k}}{\left|\widehat{h}_n\right|}\right] =& \Cov\left[\frac{\widehat{\Delta}_{n_b,h_k}}{\left|\widehat{h}_n\right|},\frac{\widehat{\Delta}_{n_b,h_k}}{\left|\widehat{h}_n\right|}\right] = \Var\left[\frac{\widehat{\Delta}_{n_b,h_k}}{h^*}\right] + o\left(n_b^{-4/5}\right)O\left(n^{1/3}\right)\\
	 =&  \frac{\sigma^2(\theta_0)}{2n_bh_k^2}\left(\frac{4nB^2}{\sigma^2(\theta_0)}\right)^{1/3} + o\left(n_b^{-4/5}\right)O\left(n^{1/3}\right).
	\end{align*} 
Therefore,
\begin{align}
&\mbox{The second term on the RHS of \eqref{eq:proof_variance_form}}\nonumber\\
=& \frac{n_b^2}{n^2}\boldsymbol{h}^{\top}\boldsymbol{P}\left[\begin{array}{ccc}
\Var\left[\frac{\widehat{\Delta}_{n_b,h_1}}{\left|\widehat{h}_n\right|}\right] & & \\
 & \ddots & \\
 & & \Var\left[\frac{\widehat{\Delta}_{n_b,h_K}}{\left|\widehat{h}_n\right|}\right]
\end{array}\right]\boldsymbol{Ph} + \frac{n_b^2}{n^2}n_b^{1/10}o\left(n_b^{-4/5}\right)O\left(n^{1/3}\right)n_b^{1/10}\nonumber\\
= & \frac{n_b}{n^2}\left(\frac{nB^2\sigma^4(\theta_0)}{2}\right)^{1/3}\boldsymbol{h}^{\top}\boldsymbol{P}\left[\begin{array}{ccc}
\frac{1}{h_1^2} & & \\
 & \ddots & \\
 & & \frac{1}{h_K^2}
\end{array}\right]\boldsymbol{Ph} + o\left(n^{-2/3}\right)\nonumber\\
=& \frac{n_b}{n^2}\left(\frac{nB^2\sigma^4(\theta_0)}{2}\right)^{1/3}\left|\left|\left[\begin{array}{ccc}
\frac{1}{|h_1|} & & \\
 & \ddots & \\
 & & \frac{1}{|h_K|}
\end{array}\right]\boldsymbol{Ph}\right|\right|_2^2 + o\left(n^{-2/3}\right)\nonumber\\
=& \frac{n_b}{n^2}\left(\frac{nB^2\sigma^4(\theta_0)}{2}\right)^{1/3}\left|\left|{\rm{Diag}}(\boldsymbol{c^{-1}}) \boldsymbol{Pc}\right|\right|_2^2 + o\left(n^{-2/3}\right),\label{eq:proof_variance_form_2}
\end{align}
where ${\rm{Diag}}(\boldsymbol{c^{-1}}) = {\rm{Diag}}\left(\frac{1}{|c_1|},...,\frac{1}{|c_K|}\right)$ denotes the diagonal matrix and $||{\boldsymbol{v}}||_2 = \left(\sum_{k=1}^{K}v_k^2\right)^{1/2}$ for any ${\boldsymbol{v}} \in \R^K$.

Then, we study the third term on the RHS of \eqref{eq:proof_variance_form}.
\begin{align}\label{eq:proof_DSR_variance_3_0}
		&\mbox{The third term on the RHS of \eqref{eq:proof_variance_form}}
		 = \frac{K^2n_b^2}{n^2}\Var\left[\widehat{\alpha}'(\theta_0) + \widehat{B}\widehat{h}_n^2\right]\nonumber\\
		=& \frac{K^2n_b^2}{n^2}\left(\Var\left[\widehat{\alpha}'(\theta_0)\right] + \Var\left[\widehat{B}\widehat{h}_n^2\right] + 2\Cov\left[\widehat{\alpha}'(\theta_0),\widehat{B}\widehat{h}_n^2\right]\right)
	\end{align}
We study the three terms on the RHS of \eqref{eq:proof_DSR_variance_3_0} one by one. Firstly, according to Corollary \ref{cor:alpha_consistence}
	\begin{align}\label{eq:proof_DSR_variance_3_1}
		\mbox{The first term on the RHS of \eqref{eq:proof_DSR_variance_3_0}} = \frac{K^2n_b^2}{n^2}\widetilde{H}_K n_b^{-4/5} = O\left(n_b^{6/5}\right)O\left(n^{-2}\right) = o\left(n^{-2/3}\right).
	\end{align}

	Secondly, we study the second term on the RHS of \eqref{eq:proof_DSR_variance_3_0}.
	\begin{align}\label{eq:proof_DSR_variance_3_2}
		&\mbox{The second term on the RHS of \eqref{eq:proof_DSR_variance_3_0}}\nonumber\\
		=& \frac{K^2n_b^2}{n^2}\Var\left[B\widehat{h}_n^2 + \left(\widehat{B} - B\right)\widehat{h}_n^2\right]\nonumber\\
		  =& \frac{K^2n_b^2}{n^2}\left(\Var\left[B\widehat{h}_n^2\right] + \Var\left[\left(\widehat{B} - B\right)\widehat{h}_n^2\right] + 2\Cov\left[B\widehat{h}_n^2, \left(\widehat{B} - B\right)\widehat{h}_n^2\right]\right)\nonumber\\
		  \leq & \frac{K^2n_b^2}{n^2}\left(\Var\left[B\widehat{h}_n^2\right] + \E\left[\left(\widehat{B} - B\right)^2\widehat{h}_n^4\right] + 2\Cov\left[B\widehat{h}_n^2, \left(\widehat{B} - B\right)\widehat{h}_n^2\right]\right)\nonumber\\
		  \leq & \frac{K^2n_b^2}{n^2}\left(\Var\left[B\widehat{h}_n^2\right] + \left(\E\left[\widehat{B} - B\right]^4\E\left[\widehat{h}_n^8\right]\right)^{1/2} + 2\Cov\left[B\widehat{h}_n^2, \left(\widehat{B} - B\right)\widehat{h}_n^2\right]\right)\nonumber\\
		  = & \frac{K^2n_b^2}{n^2}\left(n^{-2/3}\Var\left[B\left(\frac{\widehat{\bbeta}_v}{4\widehat{B}^2}\right)^{1/3}\right] + O\left(n_b^{-2/5}\right)O\left(n^{-2/3}\right) + 2\Cov\left[B\widehat{h}_n^2, \left(\widehat{B} - B\right)\widehat{h}_n^2\right]\right)\nonumber\\
		  = & \frac{K^2n_b^2}{n^2}\left(o\left(n^{-2/3}\right) + O\left(n_b^{-2/5}\right)O\left(n^{-2/3}\right)\right)\nonumber\\
		 =& O\left(n_b^{2}\right)o\left(n^{-8/3}\right) +  O\left(n_b^{8/5}\right)O\left(n^{-8/3}\right)\nonumber\\
		 =& o\left(n^{-2/3}\right),
	\end{align}
	where the third equality is because Theorem \ref{thm:B_consistence} implies that $\E\left[\widehat{B} - B\right]^4 = O\left(n_b^{-4/5}\right)$ and the fourth equality is because the covariance term is controlled by the two variance terms.
	
	According to Cauchy-Schwarz inequality,
	\begin{align}\label{eq:proof_DSR_variance_3_3}
		&\mbox{The third term on the RHS of \eqref{eq:proof_DSR_variance_3_0}}\nonumber\\
		\leq &2\sqrt{\mbox{The first term on the RHS of \eqref{eq:proof_DSR_variance_3_0}} \times \mbox{The second term on the RHS of \eqref{eq:proof_DSR_variance_3_0}}}\nonumber\\
		=& \sqrt{o\left(n^{-2/3}\right)o\left(n^{-2/3}\right)} = o\left(n^{-2/3}\right).
	\end{align}
	
	Combining \eqref{eq:proof_DSR_variance_3_0}, \eqref{eq:proof_DSR_variance_3_1}, \eqref{eq:proof_DSR_variance_3_2} and \eqref{eq:proof_DSR_variance_3_3} gives that
	\begin{align}\label{eq:proof_variance_form_3}
		\mbox{The third term on the RHS of \eqref{eq:proof_variance_form}} = o\left(n^{-2/3}\right).
	\end{align}

Next, we consider the fourth term on the RHS of \eqref{eq:proof_variance_form}. 
\begin{align*}
	&\mbox{The fourth term on the RHS of \eqref{eq:proof_variance_form}}\\
	=& \frac{2}{n^2}\left\{\E\left[\Cov\left[\sum_{i=1}^{n_2}\Delta_i\left(\widehat{h}_n\right), n_b {\boldsymbol{h}^{\top}\boldsymbol{P}\boldsymbol{\Delta_h}}\left|\widehat{h}_n,{\boldsymbol{\Delta_h}}\right.\right]\right]\right.\\
	&+ \left.\Cov\left[\E\left[\sum_{i=1}^{n_2}\Delta_i\left(\widehat{h}_n\right)\left|\widehat{h}_n,{\boldsymbol{\Delta_h}}\right.\right], \E\left[n_b {\boldsymbol{h}^{\top}\boldsymbol{P}\boldsymbol{\Delta_h}}\left|\widehat{h}_n,{\boldsymbol{\Delta_h}}\right.\right]\right]\right\}\\
	=& \frac{2}{n^2}\Cov\left[n_2 \left(\alpha'(\theta_0) + B\widehat{h}_n^2 + o_p\left(n^{-1/3}\right)\right), n_b {\boldsymbol{h}^{\top}\boldsymbol{P}\boldsymbol{\Delta_h}}\right]\\
	=& \frac{2}{n^2}\Cov\left[n_2 \left(B\widehat{h}_n^2 + o_p\left(n^{-1/3}\right)\right), n_b {\boldsymbol{h}^{\top}\boldsymbol{P}\boldsymbol{\Delta_h}}\right]\\
	=& \frac{2n_2n_b}{n^2}B{\boldsymbol{h}^{\top}\boldsymbol{P}}\Cov\left[\widehat{h}_n^2, \boldsymbol{\Delta_h}\right] + \mbox{remainder terms}\\
	=& \frac{2n_2n_b^{9/10}}{n^2}B{\boldsymbol{c}^{\top}\boldsymbol{P}}\Cov\left[\widehat{h}_n^2, \boldsymbol{\Delta_h}\right] + \mbox{remainder terms},
\end{align*}
where the first equality is due to the law of total covariance. The second equality is because when ${\boldsymbol{\Delta_h}}$ is given, $n_b {\boldsymbol{h}^{\top}\boldsymbol{P}\boldsymbol{\Delta_h}}$ is a constant and $\Cov\left[\sum_{i=1}^{n_2}\Delta_i\left(\widehat{h}_n\right), n_b {\boldsymbol{h}^{\top}\boldsymbol{P}\boldsymbol{\Delta_h}}\right] = 0$. According to the Cauchy-Schwarz inequality, 
\begin{align*}
	\Cov\left[\widehat{h}_n^2,\frac{\widehat{\Delta}_{n_b,h_1}}{\left|\widehat{h}_n\right|}\right] \leq \sqrt{\Var\left[\widehat{h}_n^2\right]\Var\left[\frac{\widehat{\Delta}_{n_b,h_1}}{\left|\widehat{h}_n\right|}\right]} = \sqrt{o\left(n^{-2/3}\right)O\left(n_b^{-4/5}n^{1/3}\right)} = o\left(n^{-1/6}\right)O\left(n_b^{-2/5}\right).
\end{align*}
Therefore, 
\begin{align}\label{eq:proof_variance_form_4}
	\mbox{The fourth term on the RHS of \eqref{eq:proof_variance_form}} = \frac{O(n_2)O\left(n_b^{1/2}\right)o(1)}{O\left(n^{13/6}\right)} = o\left(n^{-2/3}\right).
\end{align}

By Cauchy-Schwarz inequality, we have
\begin{align}\label{eq:proof_variance_form_5}
	&\mbox{The fifth term on the RHS of \eqref{eq:proof_variance_form}}\nonumber\\
	\leq & 2\sqrt{\mbox{The first term on the RHS of \eqref{eq:proof_variance_form}} \times \mbox{The third term on the RHS of \eqref{eq:proof_variance_form}}}
	= o\left(n^{-2/3}\right)
\end{align}
and 
\begin{align}\label{eq:proof_variance_form_6}
	&\mbox{The sixth term on the RHS of \eqref{eq:proof_variance_form}}\nonumber\\
	\leq & 2\sqrt{\mbox{The second term on the RHS of \eqref{eq:proof_variance_form}} \times \mbox{The third term on the RHS of \eqref{eq:proof_variance_form}}}
	= o\left(n^{-2/3}\right).
\end{align}

Combining \eqref{eq:proof_variance_form}, \eqref{eq:proof_variance_form_1}, \eqref{eq:proof_variance_form_2}, \eqref{eq:proof_variance_form_3}, \eqref{eq:proof_variance_form_4}, \eqref{eq:proof_variance_form_5} and \eqref{eq:proof_variance_form_6} gives
\begin{align*}
	\Var\left[\widehat{\Delta}_{n,\widehat{h}_n}\right] =& \left(\frac{B^2 \sigma^4(\theta_0)}{2n^2}\right)^{1/3} + \left(\frac{B^2 \sigma^4(\theta_0)}{2}\right)^{1/3}\frac{n_b}{n}\left[\left|\left|{\rm{Diag}}(\boldsymbol{c^{-1}})\boldsymbol{Pc}\right|\right|_2^2 - K\right]n^{-2/3} + o\left(n^{-2/3}\right).
\end{align*}

The proof is complete.

\subsection{Remarks on Theorem \ref{thm:DSR} and Proposition \ref{prop:DSR}} \label{app:projection_matrix}

Denote $\mathcal{C}(\boldsymbol{X}_e)$ the column space of $\boldsymbol{X}_e$. Then $\boldsymbol{P}$ is a projection matrix onto $\mathcal{C}(\boldsymbol{X}_e)^{\bot}$ which is the orthogonal complement space of $\mathcal{C}(\boldsymbol{X}_e)$, i.e., $\mathcal{C}(\boldsymbol{X}_e) \oplus \mathcal{C}(\boldsymbol{X}_e)^{\bot} = \R^{K}$. Therefore, for any ${\boldsymbol{v}} \in \R^K$, we can decompose it as ${\boldsymbol{v}} = {\boldsymbol{v}}_1 + {\boldsymbol{v}}_2$, where ${\boldsymbol{v}}_1 \in \mathcal{C}(\boldsymbol{X}_e)$ and ${\boldsymbol{v}}_2 \in \mathcal{C}(\boldsymbol{X}_e)^{\bot}$. Subsequently, applying $\boldsymbol{P}$ to ${\boldsymbol{v}}$ results in $\boldsymbol{P}{\boldsymbol{v}} = {\boldsymbol{v}}_2$. If the angle between ${\boldsymbol{v}}$ and $\mathcal{C}(\boldsymbol{X}_e)$ is small, $||\boldsymbol{P}{\boldsymbol{v}}||_2$, which is equal to $||{\boldsymbol{v}}||_2$, is also small. 

According to the property of projection matrix and Cauchy-Schwarz inequality, $|{\boldsymbol{c^{\top}Pc^4}}| = |{\boldsymbol{c^{\top}P^{\top}Pc^4}}| \leq ||{\boldsymbol{Pc}}||_2 ||{\boldsymbol{Pc^4}}||_2$. If the angles between $\boldsymbol{c}$, $\boldsymbol{c^4}$ and $\mathcal{C}(\boldsymbol{X}_e)$ are small, which are typically the case, both $||{\boldsymbol{Pc}}||_2$ and $||{\boldsymbol{Pc^4}}||_2$ are small, then $|{\boldsymbol{c^{\top}Pc^4}}|$ is manageable.

For example, if ${\boldsymbol{c}} = [1,1.5,2,...,5.5]^{\top}$, then ${\boldsymbol{c}^4} = [1,1.5^4,2^4,...,5.5^4]^{\top}$. Denote the angles between ${\boldsymbol{c}}$, ${\boldsymbol{c}^4}$ and $\mathcal{C}({\boldsymbol{X}})$ by $\vartheta$ and $\widetilde{\vartheta}$, respectively. Then, we can easily calculate that $\cos\vartheta \geq 0.9693$ and $\cos\widetilde{\vartheta} \geq 0.9595$, both of which are close to 1. In other words, both $\vartheta$ and $\widetilde{\vartheta}$ are close to 0. 

\subsection{Proof of Proposition \ref{prop:DSR}} \label{app:proofcorDSR}
It is obvious that \eqref{eq:thmDSR_case1} holds when $n_b = o(n)$. In the following, we consider the bias and variance terms of cases 2 and 3 (i.e., $Kn_b/n \to r$, where $0 < r \leq 1$) in Proposition \ref{prop:DSR} separately.
\begin{itemize}
	\item {$\E\left[\widehat{\Delta}_{n,\widehat{h}_n}\right] - \alpha'(\theta_0)$:}
	\begin{itemize}
	\item [i.] If we have
		\begin{align*}
			-1 - {\rm{sign}}(B) \leq \sqrt{\dfrac{r}{K}}\dfrac{2}{\sigma(\theta_0)}D \Lambda  \leq 1 - {\rm{sign}}(B),
		\end{align*}
		then, 
		\begin{align*}
			&\left(\E\left[\widehat{\Delta}_{n,\widehat{h}_n}\right] - \alpha'(\theta_0)\right)^2\\
			 =& \left(\frac{B\sigma^2(\theta_{0})}{4n}\right)^{2/3} + \left(\frac{4B^2}{\sigma^2(\theta_0)}\right)^{1/3}\sqrt{\frac{n_b}{n}}D \Lambda \left(\sqrt{\frac{n_b}{n}}D \Lambda  + {\rm{sign}}(B)\sigma(\theta_0)\right)\\
			 =& \left(\frac{B\sigma^2(\theta_{0})}{4n}\right)^{2/3} + \left(\frac{4B^2}{\sigma^2(\theta_0)}\right)^{1/3}\sqrt{\frac{r}{K}}D \Lambda \left(\sqrt{\frac{r}{K}}D \Lambda  + {\rm{sign}}(B)\sigma(\theta_0)\right)\\
			 \leq & \left(\frac{B\sigma^2(\theta_{0})}{4n}\right)^{2/3}.
		\end{align*}
		Therefore, 
		\begin{align*}
				\left|\E\widehat{\Delta}_{n,\widehat{h}_n} - \alpha'(\theta_0)\right| \leq \left(\dfrac{B^2\sigma^4(\theta_{0})}{16n^2}\right)^{1/6} + o\left(n^{-1/3}\right).
			\end{align*}
		\item[ii.] Else, from \eqref{eq:DSR_bias} in Theorem \ref{thm:DSR}, we have
			\begin{align*}
				\left|\E\widehat{\Delta}_{n,\widehat{h}_n} - \alpha'(\theta_0)\right| = &\left(\frac{B\sigma^2(\theta_{0})}{4n}\right)^{1/3} + \left(\frac{4B^2}{\sigma^2(\theta_0)}\right)^{1/6}\sqrt{\frac{n_b}{n}}D \Lambda n^{-1/3} + o\left(n^{-1/3}\right)\\
				\leq &\left(\frac{B^2\sigma^4(\theta_{0})}{16n^2}\right)^{1/6} + \left(\frac{4B^2}{\sigma^2(\theta_0)}\right)^{1/6}\sqrt{\frac{r}{K}}\left|D \Lambda \right|n^{-1/3} + o\left(n^{-1/3}\right) \\
				 \leq &\left(1 + \sqrt{\dfrac{r}{K}}\dfrac{2}{\sigma(\theta_0)}\left|D \Lambda \right|\right)\left(\dfrac{B^2\sigma^4(\theta_{0})}{16n^2}\right)^{1/6} + o\left(n^{-1/3}\right).
			\end{align*}
		\end{itemize}
	\item {$\Var\left[\widehat{\Delta}_{n,\widehat{h}_n}\right]$:} From Theorem \ref{thm:DSR}, it suffices to prove $\left|\left|{\rm{Diag}}(\boldsymbol{c^{-1}})\boldsymbol{Pc}\right|\right|_2^2 \leq K$.
	\begin{align*}
	&\left|\left|{\rm{Diag}}(\boldsymbol{c^{-1}})\boldsymbol{Pc}\right|\right|_2^2 \leq \frac{1}{\min_{k\in \{1,...,K\}}c_k^2}\left|\left|\boldsymbol{Pc}\right|\right|_2^2 = \frac{\min_{x,y}\left|\left|x\boldsymbol{1} + y\boldsymbol{c^2} - \boldsymbol{c}\right|\right|_2^2}{\min_{k\in \{1,...,K\}}c_k^2} \\
\leq & \frac{K\left|\left|\boldsymbol{c}\right|\right|_2^2 - \left|\left|\boldsymbol{c}\right|\right|_1^2}{K\min_{k\in \{1,...,K\}}c_k^2} \leq \frac{K\max_{k\in \{1,...,K\}}c_k^2 - K\min_{k\in \{1,...,K\}}c_k^2}{\min_{k\in \{1,...,K\}}c_k^2} \leq K,
\end{align*}
where the first inequality is because that for any matrix ${\boldsymbol{A}}$ and vector ${\boldsymbol{v}}$, $||{\boldsymbol{A} v}||_2 \leq ||{\boldsymbol{A}}||_2 ||{\boldsymbol{v}}||_2$, the first equality is due to the property of projection matrix, the second inequality holds by taking $x = \left|\left|\boldsymbol{c}\right|\right|_1/K$ and $y=0$ and the last inequality is attributed to $\max_{k\in \{1,...,K\}}c_k^2 \leq 2\min_{k\in \{1,...,K\}}c_k^2$.
\end{itemize}

The proof is complete.

\subsection{The Bias and Variance of $\widehat{\Delta}_{n,\widehat{h}_n}^{new}$}\label{app:corn_MSE_reduction}

In this section, we provide the asymptotic properties of $\widehat{\Delta}_{n,\widehat{h}_n}^{new}$ as follows.
\begin{proposition}\label{pro:corn_MSE_reduced}
	Under the same conditions as those of Theorem \ref{thm:DSR}, we have
	\begin{align}
		&\E\widehat{\Delta}_{n,\widehat{h}_n}^{new} = \alpha'(\theta_0) + \left(\dfrac{B\sigma^2(\theta_{0})}{4n}\right)^{1/3} + o\left(n^{-1/3}\right),\label{eq:pro_DSR_bias}\\
		&\Var\left[\widehat{\Delta}_{n,\widehat{h}_n}^{new}\right] = \left(\dfrac{B^2 \sigma^4(\theta_0)}{2n^2}\right)^{1/3} + o\left(n^{-2/3}\right).\label{eq:pro_DSR_var}
	\end{align}
\end{proposition}

\begin{proof}

Recall that
	\begin{align}\label{eq:pro_DSR}
		\widehat{\Delta}_{n,\widehat{h}_n}^{new} = \frac{1}{n}\left(\sum_{i=1}^{n_2}\Delta_i\left(\widehat{h}_n\right) + n_b\sum_{k=1}^{K}\frac{|h_k|}{\left|\widehat{h}_n\right|}\left(\widehat{\Delta}_{n_b,h_k} - \E\Delta(h_k)\right) + Kn_b\left(1,{\widehat{h}_n}^2\right)\widehat{\bbeta}_e\right),
	\end{align}
and we prove \eqref{eq:pro_DSR_bias} and \eqref{eq:pro_DSR_var} separately.

${\boldsymbol{\mbox{Proof of \eqref{eq:pro_DSR_bias}:}}}$ According to Appendix \ref{app:prooftheorem4}, to prove \eqref{eq:pro_DSR_bias}, it suffices to prove the expectation of the second term on the RHS of \eqref{eq:pro_DSR} is $o\left(n^{-1/3}\right)$. The proof is as follows.
\begin{align*}
	&\E\left[\frac{n_b}{n}\sum_{k=1}^{K}\frac{|h_k|}{\left|\widehat{h}_n\right|}\left(\widehat{\Delta}_{n_b,h_k} - \E\Delta(h_k)\right)\right]
	 = \frac{n_b}{n}\sum_{k=1}^{K}|h_k|\E\left[\frac{\widehat{\Delta}_{n_b,h_k} - \E\Delta(h_k)}{\left|\widehat{h}_n\right|}\right]\\
	=& \frac{n_b}{n}\sum_{k=1}^{K}|h_k|\left(\E\left[\frac{\widehat{\Delta}_{n_b,h_k}}{\left|\widehat{h}_n\right|}\right] - \E\Delta(h_k)\E\left[\frac{1}{\left|\widehat{h}_n\right|}\right]\right)\\
	=& \frac{n_b}{n}\left(\frac{4nB^2}{\sigma^2(\theta_0)}\right)^{1/6}\sum_{k=1}^{K}|h_k|\left(\alpha'(\theta_0) + Bh_k^2 + Dh_k^4 - \E\Delta(h_k)\right) + o\left(n_b^{1/2}\right)O\left(n^{-5/6}\right)\\
	=& o\left(n_b^{1/2}\right)O\left(n^{-5/6}\right) = o\left(n^{-1/3}\right).
\end{align*}

${\boldsymbol{\mbox{Proof of \eqref{eq:pro_DSR_var}:}}}$ Note that
\begin{align}\label{eq:proof_pro_DSR_var}
	\widehat{\Delta}_{n,\widehat{h}_n}^{new} = \frac{1}{n}\left(\sum_{i=1}^{n_2}\Delta_i\left(\widehat{h}_n\right) + n_b\boldsymbol{h}^{\top}\boldsymbol{I}\boldsymbol{\Delta_h} +Kn_b\left(1,{\widehat{h}_n}^2\right)\widehat{\bbeta}_e - n_b\sum_{k=1}^{K}\frac{|h_k|}{\left|\widehat{h}_n\right|}\E\Delta(h_k)\right).
\end{align}

It is straightforward to show that the variance of the last term on the RHS of \eqref{eq:proof_pro_DSR_var} is $o\left(n^{-2/3}\right)$, so the essence is calculating the variance of the first three terms on the RHS of \eqref{eq:proof_pro_DSR_var}. Compared to $\widehat{\Delta}_{n,\widehat{h}_n}$ in Appendix \ref{app:prooftheorem4}, the first three terms of \eqref{eq:proof_pro_DSR_var} simply replace the matrix $\boldsymbol{P}$ with the identity matrix $\boldsymbol{I}$. Therefore, following the technique used in Appendix \ref{app:prooftheorem4}, we get
\begin{align*}
	&\Var\left[\frac{1}{n}\left(\sum_{i=1}^{n_2}\Delta_i\left(\widehat{h}_n\right) + n_b\boldsymbol{h}^{\top}\boldsymbol{I}\boldsymbol{\Delta_h} +Kn_b\left(1,{\widehat{h}_n}^2\right)\widehat{\bbeta}_e\right)\right]\\
	=& \left(\frac{B^2 \sigma^4(\theta_0)}{2n^2}\right)^{1/3} + \left(\frac{B^2 \sigma^4(\theta_0)}{2}\right)^{1/3}\frac{n_b}{n}\left[\left|\left|{\rm{Diag}}(\boldsymbol{c^{-1}})\boldsymbol{Ic}\right|\right|_2^2 - K\right]n^{-2/3} + o\left(n^{-2/3}\right)\\
	=& \left(\frac{B^2 \sigma^4(\theta_0)}{2n^2}\right)^{1/3} + o\left(n^{-2/3}\right).
\end{align*}
Using the Cauchy-Schwarz inequality, we have
\begin{align*}
	\Var\left[\widehat{\Delta}_{n,\widehat{h}_n}^{new}\right] = \left(\frac{B^2 \sigma^4(\theta_0)}{2n^2}\right)^{1/3} + o\left(n^{-2/3}\right).
\end{align*}

The proof is complete. 
\end{proof}

\section{Additional Experiment Results in Section \ref{sec:experiments}} \label{app:addition_experiments}

\subsection{Additional Results for Example \ref{exa:ksinx}}\label{app:addition_experiments1}

\begin{figure}[t]
	\centering
	\caption{Illustration of the MSE of different methods w.r.t. $r$ for cases 1 and 2 in Example \ref{exa:ksinx}.}
	\vspace{7pt}
	\hspace*{-0.2cm}
	\includegraphics[scale = 0.4]{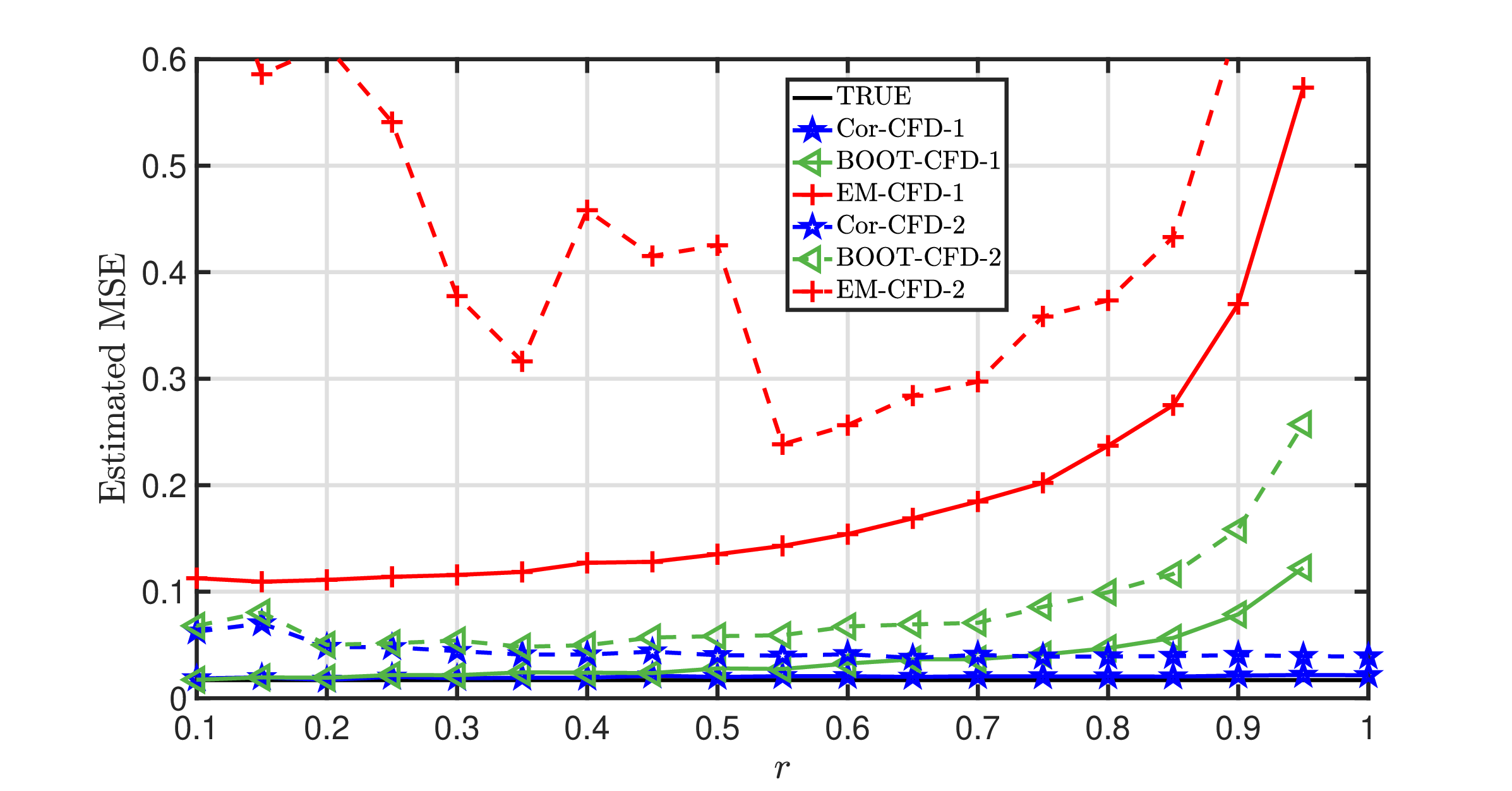}
	\label{fig:example1_robust_r}
\end{figure}

In Figure \ref{fig:example1_robust_r}, we depict the sensitivity of different methods w.r.t. $r$, where the suffixes ``1'' and ``2'' denote cases 1 and 2, respectively, in Example \ref{exa:ksinx}. We compare the performance of Cor-CFD, EM-CFD and BOOT-CFD (bootstrap-based CFD, i.e., the estimator $\widehat{\Delta}_{n_2,\widehat{h}_{n_2}}$ discussed at the beginning of Section \ref{sec:DSR}). We set the number of sample pairs to 1000. When examining the EM-CFD method, we generate $h_1, \ldots, h_{1000r} \stackrel{i.i.d.}{\sim} \mathcal{N}\left(0,1\times (1000r)^{-1/5}\right)$.

As observed in Figure \ref{fig:example1_robust_r}, the presence of heteroskedasticity complicates the estimation problem for each method, resulting in inferior estimations. However, the impact of heteroskedasticity is less pronounced in the Cor-CFD method, which consistently outperforms other methods across various scenarios. The EM-CFD method, on the other hand, exhibits less stability when $r$ is small, possibly due to less accurate perturbation estimation. Additionally, the Cor-CFD method demonstrates stability w.r.t. $r$, while the performance of EM-CFD and BOOT-CFD deteriorate as $r$ increases. Even when $r=1$, the Cor-CFD method maintains strong performance, making it particularly suitable for small sample sizes, where all samples are utilized for both constant and gradient estimation, effectively leveraging the available sample information.
 
\subsection{Additional Results for Example \ref{exa:poly}}\label{app:addition_experiments2}

We compare the estimated MSEs of our proposed Cor-CFD estimator, the EM-CFD estimator and the Opt-CFD estimator in Example \ref{exa:poly} by setting $\theta_0 = 0$ and varying the sample pairs from $n = 50$ to $n = 500$. When considering the EM-CFD algorithm, we set $r=0.1$ and generate $h_1,...,h_{rn} \stackrel{i.i.d.}{\sim} \mathcal{N}\left(0,(r n)^{-1/5}\right)$. When applying the Cor-CFD algorithm, we set $K = 10$ and $r = 1$.

\begin{figure}[t]
	\centering
	\caption{Variation of the estimated MSE by different methods with increasing sample size in Example \ref{exa:poly} $(\theta_0 = 0)$.}
	\vspace{7pt}
	\hspace*{-0.5cm}
	\includegraphics[scale = 0.58]{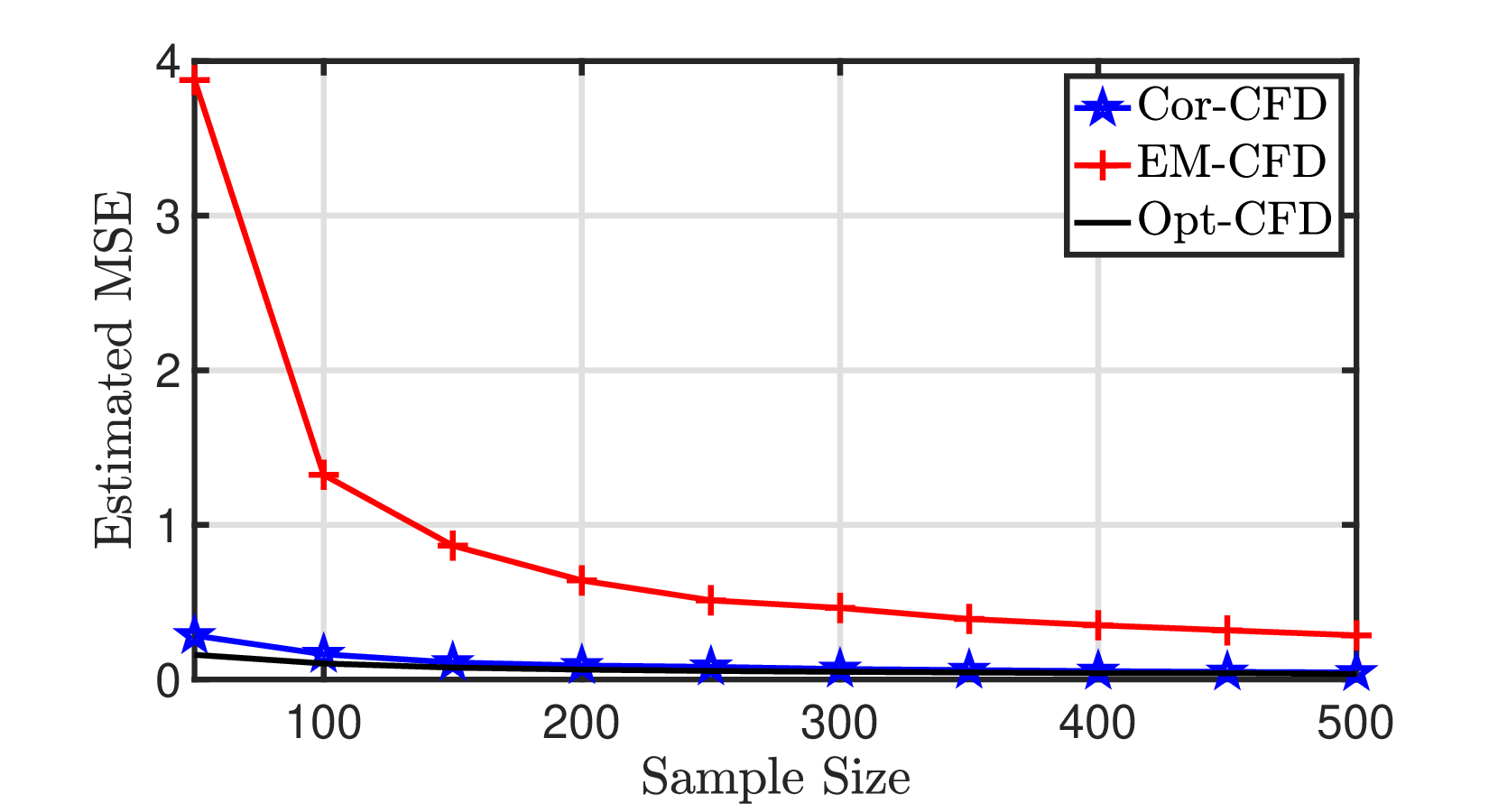}
	\label{fig:example2}
\end{figure}

Figure \ref{fig:example2} shows the comparison results. The figure illustrates that the estimated MSEs of the three estimators decrease gradually as the sample size increases. However, compared to the Opt-CFD and Cor-CFD estimators, the EM-CFD estimator exhibits a larger MSE. Moreover, the MSE of the Cor-CFD estimator is consistently close to that of the Opt-CFD estimator. For example, when the sample-pair size is 50, the MSE of the EM-CFD estimator is approximately 3.878, while the MSEs of the Cor-CFD and Opt-CFD estimators are approximately 0.284 and 0.160, respectively. When the sample-pair size is 500, the three MSEs are 0.286, 0.045 and 0.035, respectively.

\section{Simulation Settings and Additional Experiment Results in Section \ref{sec:DFO}}\label{app:setting_AdditionalResults}

\subsection{Simulation Settings in Section \ref{sec:DFO}}\label{app:setting}

The simulation settings of different methods in Section \ref{sec:DFO} are detailed as follows:

\begin{enumerate}
	\item {The CorCFD-L-BFGS algorithm.} We set the perturbation numbers $K = 5$, the initial batch size $T_0 = 20$, the proportion of the pilot samples $r = 1$, the bootstrap numbers $I = 100$, the line search parameters $(l_1, l_2) = (10^{-4}, 0.5)$, the initial step length $a_0 = 1$ and the initial perturbation generator $\mathcal{P}_0$ is identical to that in Example \ref{exa:poly}.
	
	\item {The TraCFD-L-BFGS algorithm.} We set $B = 1$ and $\sigma(\theta) = 1$ for any $\theta$ in the feasible domain, the line search parameters $(l_1, l_2) = (10^{-4}, 0.5)$ and the initial step length $a_0 = 1$.
	
	\item {The NEWUOA.} Similar to the operations in \cite{shi2023numerical}, we utilize the default parameters in R package ``\texttt{minqa}'' and repeat the NEWUOA until exhausting the allocated budget.  
	
\end{enumerate}

\subsection{Discussion on the Comparison with TraCFD-L-BFGS}\label{app:AdditionalResults1} 

\begin{figure}[t]
	\centering
	\caption{Comparison of the iteration process for $d = 10$.}
	\vspace{7pt}
	\hspace*{-0.5cm}
	\includegraphics[scale = 0.42]{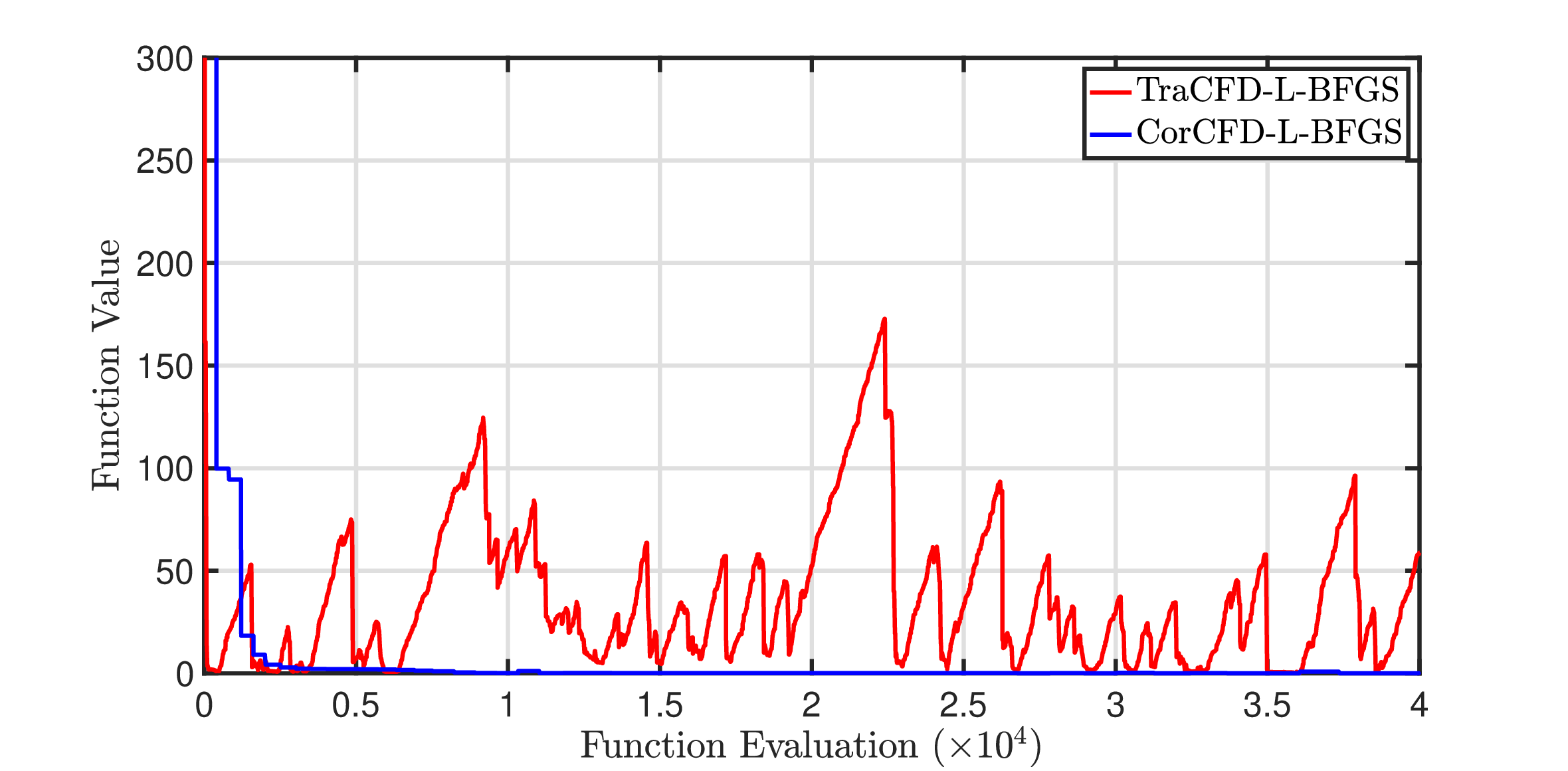}
	\label{fig:iteration}
\end{figure}

Figure \ref{fig:iteration} illustrates the iteration processes of the TraCFD-L-BFGS and CorCFD-L-BFGS algorithms in optimizing the Zahkarov function during a specific replication for $d = 10$ and $n = 10^4$. The TraCFD-L-BFGS algorithm use only one sample pair in each iteration, resulting in a significantly higher number of iterations compared to the CorCFD-L-BFGS algorithm. As shown in the figure, initially, the TraCFD-L-BFGS algorithm's function value decreases rapidly, while the CorCFD-L-BFGS algorithm's function value decreases more gradually. This is because, at the beginning, when the initial value is far from the optimum, gradient descent is more critical than precise gradient estimation. However, as iterations progress, the function value of the TraCFD-L-BFGS algorithm starts to fluctuate, whereas that of the CorCFD-L-BFGS algorithm steadily converges. Near the optimum, the true gradient value is small, making noise more significant. At this stage, obtaining an accurate gradient estimate becomes crucial for the algorithm to find the correct descent direction. Therefore, using more samples to reduce variance is more effective.

\subsection{Discussion on the Comparison with the SPSA Algorithm}\label{app:AdditionalResultSPSA}

We further evaluate the performance of the SPSA algorithm \citep{Spall1992Multivariate} using the function ``\texttt{minimizeSPSA}'' in Python library ``noisyopt''. The SPSA algorithm updates in the $k$th iteration as follows: $x_{k+1} = x_k + a_k [f(x_k + c_k \xi_k) - f(x_k - c_k \xi_k)]/c_k$, where $\xi_k$ denotes a random direction, $a_k = a (0.01*niter +k +1)^{-\alpha_0}$, $c_k = c (k+1)^{-\gamma_0}$ and $niter$ is the number of total iterations. In \texttt{minimizeSPSA}, $\alpha_0 = 0.602$ and $\gamma_0 = 0.101$, as suggested in \citep{Spall1992Multivariate}. In our experiment, $niter$ is the number of sample pairs. We manually tune the hyperparameters $a$ and $c$. 

To strike a balance between dimensionality and the complexity of hyperparameter tuning, we focus on the case where $d = 10$. We explore 20 different combinations of $a$ and $c$ with $a = \{10^{-7}, 5\times 10^{-7}, 10^{-6}, 5\times 10^{-6}, 10^{-5}\}$ and $c = \{10^{-2}, 10^{-1}, 1, 10\}$. Our findings indicate that $a = 10^{-6}$ and $c = 1$ yield the best results among all combinations. For sample-pair sizes of $10^3\times d$, $10^4\times d$ and $10^5\times d$, we observe the SGs of 1.513, 1.458 and 1.421, respectively, and the OGs of 6.581, 2.278 and 2.024, respectively. Comparing these results with those presented in Table \ref{tab:Zakharov}, we surprisingly find that our proposed algorithm significantly outperforms the SPSA algorithm. We posit several potential explanations for this observation:

\begin{enumerate}

	\item Our algorithm stems from the L-BFGS, which is a second-order algorithm, while the SPSA algorithm corresponds to a first-order algorithm. Typically, the second-order algorithm performs better than first-order one.
	
	\item Another reason is the same as that discussed in Section \ref{app:AdditionalResults1}. That is, the gradient approximated by one sample pair is less accurate than that approximated by a batch of sample pairs.
	
\end{enumerate}

It is important to note that the SPSA algorithm was originally proposed as a sample-efficient method for multi-dimensional optimization problems. However, our comparative analysis reveals an intriguing area for further investigation: the relative efficacy of using a single sample pair (as in the SPSA algorithm) versus multiple sample pairs (as in our CorCFD-L-BFGS algorithm) for gradient approximation. This discrepancy in performance suggests a valuable avenue for future research, which could potentially yield significant insights into the strengths and limitations of these approaches across diverse optimization scenarios. Such investigations may contribute to a more nuanced understanding of when and how to apply these methods most effectively, particularly in high-dimensional optimization problems.

\end{document}